# Second International EtaMesonNet Workshop

# ETA07

May 10-11, 2007, Peñiscola, Spain

**Summary of contributions**

# 1. Introduction 7

# 2. Short summaries of the talks

### I. η and η' decays, meson structure and η mass measurement

The following talks were also presented at the workshop but are missing short summary. The original presentations can be found on the EtaMesonNet webpage (http://www.isv.uu.se/etamesonnet/).

# Introduction

This workshop is recognized as part of a European Community Integrated Infrastructure Initiative devoted to Hadron Physics (I3HP). This initiative contains several activities, one of them being the network EtaMesonNet, which is created to exchange information on experimental and theoretical ongoing activities on η physics at different European accelerator facilities and institutes, and is thereby promoting the hadron infrastructure in Europe. The present workshop included the four themes

  i. η and η' decays, meson structure and η mass measurement.
  ii. Meson photo production.
  iii. Interaction of η and η' with nucleons and nuclei.
  iv. Meson production.

The different talks covered the very recent achievements in each field from the experimental facilities KLOE at DAPHNE, Crystal Ball at MAMI and WASA at COSY as well as from different theory institutes. Viewgraphs from each talk can be found on the EtaMesonNet webpage http://www.isv.uu.se/etamesonnet/.

The detailed program was arranged by a program committee consisting of:
Reinhard Beck, Caterina Bloise, Christoph Hanhart, Bo Höistad, Pawel Moskal, Eulogio Oset and Michael Ostrick

The workshop was held in May 10-11, 2007, at the resort Peniscola in Spain, enjoying kind hospitality and support from Universidad de Valencia.

The financial support from the European Community-Research Infrastructure Activity under FP6 "Structuring the European Research Area" program (Hadron Physics, contract number RII3-CT-2004-506078), is gratefully acknowledged.

*Bo Höistad and Marek Jacewicz*

# Session I:

*η and η' decays, meson structure and η mass measurement*

# Decays of $\eta$ and $\eta'$

B. Borasoy

Helmholtz-Institut für Strahlen- und Kernphysik (Theorie), Universität Bonn,
Nußallee 14-16, D-53115 Bonn, Germany

Decays of $\eta$ and $\eta'$ are investigated within a U(3) chiral unitary approach based on the chiral effective Lagrangian which includes the $\eta'$ as an explicit degree of freedom and incorporates important features of the underlying QCD Lagrangian such as the axial U(1) anomaly and explicit chiral symmetry breaking due to non-vanishing quark masses. Final state interactions are included by deriving the effective *s*- and *p*-wave potentials for meson-meson scattering from the chiral Lagrangian and iterating them in a Bethe-Salpeter equation. Resonances are generated dynamically within this approach and their importance for these decays can be investigated. This approach has been successfully applied both to the hadronic and the anomalous $\eta, \eta'$ decays [1-6]. Very good agreement with experimental data is achieved. These investigations are relevant for experiments performed at WASA at COSY, MAMI-C, KLOE at DaΦne and by the VES collaboration.

In my talk, I have focused on the hadronic decay modes of the $\eta$ and $\eta'$. First, it is discussed whether the quark mass ratio $(m_d - m_u)/m_s$ can easily be extracted from the measured decay width ratio $r = \Gamma(\eta' \to \pi^\circ \pi^+ \pi^-)/\Gamma(\eta' \to \eta \pi^+ \pi^-)$ as claimed in [7]. Our findings indicate that a determination of $r$ cannot be achieved in a simple manner and that final state interactions play an important role, particularly in $\eta' \to \pi^\circ \pi^+ \pi^-$.

Second, the inclusion of the recent VES data on the $\eta' \to \eta \pi^+ \pi^-$ spectral shape [8] into our global analysis of hadronic $\eta$ and $\eta'$ decays has important consequences also for the other hadronic decay modes. In this case, we find in $\eta' \to \pi^\circ \pi^+ \pi^-$ a strong coupling to $\rho^\pm(770)$ (the contributions from the $\rho^\circ(770)$ vanish due to *C*-invariance). This results in a rather large decay width of about 3.1 keV for $\eta' \to \pi^\circ \pi^+ \pi^-$ to be compared with the much smaller decay width $\Gamma(\eta' \to 3\pi^\circ) = 330$ eV of the neutral decay mode.

# Radiative $\eta$ and $\varphi$ decays

E. Oset[1], J. R. Pelaez[2], **L. Roca**[3], J. E. Palomar[1] and M. J. Vicente-Vacas[1]

[1] *Departamento de Física Teórica and IFIC Centro Mixto Universidad de Valencia-CSIC Institutos de Investigación de Paterna, Apdo. correos 22085, 46071, Valencia, Spain*

[2] *Departamento de Física Teórica II, Universidad Complutense. 28040 Madrid, Spain.*

[3] *Departamento de Física. Universidad de Murcia. E-30071, Murcia. Spain.*

In this contribution we summarize the work done in the last years on the $\eta \to \pi^0 \gamma\gamma$ and $\varphi \to PP\gamma$ decays by using the techniques of the chiral unitary approach.

Regarding the $\eta \to \pi^0 \gamma\gamma$ decay [1], it has attracted much theoretical attention since, in Chiral Perturbation Theory (ChPT), the calculation at $O(p^2)$ vanishes and at $O(p^4)$ is very small. The first sizeable contribution comes at $O(p^6)$. This makes this reaction to be, in principle, a good test of ChPT at $O(p^6)$ and higher. But the parameters involved in the ChPT Lagrangians at these higher orders are not easy to determine. On the other hand models using Vector Meson Dominance (VMD) have been used, but expanding the vector meson propagator to obtain the $O(p^6)$ chiral coefficients gives results $\sim 1/2$ of those obtained by keeping the full vector meson propagator which is supposed to contain higher orders. This implies that $O(p^8) > O(p^6)$. This makes us to conclude that a strict chiral counting has to be abandoned and we have to think in terms of 'relevant mechanisms' instead of 'diagrams contributing up to a certain order'. At this point is where unitary extensions of ChPT (UChPT) (for a review of UChPT see [2]) can help to solve the problem. UChPT is based on the implementation of unitary techniques to resum loops and has allowed to extend the predictions of ChPT up to $\sim 1.2\,GeV$ even generating many resonances dynamically without including them explicitly. Considering this background, we improve the calculations of the $\eta \to \pi^0 \gamma\gamma$ decay within the context of meson chiral lagrangians. We use a chiral unitary approach for the meson-meson interaction, thus generating the $a_0(980)$ resonance and fixing the longstanding sign ambiguity on its contribution. This also allows us to calculate the loops with one vector meson exchange, thus removing a former source of uncertainty. In addition we ensure the consistency of the approach with other processes. First, by using vector meson dominance couplings normalized to agree with radiative vector meson decays. And, second, by checking the consistency of the calculations with the related $\gamma\gamma \to \pi^0 \eta$ reaction. We find an $\eta \to \pi^0 \gamma\gamma$ decay width of $0.47 \pm 0.10$ eV, in remarkable agreement with the experimental measurement $0.45 \pm 0.12$ eV but in disagreement with the preliminary measurement $0.11 \pm 0.04$ eV [4].

Regarding the radiative decays of vector mesons into two pseudoscalars $V \to PP\gamma$, and particularly $\varphi \to \pi^0 \pi^0 \gamma$ and $\varphi \to \pi^0 \eta \gamma$, are a good testing ground to investigate the interesting underlying hadron dynamics in the region of $\sim 1\,GeV$, where the standard Chiral Perturbation Theory (ChPT) has no predictive power. Specially relevant is the important role played by the controversial scalar mesons, $f_0(980)$ in $\varphi \to \pi^0\pi^0\gamma$ and $a_0(980)$ in $\varphi \to \pi^0\eta\gamma$. In several works these resonances have been explicitly included using Linear Sigma Models with some ambiguities in the parameters and neglecting other relevant mechanisms that we show to be relevant in the present work. The unitary extensions of Chiral Perturbation Theory (UChPT) have manifested themselves as a very powerful tool to extend the predictions of ChPT to higher energies, even the region of low lying resonances, generating dynamically many of them by implementing the final state interaction in the meson-meson system. In the present work [1] we use the model of [6], where the unitary extensions of ChPT were used to resum the kaon loops which generate dynamically the $f_0(980)$ and $a_0(980)$ resonances, and we add other contributions, not considered previously, which have turned out to be relevant in these decays. Sequential vector meson exchange mechanisms, involving $\varphi \to \rho\pi^0, \rho \to \pi^0\gamma$ are considered, but in addition the analogous mechanisms, not OZI forbidden, leading to $K\bar{K}\gamma$, are considered, with the $K\bar{K}$ system turning into $\pi^0\pi^0$ through final state interaction. Similar mechanisms involving intermediate $K_1(1270)$ and $K_1(1400)$ axial vector mesons, not OZI forbidden, are considered in addition. These latter terms are calculated using a phenomenological Lagrangian [7] for the axial-pseudoscalar-vector vertex which makes good predictions for a large number of decays of the axial vectors. The new mechanisms had not been previously considered neither in former theoretical works nor in the data analysis of the experiments. One of the main features of our model is that there are no free parameters needed to be fitted to the final experimental results. All the used parameters are easily obtained from data published in the Particle Data Table, and their errors are carefully implemented in our model. The branching ratio obtained in the present work is $BR(\varphi \to \pi^0\pi^0\gamma) = (1.2 \pm 0.3) \times 10^{-4}$ to be compared with the most recent experimental value $(1.09 \pm 0.03 \pm 0.05) \times 10^{-4}$ [8]. For the $\varphi \to \pi^0\eta\gamma$ decay, the branching ratio obtained is $BR(\varphi \to \pi^0\eta\gamma) = (0.6 \pm 0.2) \times 10^{-4}$ to be compared with the most recent experimental value $(0.85 \pm 0.05 \pm 0.06) \times 10^{-4}$ [9].

# Dalitz plot analysis of the $\eta \to \pi^0 \pi^0 \pi^0$ and $\eta \to \pi^+ \pi^- \pi^0$ decays with KLOE.

The KLOE Collaboration[1] presented by F. Ambrosino

Dipartimento di Scienze Fisiche, Universitá degli Studi Federico II, e Sezione INFN Napoli, Italia

The decay of isoscalar $\eta$ into three pions occurs, besides a negligible electromagnetic contribution of second order [1], through isospin violation. As a consequence the decay amplitude $A(\eta \to 3\pi)$ is proportional to the $d-u$ quark mass difference; one can write:

$$A(s,t,u) = \frac{1}{Q^2} \frac{m_K^2}{m_\pi^2} \left( m_\pi^2 - m_K^2 \right) \frac{M(s,t,u)}{3\sqrt{3} F_\pi^2} \quad (1)$$

with the pion decay constant $F_\pi$ and:

$$\frac{1}{Q^2} \equiv \frac{m_d^2 - m_u^2}{m_s^2 - \frac{1}{4}(m_d + m_u)^2} \quad (2)$$

$M(s,t,u)$ is a dimensionless factor that involves exclusively measurable quantities[2].

Having a good theoretical prediction for $M(s,t,u)$, the quark mass ratio Q can be calculated from the decay rate: $\Gamma\left(\eta \to \pi^+ \pi^- \pi^0\right) \propto |A|^2 \propto Q^{-4}$. Using 17 millions $\eta$ mesons produced in 2001–2002, the dynamics of both $\pi^+ \pi^- \pi^0$ and $\pi^0 \pi^0 \pi^0$ final states has been studied through a Dalitz plot analysis. The $\eta$ mesons are clearly tagged by detecting the monochromatic recoil photon of the $\varphi \to \eta\gamma$ decay ($E_{\gamma_{rec}} \sim 363$ MeV); the background is at level of few per mill for both channels.

For the $\pi^+ \pi^- \pi^0$ final state, we choose the conventional Dalitz variables $X \propto T_+ - T_-$ and $Y = (3T_0 / Q - 1)$, where T is the kinetic energy of the pion and $Q = m_\eta - 2m_{\pi^+} - m_{\pi^0}$. The efficiency as function of Dalitz plot variables, is almost flat all over the kinematically allowed region and its mean value is about 33%. The measured distribution is parametrized as:

[1] F. Ambrosino, A. Antonelli, M. Antonelli, F. Archilli, C. Bacci, P. Beltrame, G. Bencivenni, S. Bertolucci, C. Bini, C. Bloise, S. Bocchetta, V. Bocci, F. Bossi, P. Branchini, R. Caloi, P. Campana, G. Capon, T. Capussela, F. Ceradini, S. Chi, G. Chiefari, P. Ciambrone, E. De Lucia, A. De Santis, P. De Simone, G. De Zorzi, A. Denig, A. Di Domenico, C. Di Donato, S. Di Falco, B. Di Micco, A. Doria, M. Dreucci, G. Felici, A. Ferrari, M. L. Ferrer, G. Finocchiaro, S. Fiore, C. Forti, P. Franzini, C. Gatti, P. Gauzzi, S. Giovannella, E. Gorini, E. Graziani, M. Incagli, W. Kluge, V. Kulikov, F. Lacava, G. Lanfranchi, J. Lee-Franzini, D. Leone, M. Martini, P. Massarotti, W. Mei, S. Meola, S. Miscetti, M. Moulson, S. M¨uller, F. Murtas, M. Napolitano, F. Nguyen, M. Palutan, E. Pasqualucci, A. Passeri, V. Patera, F. Perfetto, M. Primavera, P. Santangelo, G. Saracino, B. Sciascia, A. Sciubba, F. Scuri, I. Sfiligoi, T. Spadaro, M. Testa, L. Tortora, P. Valente, B. Valeriani, G. Venanzoni, R.Versaci, G. Xu

$$|A(X,Y)|^2 \simeq 1 + aY + bY^2 + cX + dX^2 + eXY + fY^3 \qquad (3)$$

the fit results are reported in table1.

| a | b |
|---|---|
| $-1.090 \pm 0.005 ^{+0.008}_{-0.019}$ | $0.124 \pm 0.006 ^{+0.010}_{-0.010}$ |
| c | d |
| $0.002 \pm 0.003 ^{+0.001}_{-0.001}$ | $0.057 \pm 0.006 ^{+0.007}_{-0.016}$ |
| e | f |
| $-0.006 \pm 0.007 ^{+0.005}_{-0.003}$ | $0.14 \pm 0.01 ^{+0.02}_{-0.02}$ |

*Table 1. Results for the slope parameter of Dalitz-plot.*

As expected from C-invariance in $\eta \to \pi^+ \pi^- \pi^0$ decay, the odd powers of $X$ are consistent with zero and can be removed from the fit without affecting the determination of the remaining parameters. We clearly observe a non zero quadratic slope in $X$, and we reach for the first time sensitivity to a cubic term of the expansion; all the cubic terms other than $f$ turn out to be zero in our fit. The $\chi^2$ probability of fit is 73%. The systematics errors take into account the efficiency evaluation, the resolution effects and the background contamination.

While the polynomial fit of the Dalitz plot density gives valuable information on the matrix element, some integrated asymmetries are very sensitive in assessing the possible contributions to C violation in amplitudes with fixed $\Delta I$. In table 2 the results for the left-right, quadrants and sextant asymmetries (for a definition see ref. [3]) are reported.

| Left-Right | $(0.09 \pm 0.10 ^{+0.09}_{-0.14}) \times 10^{-2}$ |
|---|---|
| Quadrant | $(-0.05 \pm 0.10 ^{+0.03}_{-0.05}) \times 10^{-2}$ |
| Sextant | $(0.08 \pm 0.10 ^{+0.08}_{-0.13}) \times 10^{-2}$ |

*Table 2. Results for the Dalitz plot asymmetries*

For the $\eta \to \pi^0 \pi^0 \pi^0$ decay the Dalitz plot distribution is described by a single quadratic slope parameter $\alpha$: $|A_{\eta \to 3\pi^0}(z)|^2 \sim 1 + 2\alpha z$. Here $z = \frac{\rho^2}{\rho^2_{MAX}}$ is the square ratio of the distance of a point from the Dalitz plot center ($\rho$) to the maximum kinematically allowed distance ($\rho_{MAX}$). Photons are paired to $\pi^0$'s after kinematically constraining the total 4-momentum, thus improving the energy resolution; as a second step a fit constraining also the $\pi^0$ mass is performed in order to further improve the resolution. In order to estimate $\alpha$ an unbinned likelihood function is built by convoluting the event density with the resolution function and correcting for the probability of wrong photon pairing in $\pi^0$'s. Using a sample with high purity on pairing (98%) and fitting in the range (0 - 1), we found in the past a preliminary result [4] only marginally compatible with the precise Crystal Ball [6] one. Further systematics checks have shown an unexpectedly strong dependence of the result on the numerical value of the $\eta$ meson mass used in the MonteCarlo generator.

In fact when using our official MonteCarlo production we observed a relevant dependence of $\alpha$ on the fitting range. This nonlinearity can be observed even pictorially in a low purity/ high statistics sample (see fig.1) in the Data–MC ratio of the $z$ distribution, while it was not evident in the high purity/low statistics sample used for the preliminary result quoted.

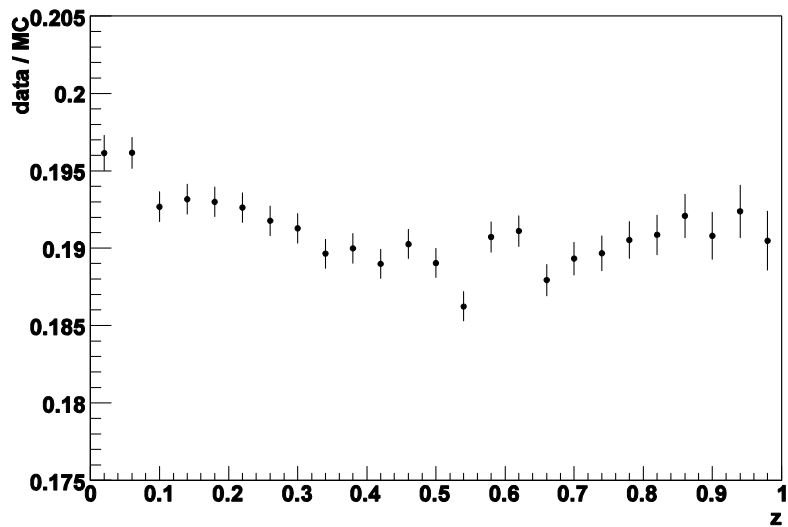

*Figure 1. Data–Monte Carlo ratio of the $z$ distribution. The MC distribution is pure phase space.*

With a dedicated simulation we have demonstrated that this nonlinearity is a kinematical effect due to a different value of invariant mass of three pions system ($\pi^0 \pi^0 \pi^0$) in the Monte Carlo generator, $M_\eta = 547.30$ MeV/c$^2$, with respect to the one recently measured [5] by our experiment:

$$M_\eta = 547.822 \pm 0.005_{stat} \pm 0.069_{syst} \text{ MeV}/c^2. \tag{4}$$

As a consequence, the accessible phase space on data is larger than the one on Monte Carlo simulation. While this effect is partially recovered if one constraints in a kinematic fit the three pions to give the same invariant mass value used in the MonteCarlo, this procedure was not used in our original fit. After applying a kinematic fit with an additional constraint on the correct $\eta$ mass, and fitting in the region $0 \leq z \leq 0.7$ we obtain the new preliminary result:

$$\alpha = -0.027 \pm 0.004\,(stat)^{+0.004}_{-0.006}\,(syst) \tag{5}$$

This results superseeds the preliminary result given in [4]. We notice that the using the correct mass we have recovered the linearity of the Data–MC ratio of the z distribution, see fig.2.

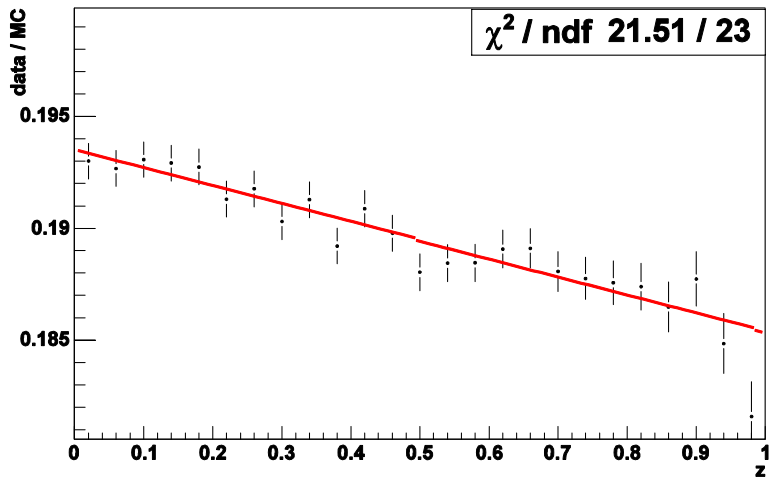

*Figure 2. Data–Monte Carlo ratio of the z distribution after the kinematic fit with $\eta$ mass constraint.*

# Recent results on the $\eta'$ photoproduction yield at MAMI-C using the Crystal Ball. Update on the analysis of the decays $\eta \to 3\pi^0$ and $\eta \to \pi^0 \gamma\gamma$

A. Starostin [*]

University of California, Los Angeles, CA 90095-1547, USA

A new high–statistics measurement of neutral $\eta'$ decays is underway at the upgraded Mainz Microtron Facility, MAMI-C. The principal decays $\eta' \to \eta\pi^0\pi^0$, and $\eta' \to 3\pi^0$ are tests of ChPT. The Dalitz plots of both $\eta'$ decays provide information on the $\pi\pi$, and $\pi\eta$ scattering lengths. The expected statistics is an order of magnitude larger than the world data. We will also test $C$–invariance by improving the upper limits on the $C$–forbidden decays $\eta' \to \pi^0 e^+ e^-$, $\eta' \to \eta e^+ e^-$, and $\eta' \to 3\gamma$. We will test $CP$–invariance by searching for $\eta' \to 4\pi^0$ [1].

The first experimental run with the new 1.507 $GeV$ electron beam has been conducted at MAMI-C in March 2007. Preliminary results indicate that the $\gamma p \to \eta' p$ events produced near the reaction threshold can be effectively detected by the combined Crystal Ball/TAPS experimental setup. All neutral $\eta'$ decays were measured simultaneously. The $\eta'$ events were observed as two–photon events ($\eta' \to \gamma\gamma$), six–photon events ($\eta' \to \eta\pi^0\pi^0$ followed by $\eta \to \gamma\gamma$), and ten–photon events ($\eta' \to \eta\pi^0\pi^0$ followed by $\eta \to 3\pi^0$). We expect about 800 hours worth of $\eta/\eta'$ data to be collected in 2007. The analysis of the currently taken data is in progress.

We continue analyzing the $\eta$ sample collected at MAMI-B in 2004-2005. That experiment was partially triggered by a statistically significant discrepancy between the early Crystal Ball results from BNL/AGS [2,3], and the results obtained by the KLOE collaboration for the decays $\eta \to 3\pi^0$ and $\eta \to \pi^0 \gamma\gamma$ [4]. Our latest developments in the $\eta \to \pi^0 \gamma\gamma$ analysis indicate a dependence of the experimental acceptance from the value of the $\gamma\gamma$ invariant mass. Currently we are investigating this effect with the calculation of the acceptance using different assumptions for the matrix element. The CB@MAMI results for the slope parameter of the $\eta \to 3\pi^0$ Dalitz plot have been presented at the previous EtaNet meeting in Mainz in 2006. Our results for the slope parameter have not changed.

[*] Email: starost@ucla.edu

# Observation of the η →π⁺π⁻e⁺e⁻ decay at KLOE

KLOE collaboration[2] presented by R. Versaci[a]

[a]INFN Laboratori Nazionali di Frascati, via Enrico Fermi 40, 00044, Rome, Italy

The study of the η→ π⁺π⁻e⁺e⁻ decay, besides the comparison among different theoretical models, could show an hypothetical CP violation. We observe ~300 events using 1/4 of the whole KLOE statistics

Apart from the comparison of different theoretical models [1–3], the $\eta \rightarrow \pi^+\pi^-e^+e^-$ decay is interesting as test of CP violation. A measurement of the asymmetry of the distribution of the angle between $\pi^+\pi^-$ and $e^+e^-$ production planes, performed with a precision higher than $10^{-2}$, will provide a stringent constraint for the CP violation mechanism [4].

The KLOE detector operates at DAΦNE, an e⁺e⁻ collider working at $\sqrt{s}$ = 1020 MeV, the mass of the φ(1020)-meson. The detector consists of a large cylindrical drift chamber (DC) surrounded by a lead/scintillating-fiber electromagnetic calorimeter (EMC). For a more detailed description refer to [5–7].

KLOE has collected about 2 fb$^{-1}$ in 2004 and 2005. Concerning the $\eta \rightarrow \pi^+\pi^-e^+e^-$ analysis, 579 pb$^{-1}$ of data have been analyzed. The Montecarlo statistics used is about 618 pb$^{-1}$ of all φ decays and about 46 fb$^{-1}$ of $\eta \rightarrow \pi^+\pi^-e^+e^-$. More MC is going to be produced and whole available data sample is going to be processed.

At KLOE the η meson is produced through the radiative decay $\varphi \rightarrow \eta\gamma$ which has a branching ratio of 1.3%. The recoil photon is monochromatic ($E_\gamma$ = 363 MeV). This feature is used for the event selection which requires 1 high energy neutral (without any track associated to it) cluster ($E_{cl} \geq$ 250 MeV) and no medium energy cluster (50 < $E_{cl}$ < 250 MeV) fired in the EMC. At least four tracks coming from a cylinder around the interaction point (IP) are also required. The cylinder has R = 4 cm and h/2 = 10 cm. A control on broken tracks has been applied. If two tracks have $\Delta p_T$ <4.5 MeV $p_z$ <3 MeV they are considered as produced by the same particle and only the one with the point of closest approach closer to the IP is taken into account. It is also required to have at least two tracks for each charge.

---

[2] F. Ambrosino, A. Antonelli, M. Antonelli, C. Bacci, P. Beltrame, G. Bencivenni, S. Bertolucci, C. Bini, C. Bloise, S. Bocchetta, V. Bocci, F. Bossi, P. Branchini, R. Caloi, P. Campana, G. Capon, T. Capussela, F. Ceradini, S. Chi, G. Chiefari, P. Ciambrone, E. De Lucia, A. De Santis, P. De Simone, G. De Zorzi, Denig, A. Di Domenico, C. Di Donato, S. Di Falco, Di Micco, A. Doria, M. Dreucci, G. Felici, A. Ferrari, M. L. Ferrer, G. Finocchiaro, S. Fiore, C. Forti, P. Franzini, C. Gatti, P. Gauzzi, S. Giovannella, E. Gorini, E. Graziani, M. Incagli, W. Kluge, V. Kulikov, F. Lacava, G. Lanfranchi, J. Lee-Franzini, D. Leone, M. Martini, P. Massarotti, W. Mei, S. Meola, S. Miscetti, M. Moulson, S. M¨uller, F. Murtas, M. Napolitano, F. Nguyen, M. Palutan, E. Pasqualucci, A. Passeri, V. Patera, F. Perfetto, M. Primavera, P. Santangelo, G. Saracino, B. Sciascia, A. Sciubba, F. Scuri, I. Sfiligoi, T. Spadaro, M. Testa, L. Tortora, P. Valente, B. Valeriani, G. Venanzoni, R.Versaci, G. Xu

The tracks are ordered by momentum and the two with the higher momentum are assigned to pions. It has been seen on MC simulation that in the 84% of cases the pion has the highest momentum.

The sources of background have been studied on MC. The most important contributions are $\varphi \to \pi^+\pi^-\pi^0$, with the neutral pion undergoing a Dalitz decay ($\pi^0 \to e^+e^-\gamma$), η decays with at least two charged pions in the final state and charged kaons decays.

Many quantities have been studied to reject the background. The following cuts have been applied: $270 < |P(p_1^+)| + |P(p_1^-)| < 370$ MeV and $450 < \sum_{i=1}^{4}|\vec{p_i}| < 600$ MeV. These two cuts allow to reject a large amount of the background, mainly from the $\varphi \to \pi^+\pi^-\pi^0$ decay. After having applied the cuts the signal over background ratio increased from about 1:60 to 1:8.

A kinematic fit is then performed on those events which have fulfilled the cuts on momenta. The kinematic fit has 22 inputs: 3 momenta times 4 tracks, position, time and energy of the neutral cluster, position of the interaction point, √s and momentum of the φ meson. Five constraints are imposed: the four momentum conservation and the photon time of flight ($t_\gamma = R_\gamma/c$). The fit is performed in all the six mass combinations and the one with the best $\chi^2$ is selected.

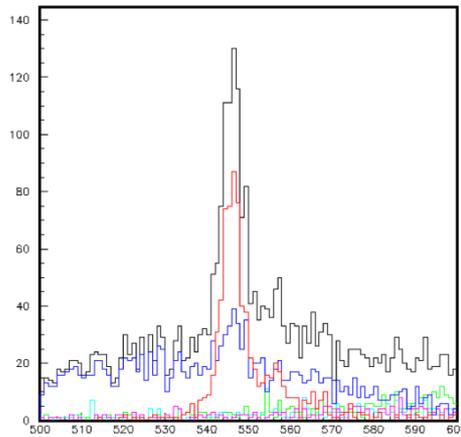

*Figure 1. $M_{inv}(\pi\pi ee)$ spectrum from MC. Red $\eta \to \pi^+\pi^-e^+e^-$; magenta $\varphi \to \pi^+\pi^-\pi^0$; blue $\eta \to \pi^+\pi^-\pi^0$ and $\eta \to \pi^+\pi^-\gamma$; green charged kaons; cyan other backgrounds; black total.*

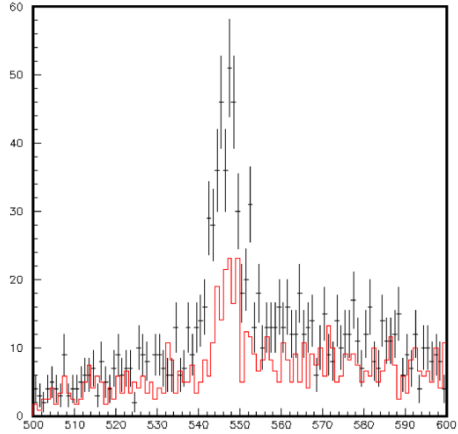

*Figure 2. $M_{inv}(\pi\pi ee)$ spectrum; black data, red background from MC*

The output of the kinematic fit is used to construct the invariant mass of the four tracks $M_{inv}(\pi\pi ee)$. In the range $M_{inv}(\pi\pi ee) \in [500,600]$ MeV the main background is due to the decays $\eta \to \pi^+\pi^-\pi^0$, where the $\pi^0$ undergoes a Dalitz decay, and $\eta \to \pi^+\pi^-\gamma$, where the $\gamma$ converts into a $e^+e^-$ pair (see fig. 1). In order to reject $\eta \to \pi^+\pi^-\pi^0$ events, a further cut on additional neutral clusters has been applied, requiring that the total energy deposited is less than 10 MeV ($\sum E_{add\,cl} < 10$ MeV). After this cut the S/B ratio is about 1:5 while in the range $M_{inv}(\pi\pi ee) \in [540,555]$ MeV it is about 2:1.

It is still present the background due to the $\eta \to \pi^+\pi^-\gamma$ events (S/B -3:1 for $M_{inv}(\pi\pi ee) \in [540,555]$ MeV). It seems possible to disentangle it applying a cut on the invariant mass of the lepton pair, which is smaller for the background. This possibility is under investigation.

This procedure has been applied both to data and MC. In fig.2 a comparison between data and the background shape from MC is shown. Using a rude background subtraction about 300 signal events are visible. The analysis is still in progress.

# Leptonic η decays from WASA experiment at CELSIUS

Marcin Berłowski for the CELSIUS/WASA Collaboration

Institute for Nuclear Studies, Warsaw

Peniscola, 10 May 2007

**Introduction**

Radiative processes are among the most common for the decays of the lightest pseudoscalar mesons $\pi^0$, $\eta$ and $\eta'$ [Fig. 1]. Therefore, as a simple consequence of the Quantum Electrodynamics (QED), they are accompanied by a process where a photon converts into an electron-positron pair. The conversion decays are suppressed by a factor of the order of the fine structure constant $\alpha$. The virtual photon probes the structure of the decaying meson and the interaction region in time-like region of momentum transfer squared, which is equal to the invariant mass squared of the lepton pair $q^2$.

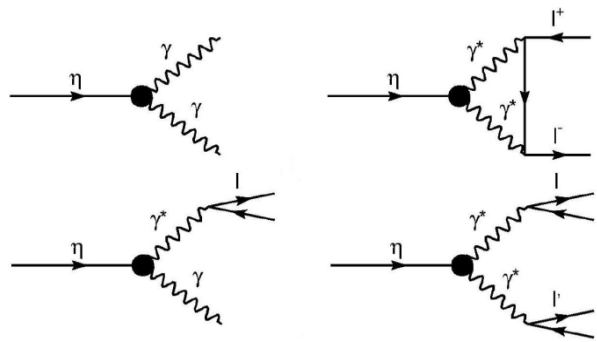

*Fig. 1. Feynman diagrams for the various η decays. Letter l represents an electron or a muon.*

The rare decays of pseudoscalar mesons $P \rightarrow l^+l^-$ represent potentially important channels to look for effects of new physics. The dominant mechanism within Standard Model is the second order in electromagnetism going through two virtual photons $P \rightarrow \gamma^*\gamma^*$. In QED this process is suppressed by $\alpha^2$ and by $(m_l/m_P)^2$ ratio from helicity suppression, and theory got expected SM BR for $\eta \rightarrow e^+e^-$ decay equal to $(5.8\pm0.2)\ 10^{-9}$ [1][2]. The present experimental limit by PDG is $<7.7\ 10^{-5}$ (90% CL) [3]. The decay $\eta \rightarrow e^+e^-$ is sometimes quoted as a sensitive probe for the possible existence of leptoquark particles since the quark-antiquark pair couples to the electron-positron pair and the SM mechanism leads to quite low BR (of the order of $10^{-9}$). An observation of a signal above this level could be evidence for an unconventional process that enhances this decay rate.

The most frequent $\eta$ decay into an $e^+e^-$ pair is $\eta \rightarrow e^+e^-\gamma$ with only a few hundred events that were collected in recent experiments and branching ratio from PDG equal to $(6.0\pm0.8)\ 10^{-3}$ [4]. As for $\eta$ double Dalitz decay no event was experimentally seen and theoretically BR limit is $<6.9\ 10^{-5}$.

**Experimental method**

The experiment was performed at the CELSIUS storage ring in Uppsala, using WASA detector setup. Proton beam interacted with frozen droplets of deuterium. The $\eta$ mesons were produced in the reaction $pd\rightarrow{}^3\text{He}\eta$. The energy of the proton beam was 893 MeV, close to the production threshold. An observation of $^3$He ion in "zero-degree spectrometer" (ZD) provided a clean, $\eta$ decay channel independent trigger. The $^3$He energy was measured and cuts on it permitted to select the $\eta$ production with background level at about 1%. The ZD detector provides trigger rate around few Hz (with proton beam luminosity $\sim 5~10^{30}\text{cm}^{-2}\text{s}^{-1}$) which yields on average one $\eta$ event per second collected in the data acquisition. During the nearly two weeks of the experiment (distributed over half a year period) nearly $3 \cdot 10^5$ $\eta$ events were collected.

The charged products of the $\eta$ decays were tracked using cylindrical drift chamber (MDC) that was placed inside thin wall superconducting solenoid, which provided magnetic field of 1T. The chamber was surrounded by plastic scintillator barrel. Measurement of photon energy and emission angle was performed in the electromagnetic calorimeter (SEC). Comparison of momentum from the curvature of the tracks and energy deposited in SEC permitted to distinguish electrons from charged pions.[5]

**Analysis and results**

Off-line, events with exactly two charged tracks reconstructed in MDC were required to form a vertex close to the intersection region of the beam and the target. Only events with equal number of positive and negative charged particles in MDC were accepted for further analysis. The results on $\eta\rightarrow\pi^+\pi^-e^+e^-$ decay channel with four charged particles were already presented in [6].

In figure [Fig. 2] the e$^+$e$^-$ invariant mass distribution was shown for events with number of neutral cluster greater than zero and less than four (left) and the e$^+$e$^-\gamma$ mass for $M_{ee}$<0.125 GeV (right). On both figures only selected background has been shown. Only events with neutral cluster with energy larger than 200 MeV were taken into account. The $\eta$ peak at about 547 MeV is seen in the right figure. After selection cuts: angle between electron and positron, invariant mass of electron-positron pair and $\gamma$, missing mass to $^3$He, angle between $\gamma$ and $\gamma^*$, emission angle of $\eta$, invariant mass of electron-positron pair and removing $\pi^0$ we were left with 479 $\eta\rightarrow e^+e^-\gamma$ event candidates.

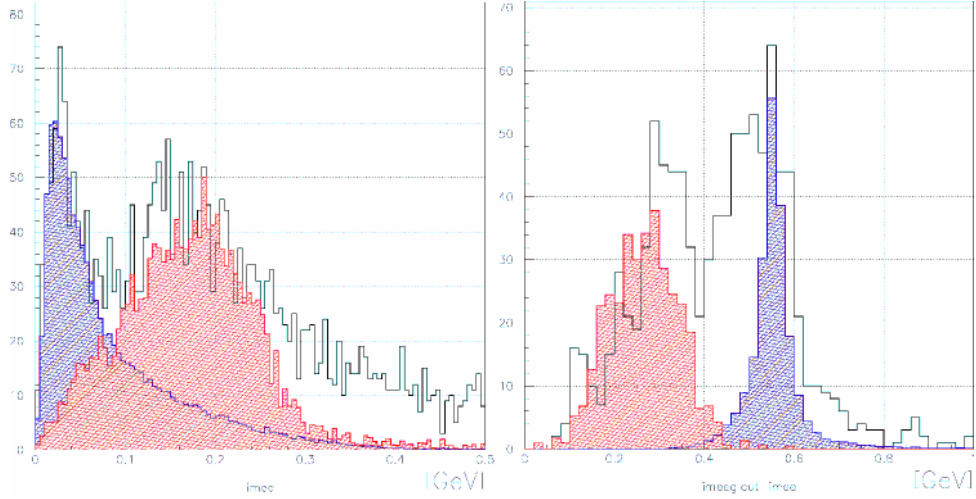

*Fig. 2. (Left) Invariant mass of electron-positron pair. (Right) Invariant mass of $e^+e^-\gamma$. Black curve represents data points, blue Monte Carlo, red curve MC for $\eta \rightarrow \pi^+\pi^-\gamma$ (Right), and $\eta \rightarrow \pi^0\pi^0\pi^0$ with $\pi^0 \rightarrow e^+e^-\gamma$ (Left)*

In search for $\eta \rightarrow e^+e^-e^+e^-$ decay only events with 4 charged tracks and equal number of positive and negative charged particles were selected. After several selection cuts performed (angle between electron-positron pairs, invariant mass of electron-positron pairs, missing mass to $^3$He, angle in each pair, invariant mass of each pair, emission angle of $\eta$) we were left with only 2 event candidates. One of them was with large number of neutral clusters that can be due to background from $\eta \rightarrow \pi^0\pi^0\pi^0$ with two neutral pions Dalitz decays (4.5 events obtained from MC simulations), and the other candidate is on [Fig. 3]. The BR limit (one event assumption) obtained was <6.5 $10^{-5}$ at CL=90%. This channel was never experimentally seen.

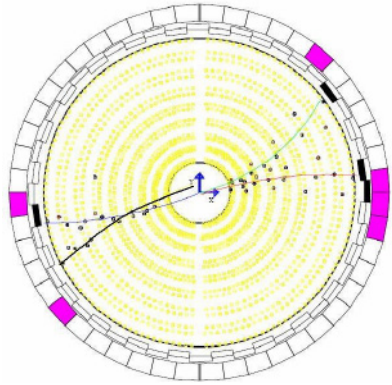

*Fig. 3. $\eta \rightarrow e^+e^-e^+e^-$ event candidate.*

The $\eta \rightarrow e^+e^-$ decay has a distinctive topology: two energetic (>150 MeV) electrons emitted at large relative angle (about 134°). The selection of events has been done by choosing two charged particles with different sign. After kinematical cuts (momentum of electron-positron pair, angle between electron and positron, invariant mass of electron-positron pair, missing mass to $^3$He, emission angle of $\eta$) we observed no events in the region populated by $\eta \rightarrow e^+e^-$ candidates in the simulation (120-160 degrees) fulfilling the identification criteria p/E <1.65. Taking into account the detector acceptance 41% estimated from MC simulations the upper limit for the BR($\eta \rightarrow e^+e^-$) <2.8 $10^{-5}$ (CL=90%) has been obtained.

**Conclusion**

In presented data from CELSIUS/WASA (~232k events – this number of $\eta$ was obtained by normalization to the $\eta \rightarrow \pi^+\pi^-\pi^0$), BR limit can be estimated only to $10^{-5}$ level. We have clear signal from $\eta \rightarrow e^+e^-\gamma$, and we can distinguish between various channels with e$^+$e$^-$ pairs and background. There is no significant background to $\eta \rightarrow e^+e^-$ after applied selection method. The BR limits for $\eta \rightarrow e^+e^-$ and $\eta \rightarrow e^+e^-e^+e^-$ were obtained.

# Theoretical calculation of the $\eta, \eta' \to \pi^+\pi^-l^+l^-$ decays.

R. Nißler

Helmholtz-Institut für Strahlen- und Kernphysik (Theorie), Universität Bonn,
Nussallee 14-16, D-53115 Bonn, Germany

The decays $\eta^{(\prime)} \to \pi^+\pi^-l^+l^-$ with $l = e, \mu$ are interesting in several respects. First, they involve contributions from the box-anomaly of quantum chromodynamics. Second, they probe the transition form factors of $\eta$ and $\eta'$. In principle, the decays are suited to test whether double vector meson dominance is indeed realized in nature, which is also an important issue for the anomalous magnetic moment of the muon and kaon decays [1].

On the experimental side, there is renewed interest in $\eta$, $\eta'$ decays which are investigated at WASA@COSY, MAMI, KLOE and by the VES collaboration. There is thus the necessity to provide a consistent and uniform theoretical description for these decays.

In this respect, the combination of the chiral effective Lagrangian which incorporates the symmetries and symmetry-breaking patterns of QCD in combination with a coupled-channels Bethe-Salpeter equation (BSE) that takes into account final state interactions in the decays and satisfies exact two-body unitarity has been proven very useful. In a series of papers, this approach has been successfully applied to the hadronic decay modes of η and η' [2-4], and the anomalous decays $\eta^{(\prime)} \to \gamma^{(*)}\gamma^{(*)}$ [5], $\eta^{(\prime)} \to \pi^+\pi^-\gamma$ [6] and $\eta^{(\prime)} \to \pi^+\pi^-l^+l^-$ [7].

In this talk I have presented the results of the last work. After constraining the parameters of the approach by experimental data of the decay modes $\eta^{(\prime)} \to \pi^+\pi^-\gamma$ we are able to predict the decay widths and spectra of the decays $\eta^{(\prime)} \to \pi^+\pi^-l^+l^-$ within this approach. We find very good agreement of our result for the branching ratio of $\eta^{(\prime)} \to \pi^+\pi^-e^+e^-$, BR($\eta^{(\prime)} \to \pi^+\pi^-e^+e^-$)$_{th}$ = $2.99^{+0.06}_{-0.09} \times 10^{-4}$, with the recent measurement at WASA@CELSIUS [8], BR($\eta^{(\prime)} \to \pi^+\pi^-e^+e^-$)$_{exp}$ = $(4.3 \pm 1.7) \times 10^{-4}$. The branching ratios of the other decay modes have not yet been determined experimentally.

# The $\eta \to \pi^0 \gamma\gamma$ branching ratio and the $\eta$ mass measurement.

B. Di Micco[a] for the KLOE collaboration[3]

[a] Università degli Studi di Roma Tre, I.N.F.N. Roma III

Here we present KLOE results on the $\eta$ meson decay into $\pi^0\gamma\gamma$ and on the measurement of the $\eta$ mass.

## 1. Introduction

The KLOE experiment[1] is performed at the Frascati $\varphi$ factory DA$\Phi$NE[2]. DA$\Phi$NE is a high luminosity $e^+, e^-$ collider working at $\sqrt{s} \sim 1020$ MeV, corresponding to the $\varphi$ meson mass. In the whole period of data taking (2001–2006) KLOE has collected an integrated luminosity of 2.5 fb$^{-1}$, corresponding to about 8 billions of $\varphi$ produced and 100 millions of $\eta$ mesons through the electromagnetic decay $\varphi \to \eta\gamma$. The main part of these events are stored on tape, the trigger efficiency ranging from 95% to 100%. The analyses described here are performed on the data collected in the years 2001-2002 corresponding to about 1/5 of all KLOE statistics.

## 2. Measurement of the $\eta \to \pi^0 \gamma\gamma$ branching ratio.

The $\eta \to \pi^0 \gamma\gamma$ branching ratio has been measured several times in the history. The observation was claimed the first time in 1966, when its presence was hypothesized to explain the spectrum shape of the $\eta \to 3\pi^0$ decay. The observed spectrum could be reproduced by a large contribution ($\sim 10-40$ %) of this decay channel. Later experiments didn't observe it imposing just upper limits to its branching ratio. The last not null measurements were published by the GAMS[3] collaboration in 1984. The AGS/CB[4] collaboration has published two different results using the same data sample but different analysis techniques: AGS/CB (1) [5] and AGS/CB (2) [6] in the table, while a preliminary KLOE[7] measurement was presented at Eta05 workshop. These measurements are shown in table 1.

| GAMS '84 | $(7.2 \pm 1.4) \times 10^{-4}$ |
|---|---|
| KLOE | $(8.4 \pm 2.7_{\text{stat.}} \pm 1.4_{\text{syst.}}) \times 10^{-5}$ |
| AGS/CB (1) | $(2.7 \pm 0.9_{\text{stat.}} \pm 0.5_{\text{syst.}}) \times 10^{-4}$ |
| AGS/CB (2) | $(3.5 \pm 0.7_{\text{stat.}} \pm 0.6_{\text{syst.}}) \times 10^{-4}$ |

*Table 1. Recent measurements of the BR($\eta \to \pi^0\gamma\gamma$).*

---

[3] F.Ambrosino, A.Antonelli, M.Antonelli, F. Archilli, C.Bacci, P.Beltrame, G.Bencivenni, S.Bertolucci, C.Bini, C.Bloise, S.Bocchetta, V.Bocci, F.Bossi, P.Branchini, R.Caloi, P.Campana, G.Capon, T.Capussela, F.Ceradini, S.Chi, G.Chiefari, P.Ciambrone, E.De Lucia, A.De Santis, P.De Simone, G.De Zorzi, A.Denig, A.Di Domenico, C.Di Donato, S.Di Falco, B.Di Micco, A.Doria, M.Dreucci, G.Felici, A.Ferrari, M.L.Ferrer, G.Finocchiaro, S.Fiore, C.Forti, P.Franzini, C.Gatti, P.Gauzzi, S.Giovannella, E.Gorini, E.Graziani, M.Incagli, W.Kluge, V.Kulikov, F.Lacava, G.Lanfranchi, J.Lee-Franzini, D.Leone, M.Martini, P.Massarotti, W.Mei, S.Meola, S.Miscetti, M.Moulson, S.Müller, F.Murtas, M.Napolitano, F.Nguyen, M.Palutan, E.Pasqualucci, A.Passeri, V.Patera, F.Perfetto, M.Primavera, P.Santangelo, G.Saracino, B.Sciascia, A.Sciubba, F.Scuri, I.Sfiligoi, T.Spadaro, M.Testa, L.Tortora, P.Valente, B.Valeriani, G.Venanzoni, R.Versaci, G.Xu

Both KLOE and AGS/CB measurements disagrees with GAMS '84, KLOE at 4.4 $\sigma$, AGS/CB (1) at 2.5 $\sigma$ and AGS/CB (2) at 2.2 $\sigma$. The preliminary KLOE measurement and AGS/CB one disagree at 1.7 $\sigma$ for the first measurement and at 2.7 $\sigma$ for the second measurement. The experimental situation is so complicated because the $\eta \to \pi^0\gamma\gamma$ signature is overcome by an huge background due to $\eta \to 3\pi^0$, $\pi^0 \to \gamma\gamma$ when two or more $\gamma$'s merge in a single cluster or escape the detector. In this situation it is quite easy to reproduce the topology of the $\pi^0\gamma\gamma$ final state. Moreover the merging probability depends heavily on the electromagnetic shower development in the calorimeter, whose simulation is hard to obtain at optimum level and also the description of the calorimeter geometry and read out is usually a not very easy task. An error on the merging probability at 5% level, that reflects in a $3\times 10^{-3}$ level probability for the merging of two photons, brings an error on the signal over background ratio at level of 100 %. For this reason direct background subtraction produces large systematic error on the measurement.

In KLOE[8] the merging of photons is treated building a likelihood distribution using informations on time, energy and position distribution of the cells.

The distributions of these variables, as given by Monte Carlo simulation, are used to build the likelihood in the hypothesis of good clusters ($L^{good}$) or merged clusters ($L^{merged}$) while the ratio $r = \frac{L^{good}}{L^{merged}}$ is used to discriminate the two type of clusters. Clusters with $r > 0$ are selected as good clusters (see fig. 1).

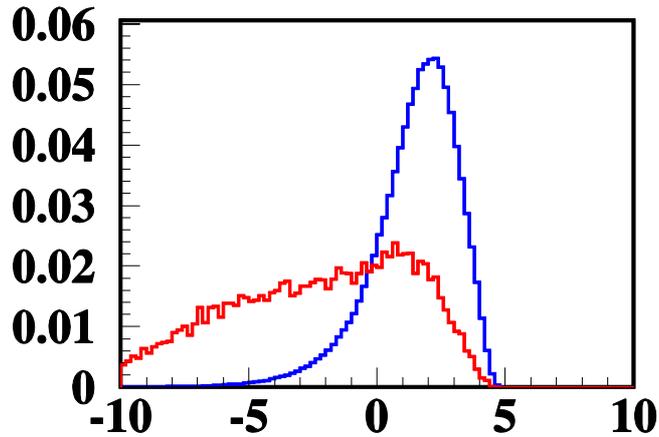

*Figure 1. $r$ distribution for good (blue) and merged (red) clusters as predicted by MC.*

The signal is selected in the 5 $\gamma$ sample against several background sources: $\varphi \to f_0\gamma$ with $f_0 \to \pi^0\pi^0$, $e^+e^- \to \omega\pi^0$ with $\omega \to \pi^0\gamma$, $\varphi \to \eta\gamma$ with $\eta \to 3\pi^0$, $\varphi \to a_0\gamma$ with $a_0 \to \eta\pi^0$ using the $\chi^2$ of a kinematic fit imposing the energy-momentum conservation (the $\varphi$ momentum being known run by run using the $e^+e^- \to e^+e^-$ events), rejecting the $2\pi^0$ and the $\eta\pi^0$ hypothesis in the final state, studying the topology of the $\eta \to 3\pi^0$ when one, two or more photons are lost in the low angular region, rejecting the clusters identified as merged clusters. The invariant mass of the least energetic photons $m_{4\gamma}$ identified as the photons coming from the $\eta$ decay is used to determine the number of the events in the final sample. The background and the signal shapes (from MC) are fitted to the DATA (see fig. 2) histogram obtaining the values for the signal and background fractions:

$$f_{signal} = 0.907 \pm 0.049$$
$$f_{background} = 0.093 \pm 0.031$$

being the total events equal to 735, we obtain: N($\eta \to \pi^0 \gamma\gamma$) = 68 ± 23.

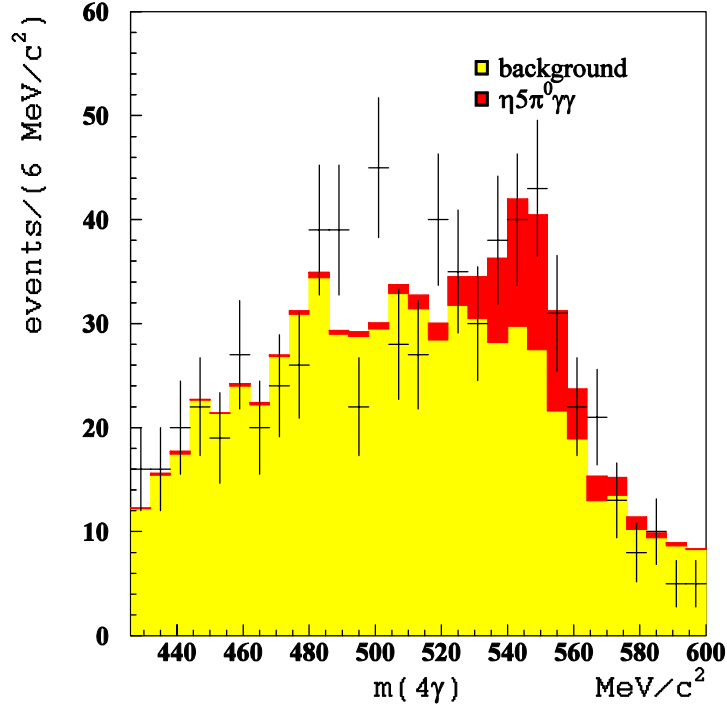

*Figure 2. $M_{4\gamma}$ distribution for DATA, points with error; signal and background (MC), full histograms. The two shapes are fitted to the DATA.*

The efficiency of the selection is evaluated using a MC generator with a flat Dalitz plot for the $\eta \to \pi^0 \gamma\gamma$ decay, it is $\varepsilon = 4.63 \pm 0.09$ %. To obtain the branching ratio we have evaluated the number of $\eta \to 3\pi^0$ events in our sample. The number of counted events is $N_{3\pi^0} = 2288882$ while the selection efficiency for this sample is $\varepsilon(\eta \to 3\pi^0)$ = 0.378 ± 0.08$_{syst}$ ± 0.01$_{stat}$, so we can write:

$$\frac{Br(\eta \to \pi^0 \gamma\gamma)}{Br(\eta \to 3\pi^0)} = \frac{N(\eta \to \pi^0 \gamma\gamma) \cdot \varepsilon(\eta \to 3\pi^0)}{N(\eta \to 3\pi^0) \cdot \varepsilon(\eta \to \pi^0 \gamma\gamma)}$$
$$= (2.43 \pm 0.82) \times 10^{-4}$$

From which we can extract: $Br(\eta \to \pi^0 \gamma\gamma) = (8.4 \pm 2.7_{stat.}) \times 10^{-5}$.

### 2.1. Systematics

The stability of the branching ratio respect to the fit window chosen and the several cuts has been checked. An important source of systematic comes from the estimate of the $\eta \to 3\pi^0$ background. It is particularly difficult because, on one side it has the same structure of the signal in the $m(4\gamma)$ invariant mass distribution, and on the other side the MC simulation is not able to reproduce the merging effect.

The most stable results are obtained using harder $r$ cut, because $r$ suppresses the $\eta \to 3\pi^0$ contamination. We give as systematic error the whole variation assuming an $\eta \to 3\pi^0$ contamination ranging from a factor 0 to 4 respect to the MC estimate.

The total systematic error is $1.4 \times 10^{-5}$. The result is thus:

$$Br(\eta \to \pi^0 \gamma\gamma) = (8.4 \pm 2.7_{stat.} \pm 1.4_{syst.}) \times 10^{-5}$$

The analysis of the remaining KLOE statistics is still on going. In fig. 3 we show the $m_{4\gamma}$ spectrum for 1.4 fb$^{-1}$ of DATA collected in the year 2005. The signal/background ratio is that obtained assuming the measured branching ratio. the new data show on one side the agreement with the given result in the quoted error, and on the other side the possibility to improve the accuracy with the full statistics. The collected DATA will allow also to study the $m_{\gamma\gamma}$ spectral shape of the two $\gamma$'s not coming from the $\pi^0$ decay. This distribution is very sensitive to the theoretical models.

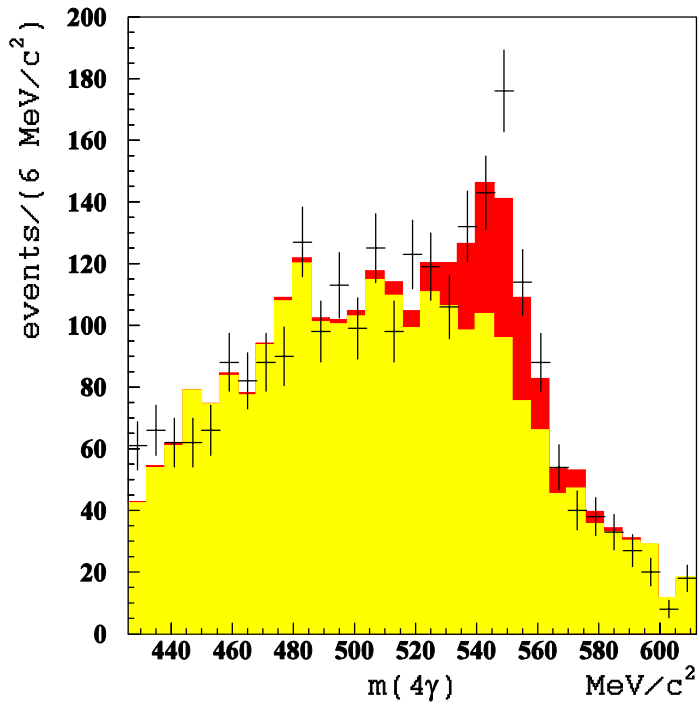

*Figure 3. $M_{4\gamma}$ distribution for 2005 DATA, points with error, signal (MC) and background (MC) full histograms. The Br measured on 2001-2002 is assumed.*

**2.2 Comparison with theoretical predictions.**

The comparison of this measurement with the theoretical predictions is shown in fig. 4.

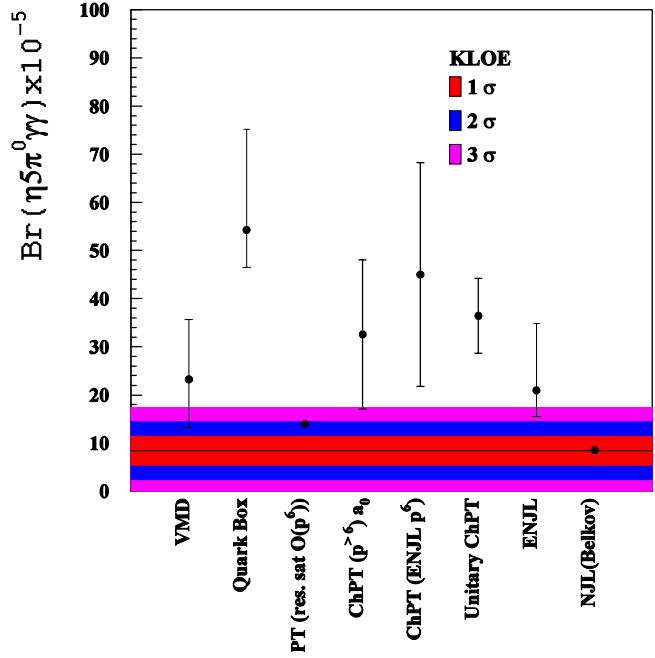

*Figure 4. Theoretical predictions of $Br(\eta \to \pi^0 \gamma\gamma)$ compared with this measurement. The references are in tab. (2).*

| Model | reference |
|---|---|
| VMD | [9] |
| Quark Box | [10] |
| CHPT (res. sat.) $p^6$ | [11] |
| ChPT (res. sat.) | [11] |
| ChPT (ENJL $p^6$) | [12] |
| Unitary ChPT | [13] |
| ENJL | [14] |
| NJL | [15] |
| NJL (Belkov) $O(p^6)$ | [16] |

*Table 2. Table of references.*

For what concern the result evaluated by [11], the ChPT $p^6$ evaluation has been reported together with the all order estimate using VMD + the introduction of scalar mesons. The $p^6$ value can be found in the article and has been reported as ChPT o($p^6$) in the figure. In [13] has been stressed that results from [11] were obtained using old values of vector mesons branching ratios. A reevaluation of this estimate using updated PDG values would be welcome.

## 3. Measurement of the $\eta$ mass.

For the measurement of the $\eta$ mass, refer to [17].

# Status report on the experiments with WASA at COSY

Magnus Wolke

*Institute for Nuclear Physics, Research Centre Jülich, D–52425 Jülich, Germany*

for the WASA–at–COSY Collaboration

**Abstract.** After installation in summer 2006 and first commissioning runs in the second half of last year, WASA at COSY has started taking data. This report updates on the status of the facility, its first production run, and on upcoming experiments.

WASA at COSY combines a unique detector concept with the availability of high quality polarised and unpolarised proton and deuteron beams at the COoler SYnchrotron COSY. Key experiments focus on symmetries and symmetry breaking mechanisms, primarily in decays of $\eta$ and $\eta'$ mesons and in dedicated production experiments, e.g. on charge symmetry breaking (CSB) in the reaction $dd \to {}^4He\pi^0$ [1].

The WASA facility has been installed during summer 2006 and commissioned in the second half of last year. To accommodate for higher beam energies available at COSY, in a first modification step the setup was extended by an additional scintillator layer for the identification of scattered projectiles and charged recoil particles via the $dE/dx$ technique, increasing the stopping power for protons from $300$ to $360\,\mathrm{MeV}$. A completely new data acquisition system with a purely hardware based module readout, buffer and event management has been designed, including the dedicated digitization hardware, i.e. TDCs based on F1 and GPX chips and Flash ADCs with FPQA integration logic. Already in the first production run the DAQ design goal was reached with $20-30\,\mu\mathrm{s}$ required for a full crate readout corresponding to $3-4$ measured events per pellet, and typical event and data rates of $10\,\mathrm{kHz}$ and $60\,\mathrm{MB/s}$, respectively, at $20\%$ deadtime. In addition, for both the central and forward straw chambers the front–end electronics were replaced. To support pellet target operation and for future developments, a laboratory for manufacturing liquid jet nozzles and vacuum injection capillaries has been established at the Research Centre Jülich. Hydrogen pellet rates of $\approx 8\,\mathrm{kHz}$, comparable with values obtained during operation at CELSIUS, have been achieved during a pellet target test in February 2007. However, due to nozzle clogging, the pellet target availability was limited to $70\%$ during the first production run. Efforts are on the way to further improve this value and to demonstrate the long term stability of deuterium pellet operation.

The experimental program has started with a first run in April 2007, focusing on a high statistics data sample of $\eta \to 3\pi^0$ decays and first background studies for a dedicated later experiment on the very rare $\eta \to e^+e^-$ decay. With luminosities slightly above $10^{31}\,\mathrm{cm^{-2}s^{-1}}$ a typical rate of $1.5\,\mathrm{Hz}$ of fully reconstructed $3\pi^0 \to 6\gamma$ events has been achieved. At present, we estimate full statistics for $\eta \to 3\pi^0$ candidates to be $1\cdot 10^6$ events, that can later be used in the Dalitz plot analysis of the $3\pi^0$ system. In addition, the experimental trigger accommodated detector signatures of further neutral and charged decays, like $\eta \to \pi^0\gamma\gamma$, the charged hadronic decay $\eta \to \pi^0 \pi^+ \pi^-$, the $\rho$ channel decays $\eta \to \pi^+ \pi^- \gamma (e^+ e^-)$, as well as single and double Dalitz decays $\eta \to e^+ e^- \gamma (e^+ e^-)$. These data are presently being analyzed.

In the second half of 2007, the WASA–at–COSY Collaboration will focus on experiments with deuteron target. A measurement of CSB in $\vec{d}d \to {}^4\text{He}\,\pi^0$ at slightly higher excess energies than reported in the pioneering IUCF experiment [2], where the final state was consistent with pure s–waves, will allow to extract the p–wave contribution. This is one essential input parameter to the effective field theory analysis [3] of the experimentally observed CSB effects in both the non–vanishing cross section for $dd \to {}^4\text{He}\,\pi^0$ [2] and the forward–backward asymmetry of $np \to d\pi^0$ [4]. In a first step, data will be taken on the breakup reaction $dd \to {}^3\text{He}\,n\pi^0$. Since the transition operators can be calculated from known amplitudes within ChPT, the CS allowed $\pi^0$ production will constrain the $dd$ inital state interaction as presently the most uncertain part of calculating $dd \to {}^4\text{He}\,\pi^0$.

In the reaction $dd \to {}^4\text{He}\,X$, double pion production is isospin allowed in the isoscalar channel. The ABC effect [5, 6, 7], a low–mass enhancement in the $\pi\pi$ invariant mass of isoscalar nature, has originally been observed in inclusive measurements of $pd \to {}^3\text{He}\,\pi\pi$. Recent exclusive results obtained at the CELSIUS–WASA facility [8] disagree with a conventional explanation of the ABC effect by $\Delta\Delta$ excitation, unless the $\Delta\Delta$ system is strongly attractive. The effect is known to be largest in the ${}^4\text{He}$ case, and exclusive data on the reaction $dd \to {}^4\text{He}\,\pi\pi$ will be taken in the second half of 2007 in an energy range $T_d = 0.7 - 1.7\,\text{GeV}$.

A measurement of the decay channels $a_0(980) \to \pi^0 \eta$ and $f_0(980) \to 2\pi$ in the reaction $pd \to {}^3\text{He}\,X$ will determine the $a_0$ and $f_0$ production cross section, which can presently not be estimated with sufficient accuracy. The knowledge about the cross section is one prerequisite for the envisaged study of the radiative decays $a_0 / f_0 \to V \gamma$ ($V = \rho, \omega$), which allows to directly measure the molecular component of the light scalars via their effective couplings to the continuum state of interest [9].

With its first production run, WASA at COSY has started to produce competitive data on $\eta$ decays. With a pellet tracking system and additional upgrades of the forward detector for higher energies, the detector performance will be further improved. In parallel, first steps are being taken to increase luminosities towards the detector design value of $10^{32}\,\text{cm}^{-2}\text{s}^{-1}$.

# Measurement of the pseudoscalar mixing angle and the $\eta'$ gluonium content with KLOE

The KLOE Collaboration[4] presented by F. Ambrosino

Dipartimento di Scienze Fisiche, Universitá degli Studi Federico II, e Sezione INFN Napoli, Italia

The value of the $\eta-\eta'$ mixing angle in the pseudoscalar meson nonet, $\theta_P$, has been discussed extensively in the last thirty five years, and nowadays it is one of the most interesting SU(3)-breaking hadronic parameters to measure [1]. In the context of Chiral Perturbation Theory, it has been demonstrated that a description of the $\eta-\eta'$ system beyond leading order cannot be achieved in terms of just one mixing angle [2]. However in the flavor basis, the two mixing angles are equal, apart from terms which violate the OZI-rule, thus working in this basis it is still possible to use a single mixing angle, $\phi_P$ [3]. The ratio $R_\varphi$ of the two branching ratios $\varphi \to \eta'\gamma$ and $\varphi \to \eta\gamma$ can be related to the $\eta-\eta'$ mixing angle in the flavor basis[3-8] and to the gluonium content of the $\eta'$ meson [5,9].

The analysis has been performed on 427 pb$^{-1}$ collected in 2001-2002. We search for two categories of events $\varphi \to \eta'\gamma$, in the $\pi^+\pi^-7\gamma$'s final state, produced through two different decay chains:

$$\varphi \to \eta'\gamma \text{ with } \eta' \to \pi^+\pi^-\eta \text{ and } \eta \to \pi^0\pi^0\pi^0 \tag{1}$$
$$\varphi \to \eta'\gamma \text{ with } \eta' \to \pi^0\pi^0\eta \text{ and } \eta \to \pi^+\pi^-\pi^0 \tag{2}$$

and $\varphi \to \eta\gamma$, with $7\gamma$'s final state, produced through:

$$\varphi \to \eta\gamma \text{ with } \eta \to \pi^0\pi^0\pi^0 \tag{3}$$

which is used for normalization of the rates and is practically background free. Processes with $\varphi \to K_S K_L$, where the $K_L$ decays near the beam interaction point (I.P.) can mimic the final state with $\pi^+\pi^-7\gamma$'s because of the presence of an additional photon due either to machine background or splitting clusters.

We select 3750 $\eta'$ candidate events and the expected background is $N_{\text{background}}$ =345. The final number of events from processes (1) and (2), after background subtraction, is $N_{\eta'\gamma} = 3405$. For the $\varphi \to \eta\gamma$ process we select $N_{\eta\gamma} = 1665000$ events. We evaluate the ratio of the two branching fractions:

$$R_\varphi = (4.77 \pm 0.09_{\text{stat}} \pm 0.19_{\text{syst}}) \cdot 10^{-3} \tag{4}$$

[4] F. Ambrosino, A. Antonelli, M. Antonelli, F. Archilli, C. Bacci, P. Beltrame, G. Bencivenni, S. Bertolucci, C. Bini, C. Bloise, S. Bocchetta, V. Bocci, F. Bossi, P. Branchini, R. Caloi, P. Campana, G. Capon, T. Capussela, F. Ceradini, S. Chi, G. Chiefari, P. Ciambrone, E. De Lucia, A. De Santis, P. De Simone, G. De Zorzi, A. Denig, A. Di Domenico, C. Di Donato, S. Di Falco, B. Di Micco, A. Doria, M. Dreucci, G. Felici, A. Ferrari, M. L. Ferrer, G. Finocchiaro, S. Fiore, C. Forti, P. Franzini, C. Gatti, P. Gauzzi, S. Giovannella, E. Gorini, E. Graziani, M. Incagli, W. Kluge, V. Kulikov, F. Lacava, G. Lanfranchi, J. Lee-Franzini, D. Leone, M. Martini, P. Massarotti, W. Mei, S. Meola, S. Miscetti, M. Moulson, S. Müller, F. Murtas, M. Napolitano, F. Nguyen, M. Palutan, E. Pasqualucci, A. Passeri, V. Patera, F. Perfetto, M. Primavera, P. Santangelo, G. Saracino, B. Sciascia, A. Sciubba, F. Scuri, I. Sfiligoi, T. Spadaro, M. Testa, L. Tortora, P. Valente, B. Valeriani, G. Venanzoni, R.Versaci, G. Xu

This result is more accurate than, and in complete agreement with our previous measurement made using a different final state but with considerably less statistics [10]. Using $BR(\varphi \to \eta\gamma) = (1.301 \pm 0.024)\%$ as from PDG [11] we find:

$$BR(\varphi \to \eta'\gamma) = (6.20 \pm 0.11_{stat} \pm 0.25_{syst}) \cdot 10^{-5} \qquad (5)$$

The value of $R_\varphi$ can be related to the pseudoscalar mixing angle in the quark-flavor basis, using the approach of ref.[7,8], where the SU(3) breaking is taken into account via constituent quark mass ratio $m_s/\bar{m}$, $R_\varphi$ is given by the following expression:

$$R_\varphi = \frac{BR(\phi \to \eta'\gamma)}{BR(\phi \to \eta\gamma)} = \cot^2\varphi_P \left(1 - \frac{m_s}{\bar{m}} \frac{C_{NS}}{C_S} \frac{\tan\varphi_V}{\sin 2\varphi_P}\right)^2 \left(\frac{p_{\eta'}}{p_\eta}\right)^3$$

where $\varphi_V = 3.4°$ is the mixing angle for vector mesons; $p_{\eta(\eta')}$ is the $\eta(\eta')$ momentum in the $\phi$ center-of-mass; the two parameters $C_{NS}$ and $C_S$ represent the effect of the OZI-rule, which reduces the vector and pseudoscalar wave-function overlap [8]. We obtain the following result:

$$\phi_P = (41.4 \pm 0.3_{stat} \pm 0.7_{sys} \pm 0.6_{th})° \qquad (6)$$

In the traditional approach the $\eta - \eta'$ mixing is parametrized in the octet-singlet basis; in this basis the value of the mixing angle becomes: $\theta_P = \phi_P - \arctan\sqrt{2} = (-13.3 \pm 0.3_{stat} \pm 0.7_{sys} \pm 0.6_{th})°$. QCD involves quanta, gluons, which may form a bound state, gluonium, that can mix with neutral mesons. While the $\eta$ meson is well understood as an $SU(3)$-flavor octet with a small singlet admixture, the $\eta'$ meson is a good candidate to have a sizeable gluonium content. If we allow for a $\eta'$ gluonium content, we have the following parametrization:

$$|\eta'> = X_{\eta'} |q\bar{q}> + Y_{\eta'} |s\bar{s}> + Z_{\eta'} |gluon>$$

where the $Z_{\eta'}$ parameter takes in to account a possible mixing with gluonium. The normalization implies $X_{\eta'}^2 + Y_{\eta'}^2 + Z_{\eta'}^2 = 1$ with

$$\begin{aligned} X_{\eta'} &= \cos\varphi_G \sin\phi_P \\ Y_{\eta'} &= \cos\varphi_G \cos\phi_P \\ Z_{\eta'} &= \sin\varphi_G \end{aligned} \qquad (7)$$

where $\varphi_G$ is the mixing angle for the gluonium contribution. Possible gluonium content of the $\eta'$ meson corresponds to non-zero value for $Z_{\eta'}^2$, that implies

$$X_{\eta'}^2 + Y_{\eta'}^2 < 1 \qquad (8)$$

If we allow for gluonium content in $\eta'$ and neglect $\eta$ gluonium content, then the theoretical expression for $R_\varphi$ is to be rewritten by multiplying it for $\cos^2\varphi_G$. We may use other relations to determine $X_{\eta'}, Y_{\eta'}$ [5,8,9] and introduce other constraints. The relations can be written as function of the angles. Apart from trivial kinematic factors one has [12]:

$$\frac{\Gamma(\eta' \to \gamma\gamma)}{\Gamma(\pi^0 \to \gamma\gamma)} \propto (5\cos\varphi_G \sin\phi_P + \sqrt{2}\frac{f_q}{f_s}\cos\varphi_G \cos\phi_P)^2$$

$$\frac{\Gamma(\eta' \to \rho\gamma)}{\Gamma(\omega \to \pi^0\gamma)} \propto \frac{C_{NS}}{\cos\phi_V}\cos^2\varphi_G \sin^2\phi_P$$

$$\frac{\Gamma(\eta' \to \omega\gamma)}{\Gamma(\omega \to \pi^0\gamma)} \propto \cos\varphi_G (C_{NS}\sin\phi_P + 2\frac{m_s}{\bar{m}}C_S \tan\phi_V \cos\phi_P)^2$$

We have therefore minimized a $\chi^2$ function with $\cos^2\varphi_G$ and $\cos^2\phi_P$, as free parameters by imposing the above constraints and properly including in the error matrix the uncertainties on the other parameters. The solution in the hypothesis of no gluonium content, i.e. $\cos^2\varphi_G = 1$ yields $\phi_P = (41.5^{+0.6}_{-0.7})°$. The $\chi^2$ quality is bad, $\chi^2/N.d.f. = 11.34/3$, with a $P(\chi^2/N.d.f.) = 0.01$ and suggests to look for a possible non-zero value for $Z_{\eta'}^2$.

The solution allowing for gluonium is $\cos^2\varphi_G = 0.86 \pm 0.04$ and $\cos^2\phi_P = 0.592 \pm 0.012$, from which $\phi_P = (39.7 \pm 0.7)°$ and $Z_{\eta'}^2 = 0.14 \pm 0.04$, which means $|\phi_G| = (22 \pm 3)°$. In this case the $\chi^2$ quality is good, $\chi^2/N.d.f. = 1.42/2$ with a $P(\chi^2/N.d.f.) = 0.49$.

Our analysis indicates a $3\sigma$ evidence for the $\eta'$ gluonium content.

# The study of the light scalars at KLOE

P.Gauzzi[5]

Dipartimento di Fisica dell' Università "La Sapienza"
e Sezione INFN, Roma, Italy

The scalar mesons $f_0$ (980) and $a_0$ (980) are not easily interpreted as ordinary $q\bar{q}$ mesons belonging to the $^3P_0$ nonet together with the $\sigma(500)$ and the $\kappa(800)$, whose existence is still questioned. Alternative hypotheses have been proposed: $q\bar{q}q\bar{q}$ states, $K\bar{K}$ bound states, or dynamically generated resonances. In order to shed some light into the structure of these mesons, the KLOE experiment at DAΦNE measured the branching ratios of $\varphi \to f_0\gamma$ and $\varphi \to a_0\gamma$, and extracted the relevant parameters of the $f_0$ and $a_0$ from their mass shapes. To this purpose two different models of the $\varphi \to$ Scalar $\gamma$ decay have been exploited: the Kaon Loop (KL) model, in which the coupling of the $\varphi$ to the scalar occurs through a charged kaon loop[1]; and another one (called "No Structure", NS) in which a pointlike coupling of the $\varphi$ to the scalar meson is considered[2].

The $f_0(980)$ is searched for in $\pi^0\pi^0\gamma$ and $\pi^+\pi^-\gamma$ final states. The main contributions to $e^+e^- \to \pi^0\pi^0\gamma$ are the non-resonant process $e^+e^- \to \omega(\to \pi^0\gamma)\pi^0$, and $\varphi \to f_0(\to \pi^0\pi^0)\gamma$. A fit of the Dalitz plot ($M_{\pi\gamma}$ vs $M_{\pi\pi}$) to a version of the KL model that includes also the $\sigma(500)$ contribution and the $f_0/\sigma$ mixing has been performed. The $\sigma(500)$ parameters has been fixed to the values: $m_\sigma$ = 462 MeV, $\Gamma_\sigma$ = 286 MeV, $g_{\sigma K^+K^-}$ = 0.5 GeV, and $g_{\sigma\pi^+\pi^-}$ = 2.4 GeV. Without including the $\sigma(500)$ the fit quality becomes very poor, as $P(\chi^2) \to 10^{-4}$. Also the fit to the NS model does not require the presence of the $\sigma(500)$. The fit results are summarized in tab.1. By integrating only the scalar component one obtains:

$Br(\varphi \to S\gamma \to \pi^0\pi^0\gamma) = (1.07^{+0.01}_{-0.03} \pm 0.06) \times 10^{-4}$, where $S = \sigma(500), f_0(980)$. The search for $\varphi \to f_0 (\to \pi^+\pi^-)\gamma$ is characterized by the presence of high rate irreducible backgrounds due to the initial state radiation (ISR), and to $e^+e^- \to \pi^+\pi^-\gamma$, occurring either through the $\varphi$ or through the $\rho^0$ tail (final state radiation, FSR). By subtracting from the invariant mass distribution of $\pi^+\pi^-$ the expected behaviour of ISR + FSR, the $f_0$ mass shape is obtained and then fit to the two models quoted before (see tab.1). In both cases the $\sigma(500)$ is not needed in the fit. From the integral of the scalar amplitude $Br(\varphi \to f_0\gamma \to \pi^+\pi^-\gamma) = (2.1 - 2.4) \times 10^{-4}$ is obtained, consistent with twice the branching ratio of the neutral channel.

---

[5] E-mail address: Paolo.Gauzzi@roma1.infn.it

Table 1: Fit results for $f_0(980) \to \pi^0\pi^0$ and $\pi^+\pi^-$

|  | $\pi^0\pi^0$ | | $\pi^+\pi^-$ | |
| --- | --- | --- | --- | --- |
| $f_0$ (980) param. | KL model | NS model | KL model | KL model |
| $m_{f_0}$ (MeV) | 976 – 987 | 981 – 987 | 980 – 987 | 973 – 981 |
| $g_{\varphi f_0 \gamma}$ (GeV$^{-1}$) | 2.7 – 4.1 | 2.5 – 2.7 | 2.9 – 4.1 | 1.2 – 2.0 |
| $g_{f_0 \pi^+\pi^-}$ (GeV) | 1.4 – 2.0 | 1.3 – 1.4 | 3.0 – 4.2 | 0.9 – 1.1 |
| $g_{f_0 K^+K^-}$ (GeV) | 3.3 – 4.9 | 0.1 – 1.0 | 5.0 – 6.3 | 1.6 – 2.3 |

The $a_0(980) \to \eta\pi^0$ decay is studied by selecting the final states corresponding to $\eta \to \gamma\gamma$ and $\eta \to \pi^+\pi^-\pi^0$ decays. The former is characterized by large statistics and by a large irreducible background (mainly from $\pi^0\pi^0\gamma$ final state and from $\varphi \to \eta\gamma$ with $\eta \to 3\pi^0$); the latter has lower statistics, but also lower background, since there are not other processes with exactly the same final state as the signal. The $a_0$ parameters are extracted from a combined fit of the two distributions of the $\eta\pi^0$ invariant mass, the results are reported in tab.2. After the background subtraction one obtains the two values: $Br(\varphi \to a_0\gamma \to \eta\pi^0\gamma) = (6.95 \pm 0.09 \pm 0.24) \times 10^{-5}$ for the neutral channel, and $Br(\varphi \to a_0\gamma \to \eta\pi^0\gamma) = (7.22 \pm 0.24 \pm 0.24) \times 10^{-5}$ for the charged one. The large values of the $Br(\varphi \to f_0/a_0\gamma)$, the large couplings of the scalars to $K^+K^-$ and to the $\varphi$, are hints of a large strange quark content in $f_0(980)$ and $a_0(980)$, and then suggest a non $q\bar{q}$ structure of the light scalar mesons.

Table 2: Fit results for $a_0(980) \to \eta\pi^0$ (systematics not included)

| $a_0$ (980) param. | KL model | NS model |
| --- | --- | --- |
| $m_{a_0}$ (MeV) | 983.9 ± 1.2 | 983.9 (fixed) |
| $g_{\varphi a_0 \gamma}$ (GeV$^{-1}$) | 1.67 ± 0.09 | 1.56 ± 0.04 |
| $g_{a_0 \eta \pi^0}$ (GeV) | 2.85 ± 0.08 | 2.10 ± 0.03 |
| $g_{a_0 K^+K^-}$ (GeV) | 2.18 ± 0.05 | 1.36 ± 0.15 |

KLOE is also studying the decay $\varphi \to K^0\bar{K}^0\gamma$, to search for $f_0/a_0 \to K^0\bar{K}^0$. The $K_S K_S \gamma$ final state with both $K_S$ decaying into $\pi^+\pi^-$ has been chosen, since it is less affected by background contamination. A preliminary analysis, based on about 1/5 of the whole KLOE statistics, showed that 7 background events are expected by Monte Carlo, and this allows to quote a "sensitivity" for the upper limit of the branching ratio of $7.5 \times 10^{-8}$ at 90% C.L., in the hypothesis that no signal is present, according to the prescriptions of ref. [3].

# Scalar and vector meson exchange in $V \to P^0 P^0 \gamma$ decays[6]

R. Escribano[†]

[†]Grup de Física Teòrica and IFAE, Universitat Autònoma de Barcelona,
E-08193 Bellaterra (Barcelona), Spain

**Summary**

The radiative decays of the light vector mesons ($V = \rho, \omega, \varphi$) into a pair of neutral pseudoscalars ($P = \pi^\circ, K^\circ, \eta$), $V \to P^\circ P^\circ \gamma$, are an excellent laboratory for investigating the nature and extracting the properties of the light scalar meson resonances ($S = \sigma, a_0, f_0$). In addition to the scalar contribution, there is also a vector meson contribution when one of the neutral pseudoscalars and the photon are produced by the exchange of an intermediate vector meson through the decay chain $V \to VP^\circ \to P^\circ P^\circ \gamma$. Fortunately, for most of the processes of interest the main contribution is by far the scalar one, thus making of the study of these radiative decays a very challenging subject in order to improve our knowledge on the lightest scalar mesons. This study also complements other analysis based on central production, $D$ and $J/\psi$ decays, etc. Particularly interesting are the so called golden processes, namely $\varphi \to \pi^\circ \pi^\circ \gamma$, $\varphi \to \pi^\circ \eta \gamma$ and $\rho \to \pi^\circ \pi^\circ \gamma$, which can provide us with valuable information on the properties of the $f_0(980)$, $a_0(980)$ and $\sigma(600)$ resonances, respectively.

This presentation is based on a series of works done in part in collaboration with G. Pancheri, A. Bramon, J. L. Lucio, and M. Napsuciale [1]–[5].

As an introduction to the subject, I present the most recent experimental data accounting for these processes and a theory review with particular emphasis in the scalar models market. The calculation of the scalar contributions to the *golden processes* within the framework of the Linear Sigma Model (L$\sigma$M) is the main topic of discussion along the presentation. The L$\sigma$M is a well defined U(3) x U(3) chiral model which incorporates *ab initio* the pseudoscalar nonet together with its chiral partner the scalar nonet. The complementarity between Chiral Perturbation Theory (ChPT) and the L$\sigma$M will be used for including the scalar meson poles while keeping the correct behaviour at low dimeson invariant mass expected from ChPT. Other processes, such as $\omega \to \pi^\circ \pi^\circ \gamma$ and $\varphi \to K^\circ \bar{K}^\circ \gamma$, as well as the ratio $\varphi \to f_0 \gamma / a_0 \gamma$ are also discussed.

[6] This work was supported in part by the Ramon y Cajal program (R.E.), the Ministerio de Educación y Ciencia under grant FPA2005-02211, the EU Contract No. MRTN-CT2006-035482, "FLAVIAnet", and the Generalitat de Catalunya under grant 2005-SGR00994

A summary of the results obtained is presented in Table 2 of Ref. [2]. Our theoretical predictions are all compatible with experimental data. This nice agreement is achieved not only for the branching ratios but also for the available mass spectra. In general, we have shown that $\varphi \to \pi°\pi°\gamma$, $\rho \to \pi°\eta\gamma$ and $\rho \to \pi°\pi°\gamma$ can be used to extract relevant information on the properties of the $f_0(980)$, $a_0(980)$ and $\sigma(600)$ scalar mesons, respectively, while $\omega \to \pi°\pi°\gamma$ and $(\rho, \omega) \to \pi°\eta\gamma$ cannot be used due to the smallness of the scalar contribution. $\varphi \to K°\bar{K}°\gamma$ could also be used in the near future for extracting the properties of the $f_0$ and $a_0$ resonances. Comments for each process in particular can be found in the conclusions of Ref. [2]. Higher accuracy data and more refined theoretical analyses would contribute decisively to clarify the sector of the lowest lying scalar states. In this line, our work aims to be one step forward.

**Acknowledgements**

I would like to express my gratitude to the ETA07 Workshop Organizing Committee for the opportunity of presenting this contribution, and for the pleasant and interesting conference we have enjoyed.

# How to identify hadronic molecules

C. Hanhart

Institut für Kernphysik, Forschungszentrum Jülich, D-52425 Jülich, Germany

The hadronic spectrum still contains many obstacles. One is that it may or may not contain bound states of hadrons, so-called hadronic molecules. The goal is to present criteria of how and under what circumstances one can quantify the molecular component of a particular state. The idea is outlined in detail in Refs. [1]. Although the method is more general, we focus on the sector of the light scalar mesons.

There are various observables that were claimed to allow for a distinction of molecules from quark states. In the sector of the scalar mesons this is, e.g., the radiative decay $\phi \to \gamma \pi^0 \pi^0 / \pi^0 \eta$ [2]. However, so far not a single observable exists that could not described within models designed in the spirit of either a molecular picture or a quark model picture. E.g. the $\phi$ radiative decay is studied theoretically from the point of view of molecular scalars in Refs. [3]. For experimental data see Ref. [4] and references therein.

The central result of Ref. [1] was to establish a relation between the effective coupling of a resonance to a particular continuum channel and its molecular content with respect to this channel. This relation holds, if the binding energy of the bound system is significantly smaller than any other intrinsic scale of the system. In case of unstable states this can be generalized to the distance of the corresponding singularity from the relevant threshold. Note that the effective coupling constant can in principle be extracted from data since it revers to the relevant residue at the pole of the state. The principle behind is that under this condition the hadron loop will be driven by its non-analytic piece, originating from the unitarity cut, that is trivially absent in any quark loop.

The central statement is that whatever model one designs -- using *s*-channel poles or just hadron-hadron non-pole interactions -- what matters at the end is the effective coupling one extracts via analytic continuation to the complex plane after fitting appropriate data. Regardless the dynamical content of the model, a large effective coupling with refer to predominantly molecular state and a small one to a predominantly non-molecular state --small and large are quantified in Refs. [1]. In this sense the data on $\phi \to \gamma \pi^0 \pi^0$ as well as those of $f_0 \to \gamma\gamma$ studied within the scheme just described in Ref. [5], strongly support a molecular assignment for the $f_0(980)$. For the $a_0$, on the other hand, the picture seems to be not as clear. A possible explanation could be that either there is some quark state admixture in this state or it is not a bound but a virtual state.

# Session II:

*Meson photo production*

# $\eta$ and $\eta'$ Photo- and Electroproduction on the Nucleon

L. Tiator

Institut für Kernphysik, Johannes Gutenberg-Universtät, Mainz, Germany

Most experimental data on photo- and electroproduction of $\eta$ mesons on the proton are well described with isobar models [1-3], lagrangian models [4,5] or in coupled-channels approaches [6,7]. There are two cases, however, which deserve special consideration:

Near threshold, in 1998 the target polarization asymmetry was measured at Bonn and to a big surprise it showed a nodal structure with negative values for backward angles. This could not be explained by the EtaMaid model and all other calculations since then were also not able to describe these data. Therefore, the data were mostly ignored. In a new electroproduction experiment at Mainz, using beam-recoil double polarization, the same response function could now be tested and the old target polarization asymmetry is practically confirmed [8]. A model-independent analysis in 1999 has found that this unexpected result is due to a strong energy-dependent $s-d$ phase rotation between the two resonant partial waves $S_{11}$ and $D_{13}$, leading to an additional phase of almost $90°$ near threshold [9]. In single-channel approaches, such a phase could only occur under the presence of a strong background contribution. But the background contribution in $\eta$ photoproduction is especially small in the threshold region [10]. It would be very interesting to see, if a dynamical calculation of these partial waves could solve this problem. So far, these calculations have concentrated on the $S_{11}$ channel only [11-13].

The second case is a bump in the differential cross section on the quasi-free neutron around $W_{cm} = 1670$ MeV, which is not observed on the proton. The structure which was first reported at GRAAL [14] was later confirmed at CB-ELSA [15] and recently at LNS-Tohoku [16]. The EtaMaid model [1] is able to explain this structure with a strong $D_{15}(1675)$ excitation. However, the $\eta N$ branching ratio of 17% ia almost a factor of 10 larger than expected [17,18]. In an alternative approach (ReggeMaid [2]), with a reggeized vector meson exchange, the $D_{15}$ does not play any significant role and the bump in the neutron can be described with a narrow $P_{11}(1670)$ resonance [19], that could be a non-strange member of the pentaquark anti-decuplet [20].

In a comparison with preliminary data of CB-ELSA with differential cross sections and photo asymmetry data of GRAAL both models can only describe part of the data. The EtaMaid2001 fails in the comparison with $\Sigma(\theta)$ [21] and the ReggeMaid shows much less structure in the unpolarized angular distributions as seen in the CB-ELSA data [22].

At this Eta07 workshop two other calculations have been presented. With the Giessen model, Shklyar [23] has given an explanation of the bump in terms of coupled channels and found an explanation with $S_{11}$ and $P_{11}$ partial waves. But as in the case of ReggeMaid, these dominant partial waves lead to a different angular distribution as seen in the experiment. Finally Nakayama [24] showed first results with a lagrangian model that can very well describe the proton data and finds an important $D_{13}(1700)$ contribution that was not observed in the other models. But this model has not yet been applied to the neutron target. However, from the strong $D_{13}$-wave contribution a different angular dependence can be expected.

The situation of $\eta'$ photoproduction is still in an early stage. With the CLAS experiment [25] a large data set of differential cross sections is available, covering a wide range of energies and angles. But without additional polarization data, a series of equivalent fits are possible with EtaprimeMaid [2] or with the lagrangian models [4,5,24], leading to different sets of resonance parameters for the four most likely nucleon resonances that can be expected in this reaction: $S_{11}(2090), P_{11}(2100), P_{13}(1900), D_{13}(2080)$. Since not only the electromagnetic couplings are unknown, but also the hadronic parameters as mass and width are very uncertain, a unique solution is not possible without further polarization observables. A first attempt in this direction should be an experiment to measure the photon beam asymmetry.

Email: tiator@kph.uni-mainz.de

# The Reaction $\gamma p \to p\pi^0\pi^0$

Michael Lang[7] *for the CBELSA / TAPS Collaboration*

## 1. Introduction and Set-Up

The CBELSA / TAPS Experiment was carried out at the Electron Stretcher Accelerator (ELSA) facility in Bonn, Germany [1]. The detector set-up mainly consists of the *Crystal Barrel Detector* [2] which offers a high detection efficiency for neutral reaction products, and a dedicated arrangement of the *Two Arm Photon Spectrometer (TAPS)* [3] in forward beam direction to provide an angular space coverage of more than 0.96 $4\pi$ sr. As target, a cell containing liquid hydrogen was used. The real photon beam, required for this experiment, was produced via the Bremsstrahlung process on a 50 $\mu$m copper foil, placed into the extracted electron beam of ELSA offering a total range of photon energies between 800 MeV and 3.5 GeV.

With this experiment set-up, data for the reaction $\gamma p \to p\pi^0\pi^0$ was acquired. Goals were to determine resonance contributions and to find missing resonances contributing to this photoproduction process as predicted by the Bonn-Model [4]. This reaction is additionally useful to measure the decays into $N\pi^0\pi^0$ from higher energy resonances such as $N^*$ and $\Delta^*$. In addition, the neutral pion photoprodcution reduces the number of possible Feynman graphs as Kroll-Rudermann, $\Delta$-Kroll-Rudermann and other terms with a photon-pion-vertex do vanish [5].

## 2. Results

To determine the total cross section of $\gamma p \to p\pi^0\pi^0$, the acceptance of the detector set-up had to be evaluated [6]. As model to compare the experimental data with, a partial wave analysis model with resonances, developed by Gatchina & Bonn, was used [7]. The acceptance corrections applied use the dynamics as determined by the PWA in the Monte-Carlo simulation. This correction to the experiment data revealed a very good agreement with previous data from GRAAL and TAPS for the total photoabsorption cross-section.

Interpreting the data with this PWA revealed for the invariant mass distributions for $p\pi^0\pi^0$ in the range between 1550 MeV/c$^2$ and 1800 MeV/c$^2$ that the main contribution in this reaction is given by the $\Delta^+\pi^0$ decay into $p\pi^0\pi^0$. This is visible in the invariant mass distribution of $\pi\pi$ as also in the invariant mass distribution of $p\pi$.

Comparing the resonance contributions set in the PWA to reproduce the experiment data to values given by the PDG, the major discrepancy was found for the Breit-Wigner width of $P_{13}(1720)$, where the PWA delivers a value of $481\pm 65$ MeV and the PDG gives as value $100-200$ MeV. In addition, the total width of the $S_{31}(1620)$ was found to be $200\pm 30$ MeV while the PDG gives $100-130$ MeV as range.

---

[7] mlang@hiskp.uni-bonn.de

## 3. Summary

The reaction $\gamma p \to p\pi^0\pi^0$ was measured with the CBELSA / TAPS experiment at ELSA, Bonn, Germany. The PWA with resonances model used is able to describe the measured data. Several $p\pi^0\pi^0$ decays were determined for the first time, but ambiguities occur such as contributions to $D_{33}(1700)$. To determine this, polarized data is required. The properties found for the $P_{13}(1720)$ resonance are different from the PDG. The decay of $N^*/\Delta^*$ into $\Delta\pi$ or $N^*/\Delta^*$ was additionally determined by using the data of this photoproduction reaction. A very preliminary assumption is that cascade decays of higher energy resonances were observed into $\Delta\pi^0$, $D_{13}(1520)\pi$, $X(1660)\pi$.

# Investigation of $\gamma p \rightarrow p\pi^0\eta$ and related reactions with CBELSA/TAPS[8]

M. Nanova[(a)]

for CBELSA/TAPS Collaboration,

[(a)] II. Physikalisches Institut, University of Giessen, Germany

The photoexcitation of baryon resonances has been extensively studied in many experiments to investigate and test the nucleon structure. Baryon structure calculations predict more resonant states than have been observed so far. Therefore, more effective theories and models are necessary in order to determine the masses, couplings and decay widths of resonances. On the other hand more experimental data are needed. Baryon resonances have often large widths and overlap largely, which makes the study of the excited states more difficult. It is possible to overcome this problem by looking at specific decay channels. Now most existing data result from $\pi N$ elastic scattering experiments. Thus, photoproduction experiments offer new possibilities. Particularly η-meson photoproduction provides such a selectivity for coupled channel investigation, since, the η-meson acts as an isospin filter. The decay chain $\gamma p \rightarrow \Delta^* \rightarrow \Delta\eta \rightarrow p\pi^0\eta$ is a suitable reaction to study $\Delta$ states and to search for $\Delta^{*}$'s. Such reactions have not been investigated well yet. Strangeness production experiments will also be an important tool to establish or disprove 'missing' resonances, since some of the resonances are predicted to decay mostly into final states containing strange particle pairs [1]. The new studies can also bring new light on the nature of some resonances.

The exclusive measurement of the reactions $\gamma p \rightarrow p\pi^0\eta$, $\gamma p \rightarrow K^0\pi^0\Sigma^+(1189)$ and $\gamma p \rightarrow K^{*0}\Sigma^+(1189)$ with p $4\pi^0$ final states has been presented. The experiment has been performed at the tagged photon facility of the ELSA accelerator (Bonn) in the beam energy range up to 2.6 GeV. The Crystal Barrel and the photon spectrometer TAPS were combined to a detector system providing an almost $4\pi$ coverage of the geometrical solid angle, very well suited for studying photoproduction with multi-photon final states. The events of the investigated reactions have been reconstructed from the measured nine particles in the final channel. A kinematic fit has been applied to the selected events. Recent results for the differential and total cross sections have been presented and discussed. The experimental data have been compared with the theoretical predictions for these channels [2-6].

---

[8] Supported by DFG (SFB/TR-16)

# Theoretical aspects of the γp → π⁰ηp and related reactions

M. Döring,* E. Oset, and D. Strottman

*Departamento de Física Teórica and IFIC, Centro Mixto Universidad de Valencia- CSIC, Institutos de Investigación de Paterna, Aptd. 22085, 46071 Valencia, Spain*

We discuss recent work about the photoproduction of ηπ⁰ on the proton and a set of different related photon- and pion-induced reactions. Resonances such as the Δ*(1700) and the N*(1535) play an important role. In the picture of dynamical generation of these resonances in a unitary chiral scheme, their predicted couplings can be tested in the set of reactions. A good global agreement with reactions spanning nearly two orders of magnitude is obtained. Additional consistency could be found in the radiative decay of the Δ*(1700).

The history of the dynamically generated resonances, which appear in the solution of the meson-meson or meson-baryon coupled channel Lippmann-Schwinger equation (LSE) with some interaction potential, is quite old. Recent work has extended the number of dynamically generated resonances to the low lying $3/2^-$ resonances which appear from the interaction of the octet of pseudoscalar mesons with the decuplet of baryons [1, 2]. A resonance which appears in this scheme is the Δ*(1700), and in [2] the couplings of the resonance to the coupled channels $\Delta\pi$, $\Sigma^*K$, and $\Delta\eta$ were predicted.

In this talk, some extensions and applications [3– 5] of the chiral unitary scheme are summarized. The Δ*(1700), together with the N*(1535) which also appears as dynamically generated in Ref. [6], are combined in order to describe $\gamma p \to \pi^0 \eta p$; see [3] for details. Also, other terms from the chiral interaction and explicit resonances are taken into account. However, it is the Δ*(1700) which contributes dominantly through its decay into $\Delta\eta$, and thus this reaction provides a good testing ground for the large resonance coupling predicted by the model. The total cross section is shown in Fig. 1. This reaction, together with $\gamma p \to \pi^0 K^0 \Sigma^+$ is currently being analyzed at ELSA. Preliminary results [7] indicate good agreement for the cross sections in both reactions.

In order to further test the properties of the dynamically generated Δ*(1700), we have applied the model in other photoproduction reactions. The scheme for the reactions from Fig. 2 is similar as in the previously studied reactions: The Δ*(1700) is excited by the photon and decays in s-wave into $K\Sigma^*(1385)[\to K\Sigma\pi, K\Lambda\pi]$. Again, the agreement is good as Fig. 2 shows.

In addition, one can test the consistency of the model also in pion-induced reactions. The results for various channels are shown in Fig. 3. One should note the scale of the various cross sections, which are correctly reproduced by the model even if there may be minor deviations at some energies.

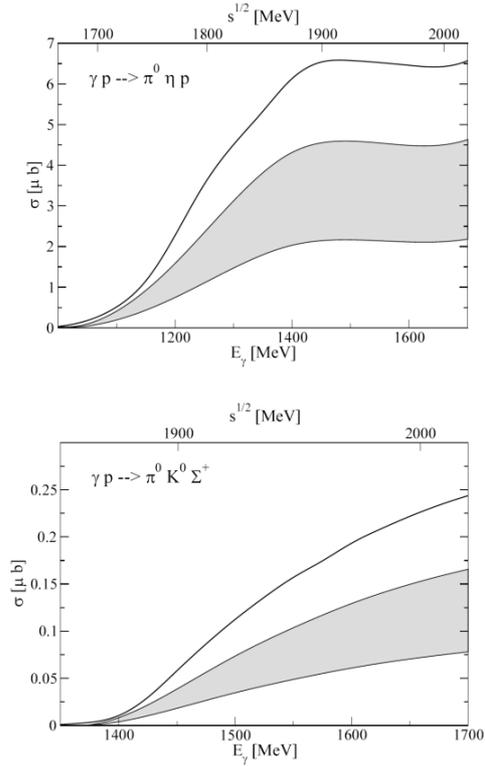

FIG. 1: Update in Ref. [4] of the predictions from Ref. [3] (gray bands) compared to the results from [3] (solid lines).

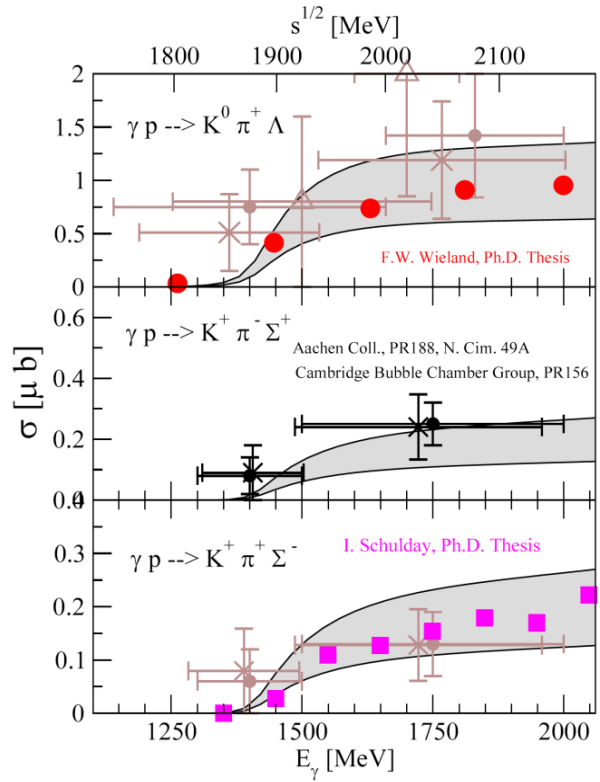

FIG. 2: Additional photon-induced reactions and theoretical predictions. The preciser data are from the Saphir experiment (see [4] for a description).

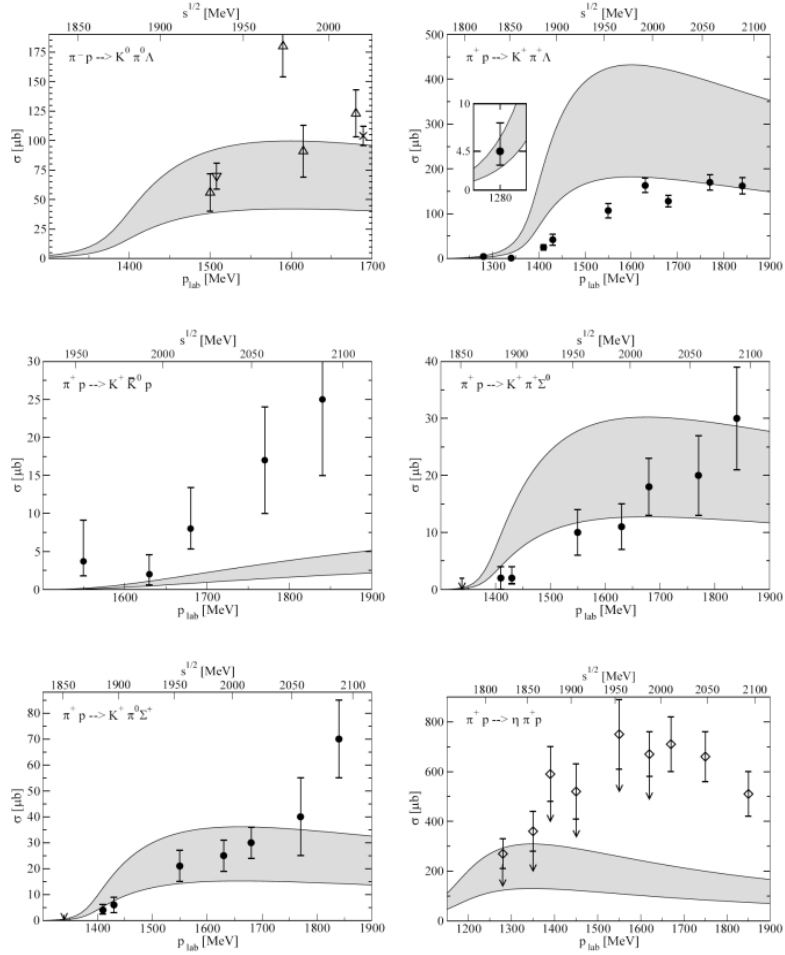

*FIG. 3: Total cross sections for the pion-induced reactions. For a description of the data, see Ref. [4].*

In all the results for the photon-induced reactions, the strength of the initial excitation of the $\Delta^*(1700)$ has been taken from the PDB. However, within the framework of dynamical generation, the phototransition, or radiative decay, can be predicted in a parameter free way. This is because the photocouplings to the constituents of the resonance (the $0^-$ meson octet and the $3/2^+$ baryon decuplet) are well known. The radiative decay has been studied in Ref. [5] with the result

$$\Gamma = 602 \pm 140 \text{ keV}$$

which should be compared to the PDG value of $570 \pm 254$ keV.

# Eta photoproduction in the nucleon resonance energy region[9]

Vitaly Shklyar[a][10], Horst Lenske[a], Ulrich Mosel[a]

[a]Institut fuer Theoretische Physik, Universitaet Giessen, D-35392 Giessen, Germany

The total $\gamma n^* \to \eta n$ cross section extracted from the γd scattering data exhibits two resonance-like bumps at W= 1.54 GeV and 1.67 GeV [1,2]. While the $S_{11}(1535)$ resonance is responsible for the first peak, there is no consistent explanation for the second structure at 1.67 GeV. The earlier prediction [3] for $\gamma n^* \to \eta n$ reaction is base on a conjecture about a narrow $P_{11}$ state with strong coupling to the $\eta N$ final state and with large $A_{1/2}^n$ helicity amplitude. Another suggestion has been made in [4] that the $D_{15}(1675)$ resonance is giving rise to $\gamma n^* \to \eta n$ close to its resonance mass.

On other hand, there are two resonances $S_{11}(1650)$ and $P_{11}(1710)$ which might also contribute to the eta photoproduction. However, since the properties of these two state are rather uncertain (even the existence of $P_{11}(1710)$ is questioned), coupled-channel calculations are crucial in order to check contributions from these states to the eta photoproduction on the neutron. It is also important that different reactions $\pi N \to \pi N$, $2\pi N$, $\eta N$, $\omega N$, $K\Lambda$, $K\Sigma$ and $\gamma N \to \gamma N$, $\pi N$, $\eta N$, $\omega N$, $K\Lambda$, $K\Sigma$ must be simultaneously analyzed in order to fix resonance properties. Here we report on our updated calculations within the effective Lagrangian coupled-channel Giessen Model. The parameters of the $S_{11}(1650)$ resonance are found to be close to that reported by the Particle Data Group (PDG). The properties of the $P_{11}(1710)$ are less known which manifest itself in large uncertainties for the branching ratios and the Breit-Wigner mass given in PDG.

We find strong contributions to the $\gamma p \to \eta p$ reaction from $S_{11}(1535)$ and $S_{11}(1650)$ resonances. The results of our calculations are compared with the experimental data in Fig 1.

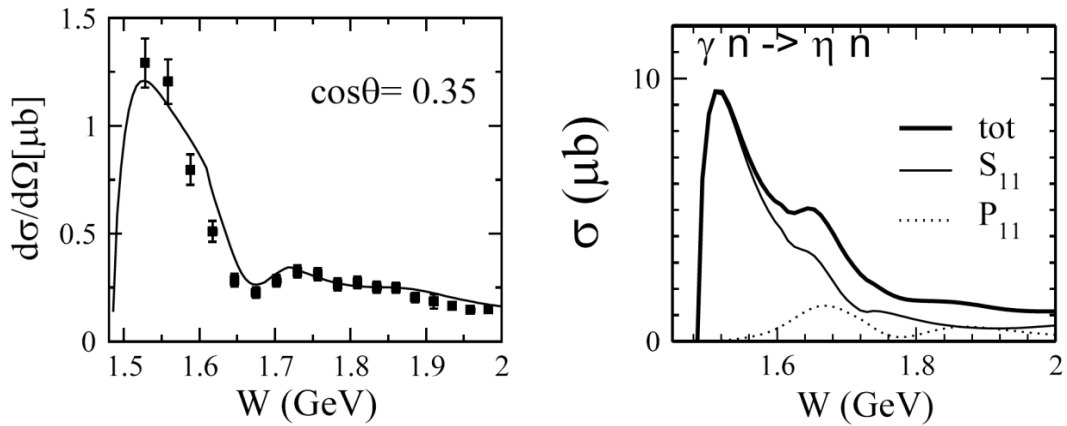

---

[9] Work supported by FZ Julich
[10] e-mail: shklyar@theo.physik.uni-giessen.de

*Figure 1. Left: result for the γp→ηp reaction. Right: total and partial wave cross sections for the γn→ηn scattering. Solid curve in bold: total cross section, solid curve: $S_{11}$-partial wave cross section, dotted curve: $P_{11}$-partial wave cross section*

We do not need an exotic state in order to describe the CB-ELSA data around 1.67 GeV. The kink structure seen in differential cross section data at this energy is driven by the $S_{11}(1650)$ contribution. Despite on small coupling to final $\eta N$ the excitation of this resonance gives a large interference effect in the $S_{11}$-partial wave.

A strong contribution from the $S_{11}(1535)$ resonance to the γn→ηn reaction is found. This result is in agreement with findings of other works. Two resonances $S_{11}(1650)$ and $P_{11}(1710)$ are responsible for the bump structure around 1.67 GeV. However, the uncertainties in electromagnetic helicity decay amplitudes $A_{1/2}^n$ do not allow to distinguish between $S_{11}(1650)$ and $P_{11}(1710)$ states. We find a strong dependence on choice of electromagnetic resonance parameters on final results [5].

In order to compare the predicted differential cross section with the experimental data extracted from γd scattering the effect of deuteron structure should be properly treated. In the present work a simplified procedure is chosen by taking only the Fermi smearing into account. We perform a Monte-Carlo simulation of the γn*→ηn scattering in the lab system where the three-momenta of the initial neutrons are distributed according to the deuteron wave function calculated in the Paris potential. The probability $A^{lab}$ to find the final eta-meson in an angular range $d(\cos\theta_\eta)$ is equal to the differential cross section $A^{lab} = \frac{d\sigma^{lab}}{d\cos\theta_\eta}(E_\gamma, \mathbf{p}_f)$, where $E_\gamma$ and $\mathbf{p}_f$ are the energy of incoming photon and the Fermi-momentum of the initial neutron respectively. Ultimately, $A^{lab}$ can be related to the invariant cross section $\frac{d\sigma}{dt}(s,t)$ which is calculated from our differential cross section in c.m.s. All events are simultaneously boosted along z-axis with $|\mathbf{v}| = \frac{E_\gamma}{E_\gamma + m_N}$ in the direction opposite to the incoming photon momenta. This procedure follows the data analysis used in CBELSA-TAPS experiment. The differential cross section in the new reference frame can be found as $\frac{d\sigma}{d\cos\theta_\eta} = \frac{\Delta N}{\Delta\cos\theta_\eta} \frac{2\pi}{N_{flux}}$, where $\Delta\cos\theta_\eta$ is the range of the bin interval and $\Delta N$ is a number of eta-mesons in the corresponding bin. The total cross section calculated using the above procedure coincides with that obtained in [5] by simple integration over the Fermi-momentum.

The results are shown in Fig. 2 for the quasi-free neutron and proton targets. One can see that the effect of Fermi-smearing can be very important even at energies significantly above the $\eta N$ threshold. The effect of the Fermi smearing enhances the differential cross section on the proton at all angles. This is due to the strong $S_{11}(1535)$ resonance contributions to $\gamma p \to \eta p$ reaction. In the $\gamma n^* \to \eta n$ scattering the only modification at forward and backward angles is found.

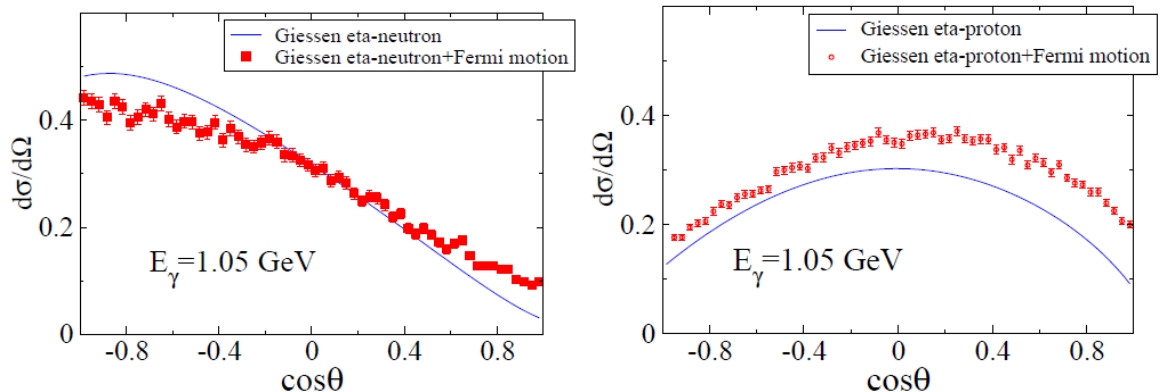

Figure 2. Effect of the Fermi smearing to the eta photoproduction on the neutron (left) and on the proton (right). Blue lines are results from Giessen Model [5], red squares are results of Monte Carlo simulation - differential cross section smeared out over the Fermi-motion

The simplified procedure described here to account for the Fermi smearing might not fully take into account the effect of the deuteron structure. To check the results a direct comparison of the calculated $\gamma p^* \to \eta p$ cross section with the experimental data is very important. More detailed calculations accounting for the deuteron structure, including the three-particle final state explicitly, interference between proton and neutron channels, are envisaged.

# Session III:

*Interaction of η and η' with nucleons and nuclei*

# Photoproduction of η-mesons off heavy nuclei and off the deuteron

B. Krusche[(a)],

[(a)] Department of Physics and Astronomy, University of Basel, Ch-4056 Basel, Switzerland

This contribution summarizes new experimental results for the photoproduction of ηmesons off nuclei that have been obtained with the CBELSA/TAPS setup at the Bonn ELSA accelerator. Two different topics are addressed:

1) Photoproduction of η-mesons off heavy nuclei in view of η-nucleus FSI (η - nucleon absorption cross section) and the in-medium properties of the $S_{11}$ resonance.

2) Resonance contributions to η-photoproduction off the neutron from quasifree photo-production off neutrons bound in the deuteron.

The motivation for the study of photoproduction of η-mesons off heavy nuclei is twofold. In the second resonance region it is an excellent tool for the study of the $S_{11}(1535)$ resonance, which completely dominates this reaction [1, 2]. The search for possible in-medium modifications of this resonance was originally triggered by the experimental observation that the second resonance bump is almost invisible in photoabsorption off nuclei. Since already for the free proton the second resonance region is a complex structure, the investigation of this excitation energy range with a reaction sensitive to a particular resonance is very attractive. The second motivation is the study of η-nucleus final state interactions (FSI), aiming at the extraction of the η-nucleon absorption cross section, respectively the mean free path of η-mesons in nuclei.

A first search for possible in-medium effects of the $S_{11}$ spectral function was done with the TAPS experiment at MAMI [3]. However, the experiment covered only incident photon energies up to 800 MeV, i.e. approximately up to the peak position of the resonance. The experimental results were in good agreement with BUU-model calculations (see e.g. [4, 5]). Furthermore, an almost perfect scaling of the cross sections with $A^{2/3}$ (A = nuclear mass number) was found, which indicates strong final state interaction effects. Subsequently, measurements at KEK [6] and Tohoku [7] extended the energy range up to 1.1 GeV. The KEK experiment reported some collisional broadening of the $S_{11}$ resonance, however possible background from ηπ final states, which can fake such a result, was neglected. The Tohoku experiment pointed to a significant contribution of a higher lying resonance to the $\gamma n \rightarrow n\eta$ reaction. However, none of these experiments covered the full line shape of the $S_{11}$.

The CBELSA/TAPS experiment has recently measured this reaction for carbon, calcium, niobium, and lead nuclei for incident photon energies up to 2.2 GeV, i.e. throughout and beyond the second resonance region. The main findings of the experiment are:

- The results agree at low incident photon energies (below 800 MeV) with the previous TAPS data [3], where they show the $A^{2/3}$ scaling typical for strong FSI.

- For inclusive η production the scaling breaks down at higher incident photon energies and approaches almost an A scaling. An investigation of the reaction kinematics and comparisons to BUU model results show that at this energies contributions from ηπ final states and secondary production of η-mesons (e.g. via the reaction chain $\gamma N \rightarrow N\eta$, $\pi N \rightarrow N\eta$) become dominant. In particular, the latter one has a completely different mass number dependence since the feeding of the η channel from pion induced reactions increases with mass number.

- The excitation function for single, quasi-free η production has been determined with cuts on the reaction kinemtaics which suppress ηπ- final states and secondary production processes. The $S_{11}(1535)$ excitation functions extracted from this data show no significant in-medium effects for the which is in agreement with recent model predictions [8].

- An analysis of the scaling properties of the η-photoproduction cross sections for kinematics which are not affected by ηπ- final states or secondary processes allows the extraction of the η-nucleon absorption cross section as function of the momentum of the η mesons. It is found almost constant at a value of 30 mb for kinetic η-energies from 20 MeV up to the maximum accessible energies of 1.2 GeV.

The study of η photoproduction off the neutron aims at the investigation of the isospin structure of the electromagnetic excitation of resonances. Due to the non-availability of free neutron targets, photoproduction off light nuclei, in particular the deuteron, must be explored. The investigation of quasifree and coherent η photoproduction off $^2$H and $^{3,4}$He (see [9] for a summary) has previously clarified the isospin structure of the $S_{11}(1535)$ electromagnetic excitation, which was found to be dominantly iso-vector. At higher incident photon energy until very recently almost nothing was known about η-photoproduction off the neutron. However, there were predictions e.g. from the η-MAID model that the ratio of neutron and proton cross sections should rise at higher incident photon energies. It should be influenced by the interference pattern of the $S_{11}(1535)$ and $S_{11}(1650)$ resonances (strongly destructive for the proton, possibly less important for the neutron) and by contributions from the $D_{15}(1675)$ resonance (stronger electromagnetic coupling to the neutron than to the proton). Furthermore, also in the framework of the chiral soliton model a state is predicted in this energy range, with has much stronger photon couplings to the neutron than to the proton and a large decay branching ratio into $N\eta$. This state is the nucleon-like member of the predicted anti-decuplet of pentaquarks, which would be a $P_{11}$ state. A first experimental result for the neutron cross section from the GRAAL experiment [10] showed a bump-like structure in the neutron excitation function around an invariant mass of 1675 MeV (incident photon energy around 1 GeV), which is not seen for the proton target.

The CBELSA/TAPS experiment has measured quasi-free η-photoproduction off protons and neutrons bound in the deuteron for incident photon energies up to 2 GeV. The data have been analyzed in two different ways in order to control systematic uncertainties arising from the detection efficiencies for the recoil nucleons (systematic uncertainties in the detection of the η-mesons can be much better controlled and in addition cancel in the neutron/proton cross section ratio). In the first analysis the neutron cross section was determined from events were the recoil neutron has been detected, and the events were corrected for the detection efficiency for recoil neutrons (which is much smaller than for recoil protons). In the second analysis, the inclusive cross section without any conditions for recoil nucleons and the cross section for events with detected recoil protons corrected for the proton detection efficiency were constructed. The difference of the two must be equal to the neutron cross section since contributions from coherent production off the deuteron are completely negligible. The two independent results for the neutron cross section are in very good agreement, setting stringent limits for systematic effects in the detection of the recoil nucleons. The resulting neutron cross section shows indeed a pronounced bump around incident photon energies of 1 GeV and a second structure around incident photon energies of 1.8 GeV, while at the same time the quasi-free proton cross section is much flatter throughout this energy region and in good agreement with the results off the free proton. The analysis of the angular distributions in view of possible resonance contributions is still under way.

Finally, it should be mentioned, that a similar analysis of meson production cross sections off the quasi-free neutron for the double $\pi^0$, $\eta'$, $\eta\pi^0$ and $\omega$ channels is also under way.

# Theoretical aspects of $\eta$ mesic nuclei and formation reactions

Satoru Hirenzaki

Depertment of Physics, Nara Women's University

Contact e-mail: zaki@cc.nara-wu.ac.jp

In the contemporary hadron physics, the light pseudoscalar mesons ($\pi$, K, $\eta$) are recognized as the Nambu-Goldstone bosons associated with the spontaneous breaking of the QCD chiral symmetry. In real world, these mesons, together with heavier $\eta'$(958) meson, show the involved mass spectrum, which are believed to be explained by the explicit flavor $SU$(3) breaking due to current quark masses and the breaking of the axial $U_A$(1) symmetry at the quantum level referred as the $U_A$ (1) anomaly [1, 2]. One of the most important subjects in hadron physics at present is to reveal the origin of the hadron mass spectra and to find out the quantitative description of hadron physics from QCD [3].

Recently, there are several very important developments for the studies of the spontaneous breaking of chiral symmetry and its partial restoration at finite density. To obtain deeper insights on the in-medium behavior of spontaneous chiral symmetry breaking, the hadronic systems, such as pionic atoms [4, 5, 6], $\eta$-mesic nuclei [7, 8, 9, 10] and $\omega$-mesic nuclei [7, 8, 11, 12, 13], have been investigated in both of theoretical and experimental aspects. Especially, after a series of deeply bound pionic atom experiments [14, 15], K. Suzuki *et al.* reported the quantitative determination of pion decay constant $f_\pi$ in medium from the deeply bound pionic states in Sn isotopes [5, 16] and stimulated many active researches of the partial restoration of chiral symmetry at finite density [4, 6, 17, 18, 19].

The $\eta$-mesic nuclei were studied by Haider and Liu [20] and by Chiang, Oset and Liu [21]. As for the formation reaction, the attempt to find the bound states by the ($\pi^+$,p) reaction led to a negative result [22]. In our study we use the theoretical models for $\eta$-nucleus interaction, as described in Refs. [9, 23] in further detail, to investigate the structure and formation of the $\eta$ mesic nuclei. In the $\eta$-nucleon system, the $N$(1535) resonance ($N^*$) plays an important role due to the dominant $\eta NN^*$ coupling. We evaluate the $\eta$-nucleus optical potential $V\eta(\omega, \rho(r))$ in the two different models which are based on distinct physical pictures of $N^*$. One is the chiral doublet model. This is an extension of the linear sigma model for the nucleon and its chiral partner [24, 25, 26]. The other is the chiral unitary model, in which $N^*$ is dynamically generated in the coupled channel meson-baryon scattering [10, 27].

The optical potentials and the calculated formation spectra show much different behaviors for these two models as shown in Refs. [9, 23]. In addition, we can also consider the quark degrees of freedom to study the $\eta$ meson in nuclear medium [28, 29]. Thus, it should be interesting and important to observe the spectra in experiments and to distinguish the theoretical models which have much different pictures for the structure of $\eta$ and N* (1535).

We believe meson-nucleus bound systems are fruitful to investigate meson properties in nuclear medium and to get new information on symmetry restoration at finite density. We also think that the realistic calculations of the structure and formation spectra are necessary for all observed results to study the meson properties in nuclear medium, and to get the decisive conclusions.

We sincerely thank to Hideko Nagahiro and Daisuke Jido for their active and fruitful collaborations on $\eta$ physics. We also thank to ETA07 workshop as a very good opportunity to get new information on current research activities on $\eta$ meson, and thank to hospitalities of Valencia university group during this stay.

# Precision Study of the $\eta^3$He System using the $dp\rightarrow{}^3$He$\eta$ Reaction at ANKE

A. Khoukaz[1]

*[1]Institut für Kernphysik, Universität Münster, 48149 Münster, Germany*

The available data sets for $pd\rightarrow{}^3$He $\eta$ production in close vicinity of the threshold [1, 2] expose discrepancies, which currently prohibit the extraction of scattering length information with sufficient precision. Therefore, the reaction $dp\rightarrow{}^3$He $\eta$ has been investigated at the ANKE spectrometer using a continuously ramped deuteron beam at excess energies ranging from below threshold up to Q=+11 MeV. Due to the full geometrical acceptance of the ANKE spectrometer high statistics data for this reaction have been obtained. The results have recently been accepted for publication and are discussed in detail in [3].

Relevant for the interpretation of total and differential cross section data very close to threshold is a precise knowledge of the apparent experimental uncertainties as well as a consideration of these effects in the later description of the data. In order to extract high precision data at ANKE, detailed studies of various topics have been performed. In detail, for each one second time bin during the COSY ramp the $\eta$ c.m. momentum $p_\eta$ was found from the size of the $^3$He momentum locus in the c.m. frame and compared with analytic formulae and simulations. Additionally, as a consistency check, for each time bin and thus each excess energy bin it has been shown that the extracted value of $p_\eta$ is independent of the azimuthal emission angle $\varphi$. As demonstrated in Fig. 1 the values of $Q=p^2_\eta/2m_{red}$, where $m_{red}$ is the $\eta^3$He reduced mass, follow the expected linear variation of the beam energy with ramp time.

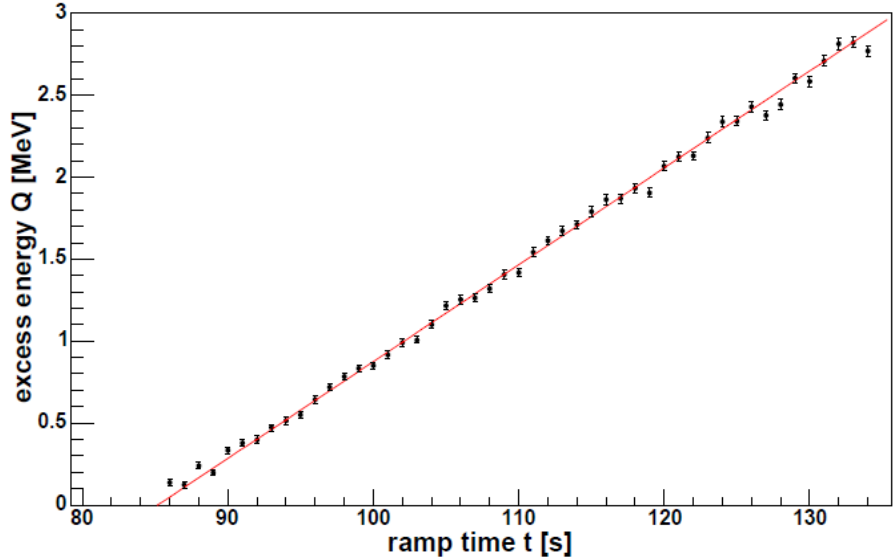

*FIG. 1: Reconstructed excess energy Q as a function of the timing information of the linearly ramped beam.*

Shown are here only data for the first three MeV of excess energy. The perfect linearity of the ramp is also visible for the complete range of excess energies as can be seen in [3]. The deviations from the linear fit are consistent with a statistical distribution of width $\sigma_{\delta Q}$ = (72 ± 11) keV. From this distribution the uncertainty in the determination of the mean value of the excess energy $Q$ itself was derived. It was found that for each time bin the mean value of $Q$ is known with a precision of ±9 keV. This excellent result is of high importance for the precise extraction of the scattering length information as will be discussed below.

The $dp\rightarrow{}^3$He $\eta$ total cross sections obtained at 195 bins in excess energy $Q$ are broadly compatible with those of SPESII [2] and any global difference is within our overall normalization uncertainty. However, in contrast to our data at low excess energies presented in Fig. 2, the SPESII results do not define firmly the energy dependence in the near-threshold region. At ANKE it was found that the total cross section reaches its maximum within 0.5 MeV of threshold and hardly decreases after that. This behavior is in complete contrast to phase–space expectations and indicates a very strong final state interaction [4].

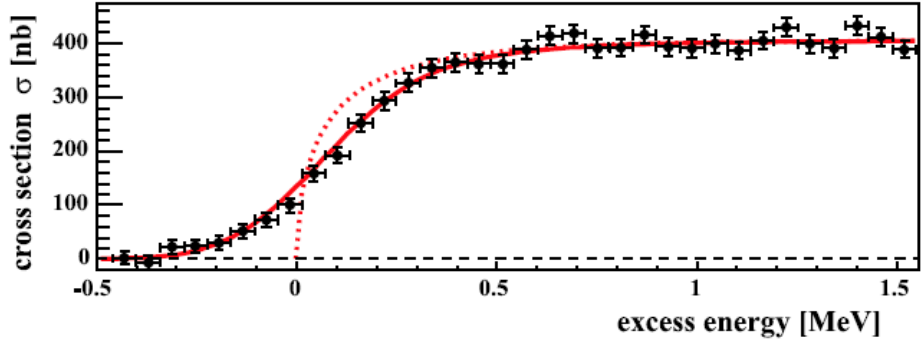

FIG. 2: Total cross section data extracted in the very near vicinity of the threshold. Equation (2) with the parameters given in Eq. (3) represents well all our results (solid line). The dotted curve is the result to be expected without the 171 keV smearing in Q. The comparison of both curves clearly displays the relevance of the consideration of the finite beam energy width for the description of the very near threshold data.

The angular average of the amplitude–squared is derived from the total cross section $\sigma_{tot}$ through

$$\overline{|f|^2} = \frac{p_d}{p_\eta} \cdot \frac{\sigma_{tot}}{4\pi} \tag{1}$$

where $p_d$ is the initial c.m. momentum. We parametrize the s–wave component $f_s$ in the form

$$f_s = \frac{f_B}{\left(1 - {p_\eta}/{p_1}\right)\left(1 - {p_\eta}/{p_2}\right)} \tag{2}$$

In the presence of a strongly attractive final state interaction, there should be a pole of $f_s$ where $|p_1|$ is very small. The second singularity of Eq. (2) has no direct physical meaning since it represents an attempt to reproduce the rest of the energy dependence, including that residing in the reaction mechanism $f_B$, in terms of a simple pole. In the scattering–length fit of SPESII [2], $p_2 = \infty$ and $p_1 = -i/a$.

The shape of the $\eta$ production below the nominal threshold shown in Fig. 2 is a very sensitive measure of the momentum width of the COSY beam, and this resolution has to be taken into account in any phenomenological analysis. Details can be found in [3]. The effect of the smearing in $Q$ is illustrated in Fig. 2, where the unsmeared parametrization is shown by the dotted curve.

It was found that any fit of Eq. (2), considering the finite COSY beam width, to the present data that neglects $p_2$ fails to satisfy simultaneously the data in the proximity of threshold ($Q \leq 1$ MeV) and at the higher energies ($Q \geq 3$ MeV). This effect becomes visible here for the first time because of the quality and extent of the data. The retention of the $p_2$ term in Eq. (2) results in the significantly better description of the whole data (solid line in Fig. 2). The higher $Q$ data might be contaminated by non-s-wave contributions. However, these have only a small effect on the value extracted for $p_1$ though they would influence the position of the effective pole $p_2$.

The fit for $Q \leq 11$ MeV gave

$$p_1 = [(-5 \pm 7^{+2}_{-1}) \pm i(19 \pm 2 \pm 1)] \, \text{MeV}/c \qquad (3)$$
$$p_2 = [(106 \pm 5) \pm i(76 \pm 13^{+1}_{-2})] \, \text{MeV}/c$$

with the resulting scattering length of

$$\begin{aligned} a &= -i(p_1 + p_2)/p_1 p_2 \\ &= [\pm(10.7 \pm 0.8^{+0.1}_{-0.5}) + i(1.5 \pm 2.6^{+1.0}_{-0.9})] \, \text{fm}, \end{aligned} \qquad (4)$$

where the first errors are statistical and the second (where given) systematic, including effects associated with the fitting range assumed.

The results show that $a$ is dominantly real with large errors on the imaginary part. The value of $|a|$ is much larger than that found in earlier work [2, 5] or in the later COSY-11 experiment [6]. This is due, in part, to the treatment of energy–smearing effects, whose need is very evident in our data with the fine energy steps near threshold.

In order to affect the cross section variation over a scale of less than 1 MeV, there must be a pole of the production amplitude in the complex plane that is typically only 1 MeV away from $Q = 0$. From our fit values we find a stable pole at $Q_0 = p^2{}_1/2m_{\text{red}} = [(-0.30 \pm 0.15 \pm 0.04) \pm i(0.21 \pm 0.29 \pm 0.06)]$ MeV.

In Fig. 3 the scattering amplitudes extracted according to Eq. (1) from the obtained ANKE measurements are compared with our fit. The dotted curve is the result to be expected without the smearing in $Q$.

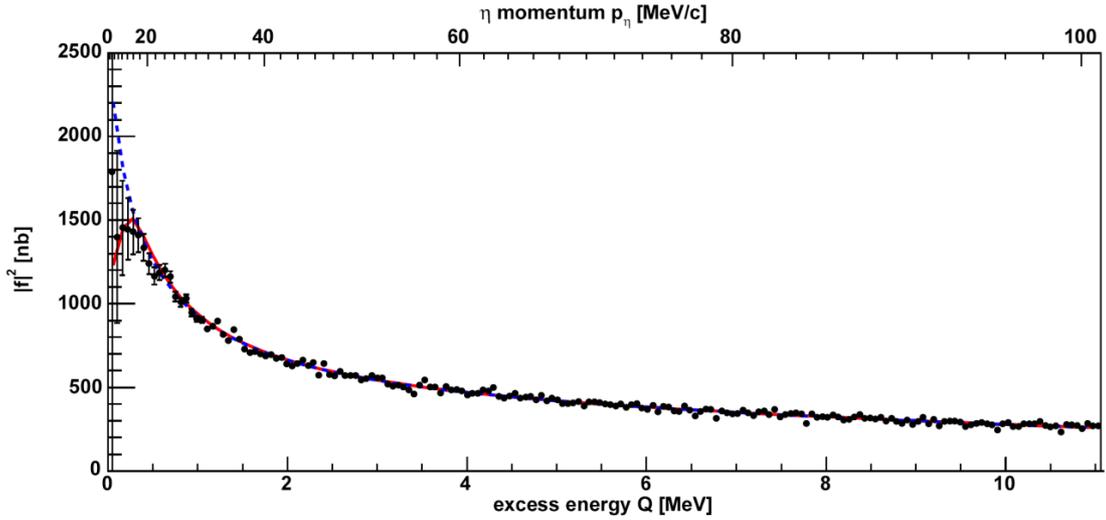

FIG. 3: *Scattering amplitudes extracted according to Eq. (1) from the obtained ANKE measurements.*

In addition to total cross section data, differential cross section have been extracted for the whole range of discussed excess energies. The observed asymmetry in the angular distribution implies that there are higher partial wave contributions even in this very near–threshold region. Defining an asymmetry parameter $\alpha$ through $(d\sigma/d\Omega)_{\text{c.m.}} = \sigma_{\text{tot}}(1 + \alpha \cdot \cos\theta_{\text{c.m.}})/4\pi$, it is seen that above an $\eta$ c.m. momentum $p_\eta$ of 40 MeV/c ($Q = 1.7$ MeV), $\alpha$ is positive and increases monotonically with $p_\eta$ but with a magnitude much larger than that found at SPESII [2]. At low momentum, both data sets show a tendency for $\alpha$ to go negative but the systematic uncertainties here are large. The slope of the cross section at $\cos\theta_{\text{c.m.}} = 0$ has the same sign as that found at higher excess energies, though the data there do not remain linear over the whole angular range [5, 7, 8]. In a recent theoretical approach this effect is addressed to a rapid variation of the phase of the near–threshold s–wave amplitude [9] and might be connected with a possible $\eta$ $^3$He quasi–bound state.

# Study of the $\vec{d}+d \to {}^4\text{He} + \eta$ reaction with polarized beam and search for $\eta$-nucleus bound state

Mariola Lesiak

for the GEM Collaboration

Institute of Physics, Jagiellonian University, Kraków, Poland

The existence of a quasi bound state of $^4$He and η meson is still an open issue [1]. The quantity, which gives indirect information about the possibility of the existence of the quasi-bound state is a complex scattering length. The sign and value of its real part is related to the binding energy of the system. The value of the imaginary part corresponds to the width of the bound state. The scattering length can be found from the s-wave amplitude [2]. Polarised data are very useful for such a study. From the measurement of the angular dependence of cross section and analyzing power, one can extract directly from the experiment contribution of *s* partial wave to the reaction amplitude.

The measurement was performed at Big Karl magnetic spectrograph [3] at the COSY accelerator in Jülich, Germany. The unpolarised as well as vector and tensor polarised deuteron beam at momentum corresponding to the excess energy of 16 MeV was used.

In the experiment charged particles were detected. The identification of the $^4$He particles was done using information about time of flight between two layers of scintillators and energy losses in one of them. The reaction of interest was identified as a peak in the missing mass spectrum.

In the measurement the total and differential cross section for unpolarised beam was measured as well as the angular distribution of analyzing power $A_{xx}$. The analysis of the obtained results is now in progress with respect to the partial wave decomposition.

# Search for η-nucleus bound states

H. Machner[a,b]

(For the GEM collaboration)

[a]Institut für Kernphysik, Forschungszentrum Jülich, Jülich, Germany
[b] Fachbereich Physik, Universität Duisburg-Essen, Duisburg, Germany

The interaction of η-meson with nucleons and nuclei and formation of η-mesic nuclei (a bound state of an η meson in a nucleus) have been topics of great interest in recent times [1]. These studies elucidate on the nature of elementary η-nucleon interaction and the behavior of an associated nuclear resonance $N^*(1535)$ in the presence of medium. In spite of various theoretical predictions an unambiguous experimental confirmation for the existence of η-mesic nucleus is still not available.

We have performed several experiments to search for the bound state of η meson in nuclei at COSY, Jülich. In one series we produce η mesons close to the production threshold with two body final states. In this region the particles in the final state are only in the *s*-state. Then one can apply a factorization for the spin averaged amplitude

$$f = \frac{f_B}{1 - iap_\eta + \frac{1}{2}ar_0 p_\eta^2} \tag{1}$$

with $f_B$ a smooth varying production amplitude and the denominator given by the final state interaction, $a$ and $r_0$ the scattering length and the effective range. One of such experiments is the reaction $\vec{d} + d \to \eta + \alpha$ using a vector and tensor polarized deuteron beam. Here we can extract the *s*-wave strength although the measurement was nor performed so much to threshold. This experiment is presented by M. Lesiak at this meeting. Another system studied is the

$$p + {}^6Li \to \eta + {}^7Be \tag{2}$$

This reaction was studied at Saclay detecting two photons [2]. Only 8 events were found which are in the range between 0 to 10 MeV excitation energy of the final nucleus. We tried to detect the recoiling $^7Be$ nucleus. since only the ground state and the first excited state at 0.7 MeV are particle stable, it is an almost exclusive experiment. Since the recoiling nucleus is highly ionizing, all detectors have to be in vacuum. A dedicated detection system for the focal plane of the magnetic spectrograph Big Karl has been built, consisting of two planes of *x* - *y* sensitive multi-wire avalanche counters followed by a $\Delta E$ - $E$ system in a huge vacuum box. The data are presently in analysis. In a further experiment the magnetic spectrometer Big Karl and a large acceptance plastic scintillator detector ENSTAR [3] have been used. Transfer reaction (p,$^3$He) on $^{27}$Al target at recoilless kinematic condition has been used to search for η-mesic nucleus in $^{25}$Mg. The momentum measurement and the identification of $^3$He particles is performed with Big Karl and its associated focal-plane detectors. In Fig. 1 the quality of particle identification in the spectrograph is demonstrated.

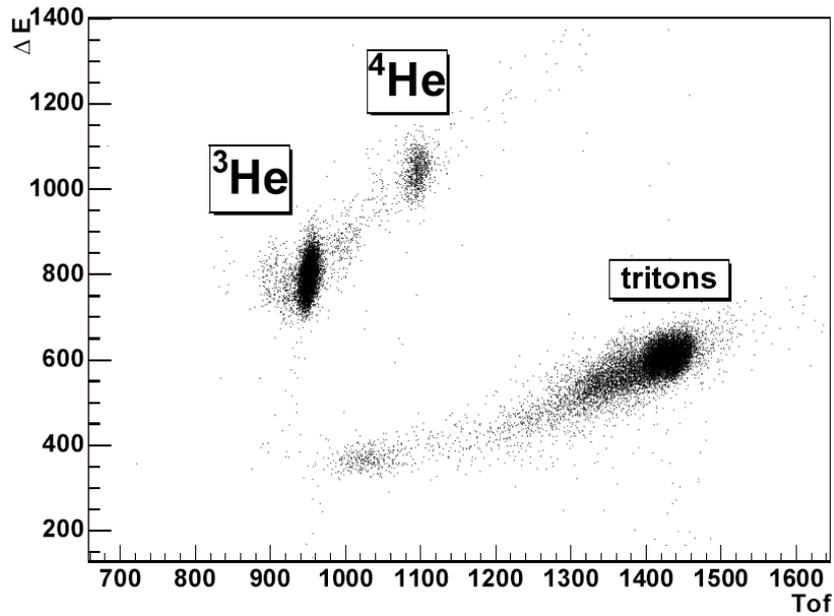

*Figure 1: Energy loss in a focal plane hodoscope of the magnetic spectrograph versus time of flight between two such hodoscopes.*

The decay products of mesic nuclei, namely protons and pions, have been measured in ENSTAR in coincidence with $^3$He particles. The events corresponding to the decay of mesic nuclei are selected by using various cuts corresponding to the correlated (p,π) decay pair. A low statistics enhancement in the missing mass spectrum of the residual "$^{25}$Mg+η"system below the threshold for the correlated events has been identified, at almost similar range of values for two different spectrometer settings.

# η- and η'(958)-mesic nuclei formation and $U_A(1)$ anomaly at finite density in Nambu-Jona-Lasinio model

H. Nagahiro[1], M. Takizawa[2] and S. Hirenzaki[3]

[1]Research Center for Nuclear Physics (RCNP), Osaka University, Ibaraki, Osaka 567-0047, Japan
[2]Showa Pharmaceutical University, Machida, Tokyo 194-8543, Japan,
[3]Department of Physics, Nara Women's University, Nara 630-8506, Japan

In the contemporary hadron physics, the light pseudoscalar mesons (π, K, η) are recognized as the Nambu-Goldstone bosons associated with the spontaneous breaking of the QCD chiral symmetry. In real world, these mesons, together with heavier η' (958) meson, show the involved mass spectrum, which are believed to be explained by the explicit flavor SU(3) breaking due to current quark masses and the breaking of the axial $U_A(1)$ symmetry at the quantum level referred as the $U_A(1)$ anomaly [1, 2].

However, as for the behavior of the $U_A(1)$ anomaly in the nuclear medium, the present exploratory level is rather poor. Although some theoretical results have been reported not only in vacuum but also at $T \neq 0$ and/or $\rho \neq 0$, there exists no experimental information on the possible effective restoration of the $U_A(1)$ anomaly at finite density. In this study, we propose the formation reaction of the η- and η'-mesic nuclei in order to investigate the η' properties, especially mass shift, at finite density [3, 4]. Since the huge η' mass is believed to have very close connection with the $U_A(1)$ anomaly, the η' mass in the medium should provide us important information on the effective restoration of the $U_A(1)$ symmetry in the nuclear medium.

We discuss the possibility of producing the bound states of the η'(958) meson in nuclei theoretically using the Nambu-Jona-Lasinio (NJL) model. We calculate the formation cross section of the η' bound states with the Green function method for the (γ,p) reaction (Fig.1) and discuss the experimental feasibility at photon facilities such as SPring-8 in Refs. [3, 4]. We conclude that we can expect to observe resonance peaks in (γ,p) spectra for the formation of η' bound states and we can deduce new information on η' properties at finite density. Detailed discussions are given in Refs. [3, 4]. These observations are believed to be essential to know the possible mass shift of η' and deduce new information on the effective restoration of the chiral $U_A(1)$ anomaly at finite density.

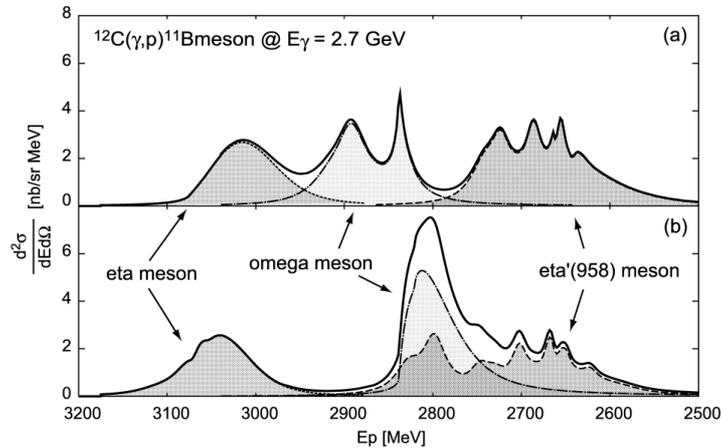

*Figure 1: Formation spectra of the η- and η'-mesic nuclei together with the ω-mesic nuclei as functions of the emitted proton energy in the $^{12}C(\gamma,p)$ reaction with (a) constant $g_D$ case and (b) energy dependent $g_D$ case.*

The experiment for the formation of the ω-mesic nucleus is already planed at SPring-8. So we expect that the η and η' mesons also can be observed by this experiment and expect to obtain new information on the η' meson and the $U_A$ (1) anomaly effect in finite density. We believe that the present theoretical results are important to stimulate both theoretical and experimental activities to study the $U_A$ (1) anomaly effect at finite density and to obtain the deeper insights of the QCD symmetry breaking pattern and the meson mass spectrum.

# The $np \to \eta d$ reaction

H. Garcilazo[1] and M. T. Peña[2]

[1] Escuela Superior de Física y Matemáticas Instituto Politécnico Nacional, Edificio 9 07738 México D.F., Mexico and Grupo de Física Nuclear and IUFFyM, Universidad de Salamanca, E-37008 Salamanca, Spain

[2] Instituto Superior Técnico, Centro de Física Teórica de Partículas, and Department of Physics, Av. Rovisco Pais, 1049-001 Lisboa, Portugal

At present the high quality data on the $np \to \eta d$ [1], and $dp \to {}^3He\eta$ reactions [2] near threshold present several challenges to theorists [3]. Namely, the $\eta$ meson interaction is stronger than expected for a pseudoscalar meson, and the $\eta N$ scattering length extracted from 2-body meson-nucleon data alone is still largely uncertain [4–7].

In a recent paper we studied the $np \to \eta d$ reaction [8]. One conclusion was that near threshold the shape of the $np \to \eta d$ cross section is determined by the three-body $\eta d$ final-state interaction. Another result was that the data implied a $\eta N$ interaction with features in between the Juelich model [4] and the Zagreb model [5], which respectively over- and underestimate the data.

Here, to determine which $\eta N$ interaction the data demands, we aim at a finer investigation by varying continuously the $\eta N$ scattering length on the $np \to \eta d$ cross-section. The method used is based on the observation that the several 2-body meson-nucleon data analyses provide strong correlations between the real part of the corresponding scattering lengths and the low-energy effective range and shape scattering parameters. Two important improvements to the calculation of Ref. [8] are now considered: 1) the unstable nature of the a meson, by including a width determined from the data for $\pi\pi$ scattering; 2) the exact treatment of the initial nucleon-nucleon interaction, by removing the approximation of freezing its value at the threshold energy.

In the figure we compare the results obtained with a dipole form for the meson-nucleon separable interaction to the ones with a gaussian form for that interaction. The results for the two possible signs for the heavy pseudoscalar meson-baryon-baryon coupling are shown in each case. We find that the real parts of the $\eta N$ scattering length and of the effective range are, respectively, Re($a_{\eta N}$) = 0.5 ± 0.1 fm and Re($r_{\eta N}$) = −2.75 ± 0.7 fm. In order to decrease even further the uncertainty in the $a_{\eta N}$ determination, it will be necessary in the future to look at other reactions, in an integrated and consistent way

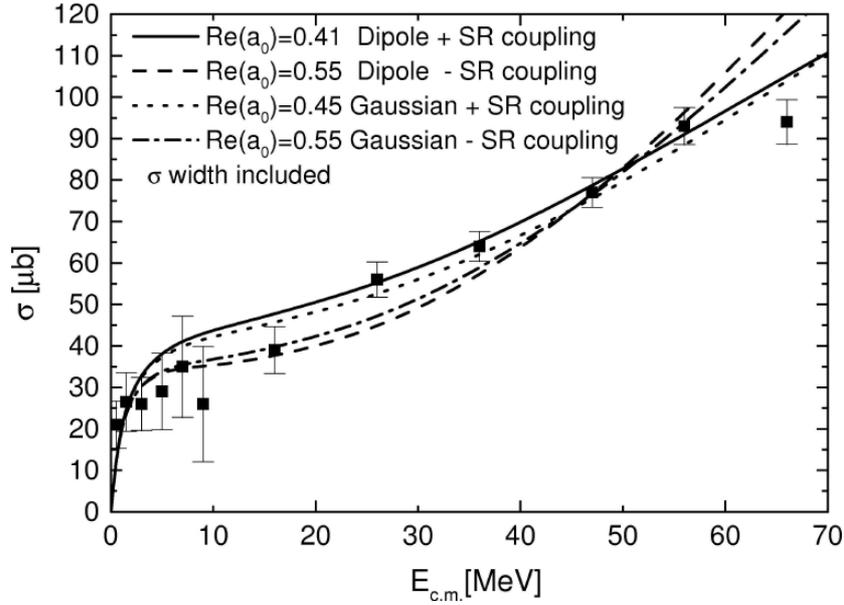

# Comparative study of proton-η and proton-η' interaction via pp and p - meson invariant mass distributions

P. Klaja for the COSY-11 collaboration

The COSY-11 collaboration continues the comparative study of interaction of the $\eta$ and $\eta'$ mesons with protons. To perform that studies the $pp \rightarrow pp\eta$ and $pp \rightarrow pp\eta'$ reactions were measured at beam momentum 2.0259 GeV/c and 3.257 GeV/c, respectively. Those momenta correspond to the excess energy Q = 15.5 MeV.

First part of investigations, namely the study of the proton-$\eta$ interaction is finished. The completed evaluation of the high-statistics measurement of the $pp \rightarrow pp\eta$ reaction was published in [1]. Qualitative phenomenological analysis of the determined differential invariant proton-proton and proton-$\eta$ mass distributions revealed an enhancement of the population density at the kinematical region corresponding to the small proton-η momentum. The effect occurs to be too large to be described by the on-shell inclusion of the proton-proton and proton-$\eta$ FSI. Also contributions from higher waves or off-shell effects of the proton-proton potential are not the sufficient explanation of that finding.

Presently, the analysis of the $pp \rightarrow pp\eta'$ reaction is in progress. We perform analysis of the $pp\eta'$ system in similar way as it has been done for the $pp\eta$ system. The determined $pp$ and *p-meson* invariant mass distributions will be used for comparative study of the interaction within *proton-meson* system.

Here, we would like to show the very preliminary results of analysis. In the figure 1 we present the invariant mass distributions for *pp* (left panel) and *pη'* (right panel). The experimental points are compared with theoretical calculations. Dotted lines correspond to the calculations where only proton-proton FSI was taken into account, whereas the solid lines depict the homogeneous phase space distribution.

In the figure 1 one can easily recognize kind of enhancement in region of small proton-$\eta'$ momentum. For the better visualisation we plotted differential cross sections as a function of the proton-proton invariant mass achieved for the *ppη'* system together with earlier results determined for the $pp \rightarrow pp\eta$ reaction. The comparison is presented in the figure 2.

The determined distributions should help to judge between various interpretations advocated in references [1, 3, 4].

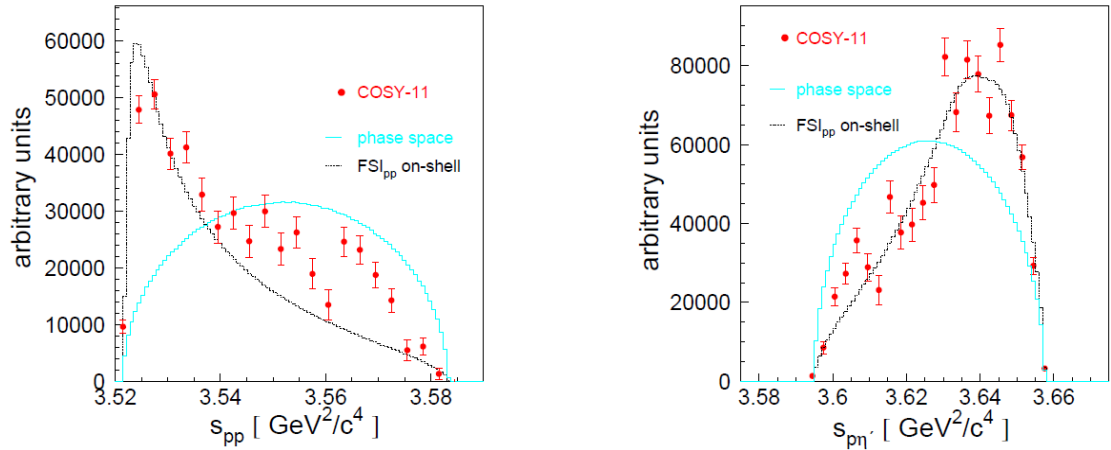

Figure 1: Distributions of the square of the proton-proton ($S_{pp}$) and proton-η' ($S_{pη'}$) invariant masses determined experimentally for the $pp \rightarrow pp\eta'$ reaction (full circles). The integrals of the phase space weighted by a square of the proton-proton on-shell scattering amplitude (dotted lines)-$FSI_{pp}$, have been normalized arbitrarily at small values of $S_{pp}$. The expectation under the assumption of the homogeneously populated phase space are shown as thick solid lines

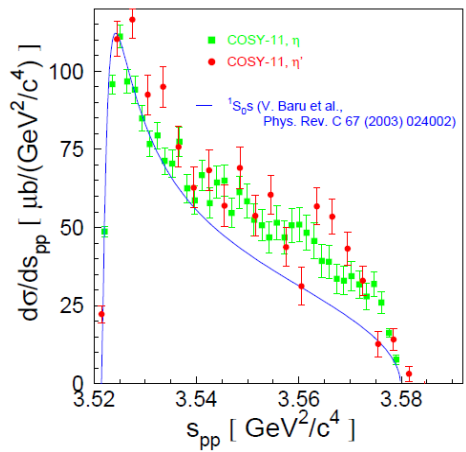

Figure 2: Distributions of the square of the proton-proton ($S_{pp}$) determined for the $pp \rightarrow pp\eta$ reaction (full squares) in comparison with results presented in the left panel of figure 1 (prescaled). Solid line corresponds to the calculations under assumption of the $^3P_0 \rightarrow {}^1S_0 s$ transition according to the model described in [2].

# Session IV:

*Meson Production*

# Production mechanisms for η and η' mesons

K. Nakayama[a]

[a]Department of Physics and Astronomy, University of Georgia, Athens, GA 30602, USA

The production of η and η' mesons in both the photonic and hadronic reactions are investigated in a combined analysis in order to learn about the relevant production mechanisms of these mesons. We consider the nucleonic, mesonic and nucleon resonance currents constructed within an effective Lagrangian approach.

The primary motivation for studying the production of mesons off nucleons and nuclei is to investigate the structure and properties of the nucleon resonances and, in the case of heavier meson productions, to learn about hadron dynamics at short range. In particular, we still lack the knowledge on the production mechanism of these heavier mesons to a large extent. Apart from pion production, the majority of theoretical investigations of meson production processes are performed within meson exchange phenomenology approach. Such an approach forces us to correlate as many independent processes as possible within a single model, if one is to extract meaningful physics information. Here, I report on our investigation of the η and η' meson production in both the photonic and hadronic induced reactions. More specifically, we perform a combined analysis of the

$$\gamma + N \to N + M$$
$$\pi + M \to M + M$$
$$N + N \to N + N + M$$
(1)

reactions, where $M$ = η, η'. The photoproduction reaction is calculated by considering the s-, u- and t-channel Feynman diagrams plus the generalized contact terms [1] which ensure the gauge invariance of the total amplitude, in addition to accounting for the final state interaction (FSI) effects [2]. The $\pi + N \to N + M$ reaction is calculated in the tree-level approximation including the s-, u-, and t-channels. For this reaction only the η-meson production is considered. To the extent that this reaction is dominated by the excitation of the $S_{11}$(1535) resonance at least for energies close to threshold, this should be a reasonable approximation if one confines ourselves to energies not too far from threshold. For higher energies, effects of the $\pi\pi N$ channel becomes important [3]. The $N + N \to N + N + M$ process is calculated in the DWBA approximation, where both the $NN$ FSI and the initial state interaction (ISI) are taken into account [4]. The $NN$ FSI is known to be responsible for the dominant energy dependence observed in the total cross section apart from that due to the phase space. As for the basic meson production amplitude, our model includes the nucleonic, mesonic and nucleon resonance currents which are derived from relevant effective Lagrangians [1, 4]. The free parameters of our model (the resonance parameters, except the resonance masses whose values are taken from the PDG, the $NN\eta$ coupling constant, and the cutoff parameter at the photon vertex in the t-channel meson exchange currents) are fixed such as to reproduce the available data in a global fitting procedure of the three reaction processes listed in (1).

## I. η MESON PRODUCTION

In this section we discuss the results of our model calculation according to the procedure outlined above. The calculation is basically the same as that reported in [1] for η', except that here we consider the production of η and that there is an additional reaction channel, $\pi N \to N\eta$.

### A. $\gamma p \to p\eta$

For photoproduction of η off nucleons, the available data basis is considerable, in particular, for the proton target. We have, not only the total and differential cross sections over a wide range of energy starting from threshold, but also the beam and target asymmetries. The recent data on the neutron target from GRAAL [5] have attracted much interest in this reaction in connection to the possibility of the existence of a narrow (exotic) baryon resonance with mass near 1.68 GeV. Here, we restrict our discussion to the $\gamma p \to p\eta$ reaction.

As far as the resonance currents are concerned, we follow the strategy adopted in [1] to include the resonances one by one until a reasonable fit is achieved. We found that a reasonable fit quality of $\chi^2/N \approx 3.7$ is achieved by considering the $S_{11}(1535), S_{11}(1650), D_{13}(1520), D_{13}(1700)$, and $P_{13}(1720)$ resonances if only the photoproduction reaction is considered. In a global fit with all the three reaction processes (1) considered, we have $\chi^2/N \approx 5.7$. Inclusion of more resonances didn't change the fit quality considerably. In particular, we found that no higher spin resonances ($D_{15}$ and $F_{15}$) were necessary to reproduce the existing data, including the beam asymmetry. However, we were unable to reproduce the measured target asymmetry. This requires a further detailed study. The spin-3/2 resonances are important in reproducing the measured angular distributions in the range of $T_\gamma$ = 1.07-1.6 GeV and the beam asymmetry. We emphasize that the resonance parameter values in our model are highly correlated to each other and that the existing data are insufficient to establish a unique set of these parameters. The relatively small cross sections measured at higher energies and backward angles constrain the nucleonic current contribution to be very small, so that the $NN\eta$ coupling constant is compatible with zero.

### B. $\pi^- p \to n\eta$

The total cross section is nicely reproduced up to W≈1.6 GeV, where it is dominated by the $S_{11}$ resonances, especially, the $S_{11}(1535)$. We underpredict the measured total cross section at higher energies due to the absence of the $\pi\pi N$ contribution via the coupled channel [3] in our model. The $P_{13}(1720)$ resonance is important in reproducing the structure exhibited by the measured angular distributions at higher energies.

### C. $NN \to NN\eta$

This process is particularly relevant in connection to the role of the $\eta N$ FSI. Most of the existing calculations take into account the effects of the $NN$ FSI in one way or another which is well known to influence the energy dependence of the cross section near threshold. Calculations which include the $\eta N$ FSI to lowest order reproduce the bulk of the energy dependence exhibited by the data. However, they fail to reproduce the $pp$ invariant mass distribution measured by the COSY-TOF [6] and COSY-11 collaborations [7]. In ref. [8], it has been emphasized the importance of the three-body nature of the final state in the S-wave in order to account for the observed $pp$ invariant mass distribution. Other authors have suggested an extra energy dependence in the basic production amplitude [9] to reproduce the existing data. Yet another possibility has been offered which is based on a higher partial wave(P-wave) contribution [10]. We observe that all what is required to reproduce the measured $pp$ invariant mass distribution is an extra $p'^2$ dependence, where $p'$ denotes the relative momentum of the final $pp$ subsystem. Obviously, this can be achieved either by an S-wave or by a P-wave contribution. Note that the $NN$ P-wave ($^3P_0$) can also yield a flat proton angular distribution as observed in the corresponding data. Although the model calculation of [10], based on a stronger P-wave contribution, reproduces nicely the shape of the measured pp invariant mass distributions, it underpredicts the total cross section data near threshold (Q <30MeV). Here we show the new results, based on a combined analysis of the $\gamma p \to p\eta$, $\pi^- p \to n\eta$ and $NN \to NN\eta$ reactions, which reproduce the currently existing data on $NN \to NN\eta$. The major difference from the previous calculation [10] is a much stronger spin-3/2 resonance contribution, in particular, those from the $D_{13}$ resonance. In contrast to the $S_{11}$ resonances, the $D_{13}$ contribution follow more closely the empirically observed energy dependence of the total cross section.

In any case, as pointed out in ref.[10], the measurement of the spin correlation function should settle down the question on the S- versus P-wave contributions in a model independent way.

## II. η' MESON PRODUCTION

In contrast to the η meson production reactions, not many data exist for the η' meson production reactions. In photoproduction, only the earlier total cross section data from ABBHHM collaboration [11] were available until late 90's. More recently, the differential cross section data from the SAPHIR [12] and the CLAS [13] collaborations became available. The latter are high precision measurements. Currently, there is some initiative to measure the beam asymmetry by the CLAS collaboration. For $NN \to NN\eta'$, there are total cross sections for excess energies up to Q≈150MeV and the differential cross sections at two excess energies. (See, however, P. Klaja's contribution to this meeting, where the preliminary data on the $pp$ and $p\eta'$ invariant mass distributions in this reaction have been reported.)

### A. $\gamma p \to p\eta'$

With the only data available so far for $\gamma p \to p\eta'$, it is not possible to have a stringent constraint on the resonance parameters. In fact, there are many sets of parameter values which can reproduce the data equally well [1]. In order to impose a more stringent constraints on these parameters, one requires more exclusive data, in particular, the spin observables, such as the beam asymmetry. A common feature of the model results corresponding to different sets of parameters is the bump structure in the total cross section around W=2.1GeV. If this structure is confirmed experimentally, it is likely to be due to the $D_{13}(2080)$ and/or $P_{11}(2100)$ resonance. The angular distributions at higher energies and larger angles restricts the value of the $NN\eta'$ coupling constant to be not much larger than $g_{NN\eta'} = 2$. The value of $g_{NN\eta'}$ is of particular interest in connection to the spin content of the nucleon.

## B. $pp \rightarrow pp\eta'$

In this reaction, the dominant production mechanism is found to be the excitation of the $S_{11}$ resonance, which give rise to an energy dependence of the total cross section close to that observed experimentally. Note that the energy dependence exhibited by the $pp \rightarrow pp\eta'$ reaction is markedly different from that of $pp \rightarrow pp\eta$. Both the (measured) total and differential cross sections are rather well reproduced.

[1]  K. Nakayama and H. Haberzettl, Phys. Rev. C 69 065212 (2004); ibid. C73 045211 (2006).
[2]  H. Haberzettl, K. Nakayama, and S. Krewald, Phys. Rev. C74 045202 (2006).
[3]  A. Gasparyan, Ch. Hanhart, J. Haidenbauer, and J. Speth, Phys. Rev. C74 045202 (2006).
[4]  K. Nakayama, J. Speth, and T-.S. H. Lee, Phys. Rev. C65 045210 (2002).
[5]  V. Kuznetsov et al., Phys. Lett. B647 23 (2007).
[6]  M. Abdel-Bary et al., Eur. Phys. J. A 16 127 (2003).
[7]  P. Moskal et al., Phys.Rev. C69 025203 (2004).
[8]  A. Fix and H. Arenhövel, Phys. Rev. C69 014001 (2004).
[9]  A. Deloff, Phys. Rev. C69 035206 (2004).
[10] K. Nakayama, J. Haidenbauer, C. Hanhart, and J. Speth, Phys. Rev. C68 045201 (2003).
[11] ABBHHM Collaboration, Phys. Rev. 175 1669 (1968); Nucl. Phys. B108 45 (1976).
[12] R. Plötzke et al., Phys. Lett. B444 555 (1998); J. Barth et al., Nucl. Phys. A691 374c (2001).
[13] M. Dugger et al., Phys. Rev. Lett. 96 062001 (2006), Erratum-ibid. 96 169905 (2006).

# Isospin dependence of the η' meson production in collisions of nucleons

Joanna Przerwa for the COSY-11 collaboration

According to the quark model, the masses of η and η' mesons should be almost equal. However, the empirical values of these masses differ by more than the factor of two. Similarly, though the almost the same quark-antiquark content, the total cross section for the creation of these mesons close to the kinematical thresholds in the $pp \to ppX$ reaction differs significantly. Using the COSY-11 detection setup we intend to determine whether this difference will also be so significant in the case of the production of these mesons in the proton-neutron scattering. Additionally, the comparison of the $pp \to pp\eta'$ and $pn \to pn\eta'$ total cross sections will allow to investigate the η' meson structure and the reaction mechanism and may provide insight into the flavour-singlet (perhaps also into gluonium) content of the η' meson.

Since the quark structure of η and η' mesons is very similar, in case of the dominant isovector meson exchange – by the analogy to the η meson production – we can expect that the ratio $R_{\eta'}$ should be about 6.5. If however η' meson is produced via its flavour-blind gluonium component from the colour-singlet glue excited in the interaction region the ratio should approach unity after corrections for the initial and final state interactions [1]. The close–to–threshold excitation function for the $pp \to pp\eta'$ reaction has already been determined [2-5] whereas the total cross section for the η' meson production in the proton-neutron interaction is still unknown.

In August 2004 –for the first time– using the COSY–11 facility we have conducted a measurement of the η' meson production in the proton-neutron collision.

A quasi-free $pn \to pnX$ reactions were induced by a proton beam in a deuteron target. For the data analysis the proton from the deuteron is considered as a spectator which does not interact with the bombarding proton. The experiment is based on the registration of all outgoing nucleons from the $pd \to p_{sp}pnX$ reaction. Protons are measured in two drift chambers and scintillator detectors, neutrons are registered in the neutral particle detector. Protons considered as a spectators are measured by the dedicated silicon-pad detector. Application of the missing mass technique allows to identify events with the creation of the meson under investigation and the total energy available for the quasi-free proton-neutron reaction can be calculated for each event from the vector of the momenta of the spectator and beam protons.

Due to the smaller efficiency and lower resolution for the registration of the quasi-free $pd \to pn\ meson$ reaction in comparison to the measurements of the proton-proton reactions, the elaboration of the data encounters problems of low statistic. However one can extract the number of registered $pd \to pn\ meson$ events from the missing mass distribution provided that the contribution of the continuous spectrum originating from the multi-pion production can be disentangled from the signal resulting from the production of the investigated meson. This can be done by comparison of the missing mass distribution for the negative values of Q, when only pions may be created, and the missing mass distribution for Q larger than zero. In case of the positive values of Q a signal from the η(η') meson is expected on the top of the multi-pion mass distribution. Example of application of this method [6] in the analysis of the quasi-free $pn \to pn\eta'$ reaction is shown in the fig 1. Figure 1 (left) presents missing mass spectra for Q larger than zero (red line) and the corresponding background histogram (blue line). Figure 1 (right) shows the spectrum of the missing mass for Q larger than zero after the subtraction of the background.

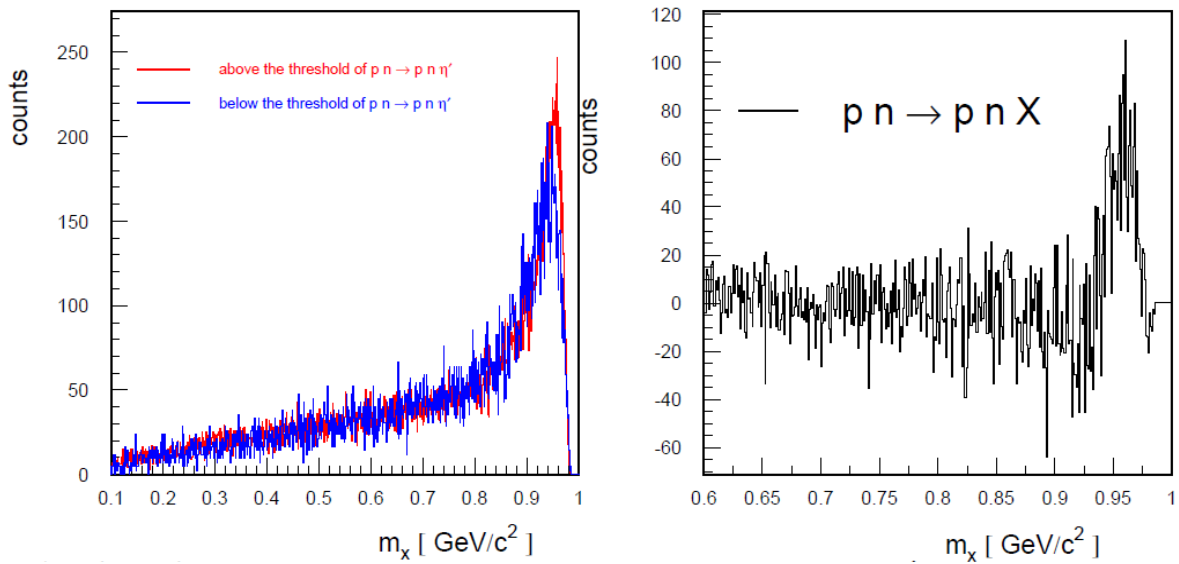

Figure 1: (left) Missing mass spectra of the $pn \to pn\eta'$ process determined for the excess energies larger and smaller than zero. (right) Missing mass spectrum for Q larger than 0 after the subtraction of the multi-pion background.

At present the analysis aiming for establishing the excitation function for the $pn \to pn\eta'$ reaction is in progress and will deliver the values for the total cross section in the excess energy range between 0 and 20 MeV.

# Direct determination of the width of η'

Eryk Czerwiński on behalf of the COSY–11 collaboration

The precise knowledge of total width of the η' meson ($\Gamma_{\eta'}$) will be important for the studies of branching ratios and in general for the investigations of the η' decays conducted for example at WASA-at-COSY or KLOE–2 experiments.

There are two direct measurements of the total width of the η' meson. First experiment took place at the NIMROD accelerator ($\Gamma_{\eta'}$= 0.28 ± 0.10 MeV/c$^2$) and the second one at the SATURNE ($\Gamma_{\eta'}$= 0.40 ± 0.22 MeV/c$^2$). In September/October 2006 the COSY–11 collaboration measured reaction pp → ppη'. The total width of the η' will be obtained via missing mass spectra established from measured four-momentum vectors of protons. Figure 1 shows preliminary spectra obtained at five different beam momenta. Presented spectra were determined with preliminary calibration of the detector system and taking into account the whole statistics. The width of signals is a sum of our experimental resolution and $\Gamma_{\eta'}$. Figure 1a shows that the achieved experimental resolution is smaller than ~0.4 MeV/c$^2$, when using stochastically cooled proton beam. Whereas the $\Gamma_{\eta'}$ remains constant, the width of the observed missing mass signals gets smaller with decrease of available energy, this is due to kinematical factors which depend on the access energy. Therefore, measurements at five different beam momenta allow to control systematical errors. From the momentum distributions of the elastically scattered protons, we observed small (10$^{-4}$) changes of beam optics which may account for abnormal flat top of signal in the figure 1d. In previous experiments with worse accuracy this effect wasn't observed.

It is worth noting that the obtained precision of the mass determination is comparable with the natural width of the η' meson, and, to our best knowledge, it is by order of magnitude better than ever achieved in the experimental studies of hadrons.

Off-line analysis is in progress and after calibrations of all detectors and taking into account all corrections (i.e. changes of beam optics) final results will be presented.

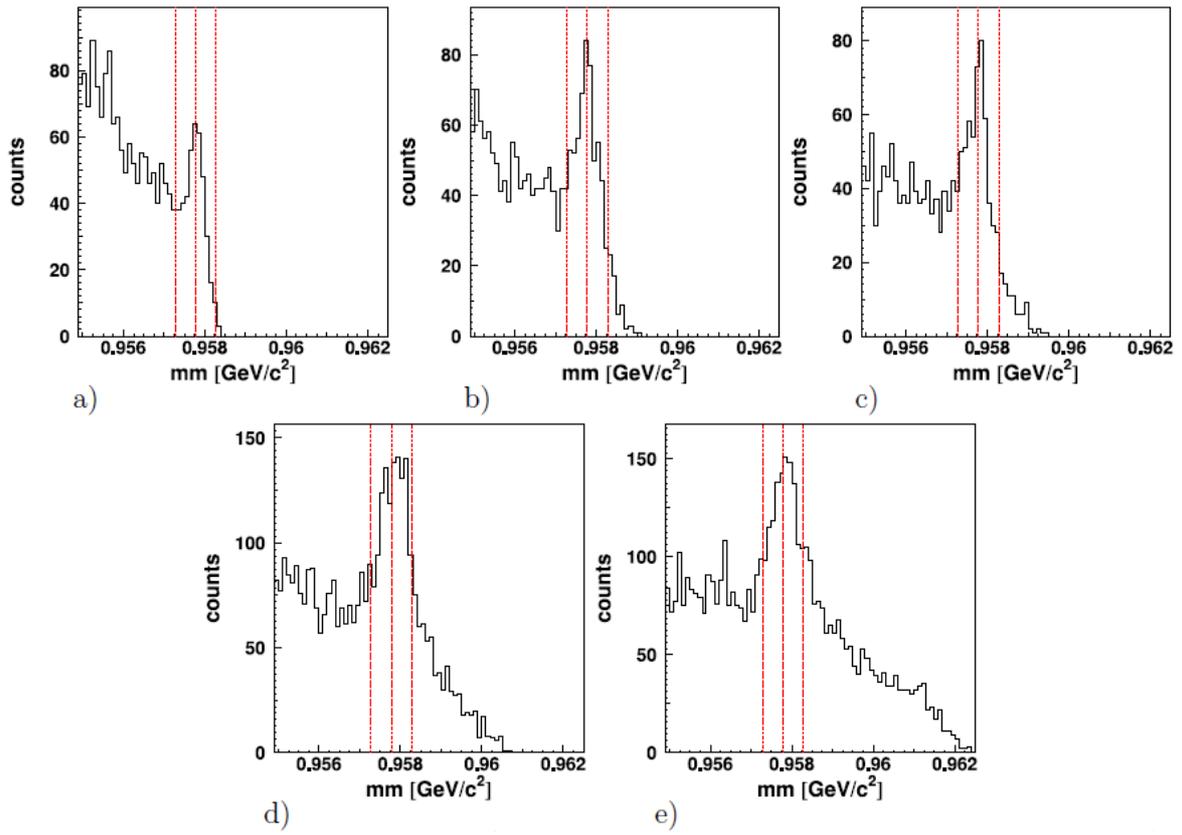

Figure 1: Missing mass spectra of pp → ppη' reaction conducted at COSY–11 detection setup for a) 3211 b) 3213 c) 3214 d) 3218 e) 3224 GeV/c beam momentum. Dashed-dotted lines represent the range of ±0.5 MeV around the mass of the η' meson.

# High statistics production of π⁰ and η via deuteron-proton reactions close to the η production threshold at COSY-11

Jerzy Smyrski

*Institute of Physics, Jagiellonian University, PL-30-059 Kraków, Poland* for COSY-11 collaboration

The $^3$He-η interaction is of attractive nature and it might be strong enough to lead to existence of $^3$He-η bound state. We studied this interaction by measurement of the $dp \to {}^3$He $X$, $X=\eta,\pi^0$ reactions near the η-meson production threshold, where the η-meson production was used to determine the $^3$He-η scattering length and the $\pi^0$ production below the η-threshold was used for a search of a signal from decay of $^3$He-η bound state. The experiment was performed during a slaw acceleration of the internal deuteron beam of COSY scattered on a proton target of the cluster jet type and the COSY-11 facility detecting the $^3$He ions. The produced neutral mesons were identified by means of the missing mass method.

Our results for the $dp \to {}^3$He$\eta$ total cross section published in Ref. 1 are shown in Fig. 1. A scattering length fit to our data using the FSI enhancement factor from Ref. 2 and a contribution from P-wave described by the penetrability factor of the centrifugal barrier results in the $^3$He-η scattering length of $|a|=4.3\pm0.5$ fm. The forward-backward asymmetries of the differential cross sections deviate clearly from zero for c.m. momenta above 50 MeV/c (see Fig. 2). The asymmetries are described assuming S- and P-wave interference in the final state.

For a search of a resonance-like structure originating from decays of $^3$He-η bound state in the $dp \to {}^3$He$\pi^0$ channel we concentrated on angular range of pions covering the forward angles in the c.m. system, where the $dp \to {}^3$He$\pi^0$ cross section is up to two orders of magnitude smaller than at the backward angles. The excitation function for the $dp \to {}^3$He$\pi^0$ reaction does not show any structure which could originate from decay of $^3$He-η bound state. This can not be treated, however, as a proof of non existence of $^3$He-η bound state, since the signal from decays of such state can be too week to be seen in the present experiment.

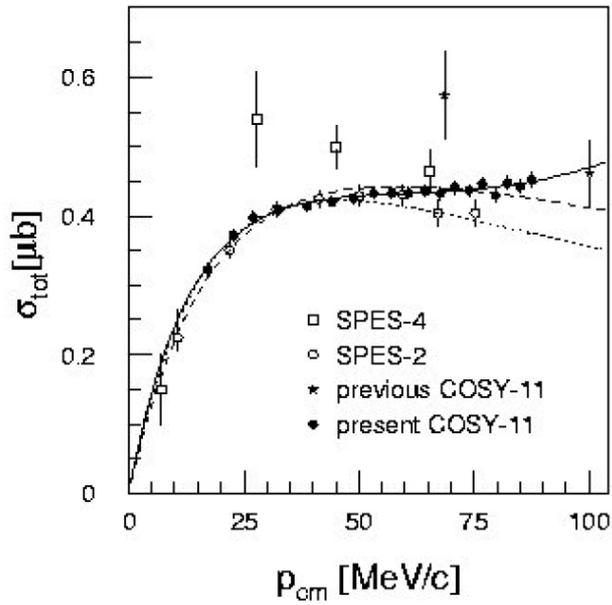

Fig. 1 Total cross section for the $dp\rightarrow{}^3$He$\eta$ reaction. The dotted line represents scattering length fit to present data for $p_{cm}$ < 50 MeV/c, the dashed line corresponds to fit to in the full momentum range, and the solid line represents fit with included P-wave contribution determined according to the centrifugal barrier model.

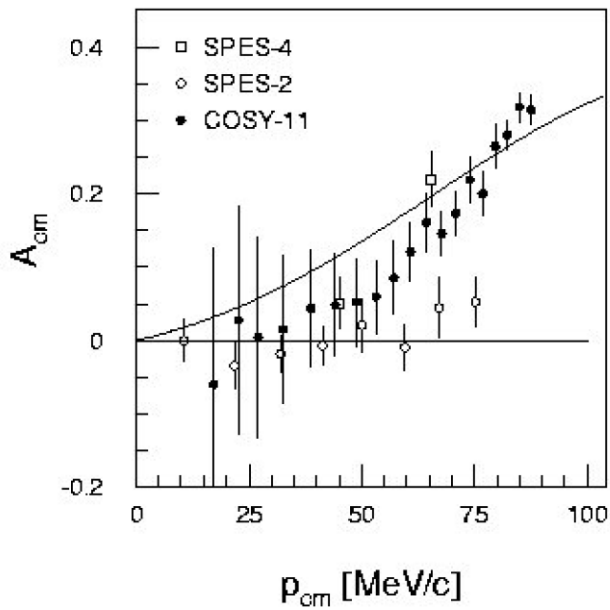

Fig. 2 Forward-backward asymmetries in the c.m. system. The solid line represents calculations including S- and P-wave interference in the final state under assumption of a constant phase between the two waves.

# Hadronic production of $\eta$ mesons: Spin degrees of freedom[11]

Colin Wilkin

*Physics & Astronomy Dept., UCL, London, WC1E 6BT, UK*

In order to explain the momentum dependence of the $dp \to {}^3\text{He}\eta$ angular distribution near threshold, a rapid variation of the phase of the near–threshold $s$–wave amplitude seems to be required. Such a variation is consistent with that expected from an $\eta\,{}^3\text{He}$ quasi–bound state. This could be clarified further through the measurement of deuteron and proton analysing powers. A brief comment is also made on the recent measurement of the analysing power in the $\vec{p}p \to pp\eta$ reaction.

New measurements of the $dp \to {}^3\text{He}\eta$ differential cross section close to threshold are now available from ANKE [1] and COSY-11 [2]. The results are broadly compatible with those published from Saturne [3,4] but they are of much higher quality and extend over a greater range in excess energy $Q$. The very rapid rise of the cross section within a fraction of an MeV of threshold implies that there is a pole in the production amplitude in the immediate vicinity, possibly associated with a quasi–bound $\eta\,{}^3\text{He}$ state [5].

The data also show a linear dependence on $\cos\theta_\eta$ that gets steadily stronger between $Q = 4$ and 11 $MeV$ but with no sign of the $\cos^2\theta_\eta$ term observed at higher energies [6]. Data with a polarised deuteron beam, and possibly a polarised hydrogen target, should soon be available from ANKE [7] and so it is important to see what extra information could be derived from a study of the analysing powers.

There are two independent $s$–wave $dp \to {}^3\text{He}\eta$ amplitudes ($A$ & $B$) [8] and five $p$–wave. In order to get a cross section linear in $\cos\theta_\eta$, we retain just two ($C$ & $D$) to leave the operator

$$\hat{f} = A\vec{\varepsilon}\cdot\hat{p}_p + iB(\vec{\varepsilon}\times\vec{\sigma})\cdot\hat{p}_p + C\vec{\varepsilon}\cdot\vec{p}_\eta + iD(\vec{\varepsilon}\times\vec{\sigma})\cdot\vec{p}_\eta, \qquad (1)$$

that has to be sandwiched between ${}^3\text{He}$ and proton spinors. Here $\vec{p}_p$ and $\vec{p}_\eta$ are, respectively, the c.m. momenta of the proton and $\eta$–meson, $\vec{\varepsilon}$ the polarisation vector of the deuteron. The corresponding unpolarised differential cross section

$$\frac{d\sigma}{d\Omega} = \frac{p_\eta}{3p_p} I, \qquad (2)$$

where

$$I = |A|^2 + p_\eta^2 |C|^2 + 2|B|^2$$
$$+ 2p_\eta^2 |D|^2 + 2p_\eta \left[ Re(A^*C) + 2Re(B^*D) \right]\cos\theta_\eta, \qquad (3)$$

has the desired linear dependence on $\cos\theta_\eta$, with a slope parameter

---

[11]Presented at the ETA07 workshop, Peñiscola, May 10-11, 2007.

$$\alpha = \frac{d}{d(\cos\theta_\eta)} \log\left(\frac{d\sigma}{d\Omega}\right)\bigg|_{\cos\theta_\eta=0}$$
$$= 2p_\eta \frac{Re(A^*C) + 2Re(B^*D)}{|A|^2 + p_\eta^2|C|^2 + 2|B|^2 + 2p_\eta^2|D|^2}. \tag{4}$$

The deuteron and proton analysing powers become [9]

$$\sqrt{2}\, I t_{20} = 2(|B|^2 - |A|^2) + (|D|^2 - |C|^2)p_\eta^2(3\cos^2\theta_\eta - 1)$$
$$+ 4p_\eta \cos\theta_\eta\, Re(B^*D - A^*C),$$
$$I t_{21} = \sqrt{3}\,(Re(A^*C - BD^*)p_\eta \sin\theta_\eta$$
$$+ (|C|^2 - |D|^2)p_\eta^2 \sin\theta_\eta \cos\theta_\eta),$$
$$2I t_{22} = \sqrt{3}\left(|D|^2 - |C|^2\right)p_\eta^2 \sin^2\theta_\eta,$$
$$I it_{11} = \sqrt{3}\, Im(A^*C - BD^*)p_\eta \sin\theta_\eta,$$
$$I t_{10} = 0,$$
$$I A_y^p = 2\, Im\left(A^*D + BD^* + CB^*\right)p_\eta \sin\theta_\eta. \tag{5}$$

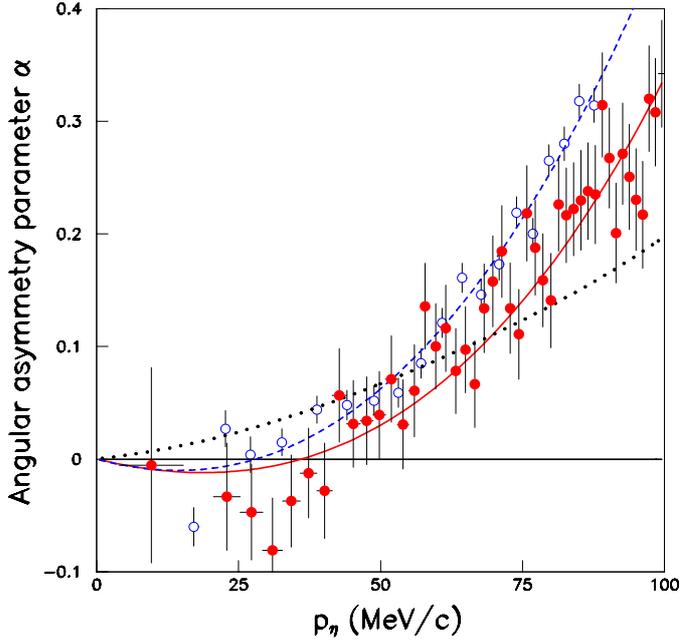

FIG. 1. Asymmetry parameter $\alpha$ measured at ANKE [1] (closed circles) and COSY-11 [2] (open circles). Solid and dashed lines are fits to these data taking the $s$-wave phase variation into account, which is neglected for the dotted line.

The SPES-IV experiment showed that $t_{20}$ is small near threshold, which means that in this region $|A| \approx |B|$ [3]. The effects of $p$-waves at slightly higher energies should be felt first through interference with the $s$-waves, not only in $t_{11}$, $t_{21}$, and $A_y^p$, but also in the angular asymmetry $\alpha$, for which there are now the good measurements presented in Fig. 1 [1,2]. These do not show the linear dependence on $p_\eta$ from the origin, that might be expected on the basis of Eq. (4), and this is due in part to the rapid decrease of $|A|$ and $|B|$ caused by the $\eta^3$He final state interaction. If one neglects any phase variation in the amplitudes, then this effect induces the small curvature illustrated by the dotted curve in Fig. 1, which does not describe well the data.

However, a quasi-bound state would also induce a sharp variation in the <u>phase</u> of the $s$-wave amplitude and this can change the momentum dependence of $\alpha$. Following Ref. [1], we assume

$$A = B = \frac{f_B}{(1 - p_\eta / p_1)(1 - p_\eta / p_2)}, \qquad (6)$$

where the values of the real parameter $f_B$ and complex pole positions $p_i$ have been taken from fits to the ANKE data [1]. The phases of the $p$-wave amplitudes are expected to vary slowly so that we take $C = D =$ complex constant. By adjusting the magnitude and phase of $C$, the values of either the ANKE or COSY-11 results can be well described though, in this preliminary analysis, the effects of the beam smearing and the distortion of the ANKE fits through the $|C|^2$ term in Eq. (4) have been neglected. The truncation in the $p$-wave amplitude basis must also be noted

It is clear that the angular asymmetry data seem to require a strong variation of both the magnitude and phase of the $s$-wave amplitudes that is consistent with the existence of an $\eta^3$He quasi-bound state very close to threshold, though the extra information afforded by the angular distributions may not necessarily lead to a better determination of the parameters of the pole. Similar signals might exist in the momentum dependence of $t_{21}$ and $t_{11}$ and the angular dependence of $t_{20}$ though, as seen from Eq. (5), the overall strength could be reduced significantly if $A \approx B$ and $C \approx D$. In this event, the proton analysing power looks more promising. Any possible cancellation between the amplitudes could be eliminated by measuring the proton analysing power for an $m = 0$ deuteron beam because this is proportional to $Im(A^*D)$.

The only other measurement of an analysing power in the hadronic production of the $\eta$-meson has been reported for the $\vec{p}p \to pp\eta$ reaction [10]. Here $A_y$ was shown to be small at both $Q = 10$ and $36$ $MeV$ but it was claimed that the results were closer to those of a $\pi$-exchange model [11] rather than $\rho$-exchange [12]. We would like to point out that the latter calculation used vector dominance and was based upon the sole measurement of the target analysing power in $\gamma\vec{p} \to \eta p$ [13]. These data are actually quite difficult to reconcile with other $\eta$-photoproduction measurements within the framework of existing models. If one takes instead the predictions from the ETA-MAID amplitude analysis [14], this suggests that the analysing power in $\vec{p}p \to pp\eta$, though still small, has quite a different form from that presented in Ref. [12]. Hence, until this situation is clarified, it must be clearly understood that there are no reliable predictions of $A_y$ for $\vec{p}p \to pp\eta$ on the market.

# The $A_{1/2}$, $S_{1/2}$ form factors of the $N^*(1535)$ as a dynamically generated resonance

D. Jido[1], M. Döring[2] and E. Oset[3]

[1] Yukawa Institute for Theoretical Physics, Kyoto University, Kyoto, 606-8502, Japan
[2] Department of Physics and Astronomy, University of Georgia, Athens, GA 30602, USA
[3] Departamento de F´ısica Te´orica and IFIC, Centro Mixto Universidad de Valencia-CSIC, Institutos de Investigaci´on de Paterna, Aptdo. 22085, 46071 Valencia, Spain

In this talk, we have discussed the electromagnetic transition form factors of the *N*(1535) resonance, assuming the *N*\* resonance is dynamically generated in the coupled channels of the meson-baryon scatterings. The theoretical tool for this picture is the chiral unitary approach [1, 2], in which the *N*\* resonance is described as a pole of the scattering amplitudes obtained by summing up a series of diagrams non-perturbatively to restore unitarity of amplitude. The model reproduces well the empirical observation that the coupling strength of the *N*\* to the *ηN* channel is larger than the that to the *πN* channel. In addition, it is important to point out that, in this approach, the coupling strength of *N*\* to the *KΣ* channel is also large in comparison with other channels. Since the *N*\* is lying below the *KΣ* threshold, the *N*\* can be regarded as the quasi bound state of the *KΣ* [1]. This implies that the *N*\* has large components of strangeness.

The transition form factors are evaluated in this meson-baryon picture of *N*\* by considering diagrams in which the photon excites *N* to *N*\* via photon couplings in the meson-baryon loop of $\pi^-p$, $\pi^\circ n$, $\eta n$, $K^+\Sigma^-$, $K^\circ\Sigma^\circ$, $K^\circ\Lambda$ for the neutron resonance (Q = 0) and $\pi^\circ p$, $\pi^+n$, $\eta p$, $K^+\Sigma^\circ$, $K^\circ\Sigma^+$, $K^+\Lambda$ for the proton resonance (Q = 1). The calculations of these diagrams are straightforward. The elementary vertexes among the meson, baryon and photon are given by chiral Lagrangians, and couplings of *N*\* to meson and baryon are given by the chiral unitary approach. We have confirmed that the divergences coming from the loop integral of each diagram are canceled after summing up all the diagrams.

We have calculated the helicity amplitudes $A_{1/2}$ and $S_{1/2}$ for the *N*\*s with Q = 1 and Q = 0. Unfortunately our present model underestimated the $A_{1/2}$ amplitude with Q = 1. One thing which we should emphasize here is that, to extract the helicity amplitudes from the experimental data, one has assumed the N* width as 150 MeV, while in this approach the width is around 90 MeV. So the normalizations of the amplitudes might be inconsistent. We have also shown the neutron-proton ratio of the helicity amplitudes. The PDG values of the helicity amplitudes with the real photon is around -1/2 with a large error. Our estimation of the ratio is around -1, which is within the error. It is interesting noting that, if we take only *πN* channels in the loops, the result becomes much worse. This implies that the strange component is important to reduce the ratio.

# Singularity structure of the η-nucleon scattering matrix

A. Švarc, S. Ceci and B. Zauner
Ruđer Bošković Institute, Bijenička cesta 54, 10000 Zagreb, Croatia.

The general structure of all coupled-channel models is identical: the same type of Dyson-Schwinger integral equation is always solved, but the channel-resonance vertex interaction is treated differently: the approach varies from phenomenological to microscopic [1]. Consequently, all models contain two similar types of scattering matrix singularities: full and bare. While the full scattering matrix singularities are unanimously identified with "measurable" scattering matrix poles, the interpretation of bare poles, related to the vertex interaction, is a matter of rousing disputes. Desirous ones attempt to identify them with quark-model resonant states [2], while the second, more cautious ones give them certain measure of physical importance, but strongly refrain from giving them such a tempting physical meaning [3]. The main reason for such a disagreement lies in the understanding that the poles of the interaction potential do arise only from the assumed model, and as such do not reveal much dynamics of the interaction [4, 5]. Despite that, the possibility that within a well defined model a link might be to established between the interaction vertex poles and quark-model resonant states, is still quite appealing.

Without taking sides we would like to present how the tempting possibility of identifying bare propagator poles with quark-model resonant states "holds water" in a case of Carnagie-Melon-Berkeley type [6] coupled channel model with limited input.

*The model:*

1) CMB model with three channels: πN elastic, πN → ηN and $\pi^2$N - effective 2-body channel
2) Input:
   a) πN elastic: VPI/GWU single energy solution [7]
   b) πN→ηN: Zagreb 1998 PWA data [8]
3) Comparison is made to Capstick-Roberts constituent quark model states [9] in a way that we have:
   a) performed a constrained fit by fixing the bare propagator poles to the values of lowest quark-model resonant states and compared the full scattering matrix poles with experiment [10]
   b) released the fit and followed the change

*Identification:*

1) bare propagator pole positions ⇐⇒ quark-model resonant state masses
2) imaginary part of the dressed propagator pole ⇐⇒ decay width

*Results:*

ranging from reasonable to speculative.

*Reasonable:*

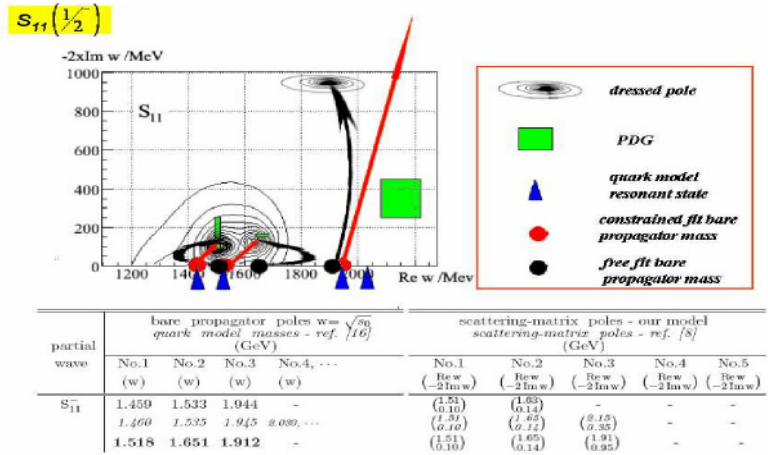

FIG. 1: Scattering matrix poles for the $S_{11}$ partial wave.

*Speculative:*

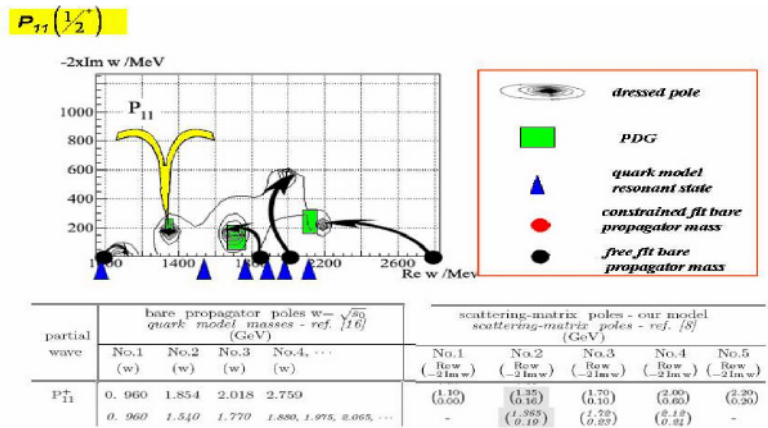

FIG. 2: Scattering matrix poles for the $P_{11}$ partial wave.

We conclude that for this model there is no obvious contradiction in understanding bare propagator poles as quark model resonant states. If such an assumption is legitimate, a mechanism is established how to account for the "missing resonance problem".

However, even if this assumption turns out to be incorrect, we have still:

i. shown the mechanism how a physical resonance gets generated from a bare propagator pole
ii. demonstrated that the well known statement: "the Roper resonance is, contrary to most other resonances, dynamically generated" remains confirmed for the CMB type model even with limited input data base.

# Production of $\omega$ in pd $\to$ $^3$He$\omega$ at CELSIUS/WASA

K. Schönning for the CELSIUS/WASA collaboration

*Department of Nuclear and Particle Physics, Box 535, S-75121 Uppsala, Sweden*
*(karin.schonning@tsl.uu.se)*

The CELSIUS/WASA collaboration has pd $\to$ $^3$He$\omega$ data at two different beam energies; at $T_p = 1450$ MeV and at $T_p = 1360$ MeV, which correspond to $p_\omega^* = 280$ MeV/c and $p_\omega^* = 144$ MeV/c, respectively. At these energies, the WASA detector[1] covers the full phase space. We have detected all final state particles and can thereby separate the two most important decay channels: $\omega \to \pi^+\pi^-\pi^0$ (BR = 89.1%) and $\omega \to \pi^0\gamma$ (BR = 8.7%).

The data analyzed in this work are from May, 2005, with a proton beam at $T_p = 1450$ MeV impinging on a deuterium pellet target[2][3]. The $^3$He's were preliminary identified by the simple dE/E method described in [4]. The $^3$He hypothesis was then tested by comparing the expected energy deposits in the detector layers traversed by the particle to the energy deposits that were actually measured.

Let's start by looking at all events containing one good $^3$He candidate and at least two detected photons. The acceptance for pd $\to$ $^3$He$\omega$, taking both geometry and detector efficiencies into account, is 49%. The event reduction is mainly due to nuclear interactions that the $^3$He's undergo when traversing the detector layers. By reconstructing the missing mass of the $^3$He and fitting a gaussian peak on top of a polynomial background to our data, we get 5900 $\omega$ candidates.

We consider first the decay channel with the largest branching ratio, $\omega \to \pi^+\pi^-\pi^0$, and apply constraints optimised for $\omega \to \pi^+\pi^-\pi^0$ selection[12]. This reduces the background significantly and gives a pd $\to$ $^3$He$\omega, \omega \to \pi^+\pi^-\pi^0$ acceptance of 39%. Thus we expect to find $5900 * 0.891 * 0.39 / 0.49 = 4200$ $\omega \to \pi^+\pi^-\pi^0$ candidates in our event sample. From fitting a gaussian peak on top of a polynomial background in the $^3$He missing mass spectrum, we find 3800 $\omega$ candidates, which differs from what we expect from the full $^3$He sample with not more than 11%.[13] The background under the peak can be fairly well reproduced with simulated pd $\to$ $^3$He$\pi^+\pi^-\pi^0$ data assuming phase space production, as shown in figure 1. To make sure that the systematic uncertainties are reasonably small, we also look into the $\omega \to \pi^0\gamma$ decay channel. The acceptance with the cuts optimized for selection of this channel is 20%, meaning that having 3800 $\omega$ candidates in the $\omega \to \pi^+\pi^-\pi^0$ case, we expect to find $3800 * 0.2 * 0.087 / (0.891 * 0.39) = 190$ $\omega \to \pi^0\gamma$ candidates. After subtracting the background, by fitting a gaussian peak on top of a polynomial, we find 160 candidates in our $^3$He missing mass

---

[12]for details, see the slides from the presentation or contact the author

[13]The systematic uncertainty in the number of extracted $\omega$ candidates is larger when we use the whole $^3$He sample and not the $\omega \to \pi^+\pi^-\pi^0$, since the background is better known in the latter case

spectrum. This agrees with what we expect within 19%. The background under the $\omega$ peak is fairly well consistent with phase space $pd \to ^3He2\pi^0$ (figure 2).

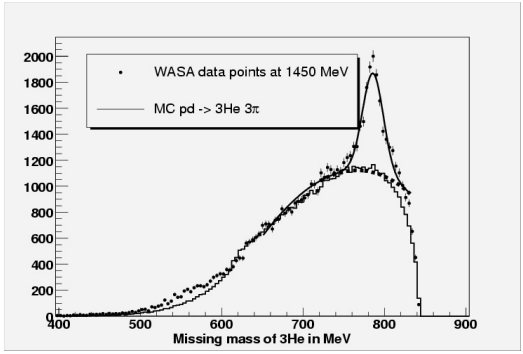

Figure 1: WASA data taken at $T_p = 1450 MeV$ with cuts optimised for $pd \to ^3He\omega, \omega \to \pi^+\pi^-\pi^0$ (dots), simulated $pd \to ^3He\pi^+\pi^-\pi^0$ (histogram), gaussian + polynomial fit (line) and polynomial (dashed line).

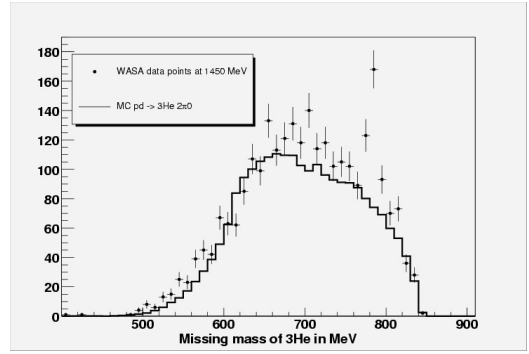

Figure 2: WASA data with cuts optimised for $pd \to ^3He\omega, \omega \to \pi^0\gamma$ and simulations of $pd \to ^3He2\pi^0$ (histogram)

Now we can confidently return to the $\omega \to \pi^+\pi^-\pi^0$ sample to extract the differential cross section. The data are divided into 14 bins with respect to $cos\theta_\omega^*$ [14]. In each $cos\theta_\omega^*$ region, the $^3He$ missing mass distribution is reconstructed. The number of $\omega$'s in each region is obtained in two ways: by fitting a polynomial to the background and subtracting from the data, and by fitting and subtracting simulated $pd \to ^3He\pi^+\pi^-\pi^0$ data. The difference in the number of $\omega$'s obtained in each way is a measure of the systematic uncertainty from the background. Summing up the number of $\omega$'s in each bin gives 3600 $\omega$'s in the entire $cos\theta_\omega^*$ range, which is in good agreement with the 3800 $\omega$'s extracted from the missing mass distribution for the entire $\omega \to \pi^+\pi^-\pi^0$ sample. This shows that our data are reliable with systematic uncertainties well under control. The acceptance is rather smooth in $cos\theta_\omega^*$ and fairly model-independent. After correcting for the acceptance, the absolute normalisation of the data is performed using the data point at $cos\theta_\omega^* = -0.65$ from SPES3 [5]. The result is shown in figure 3.

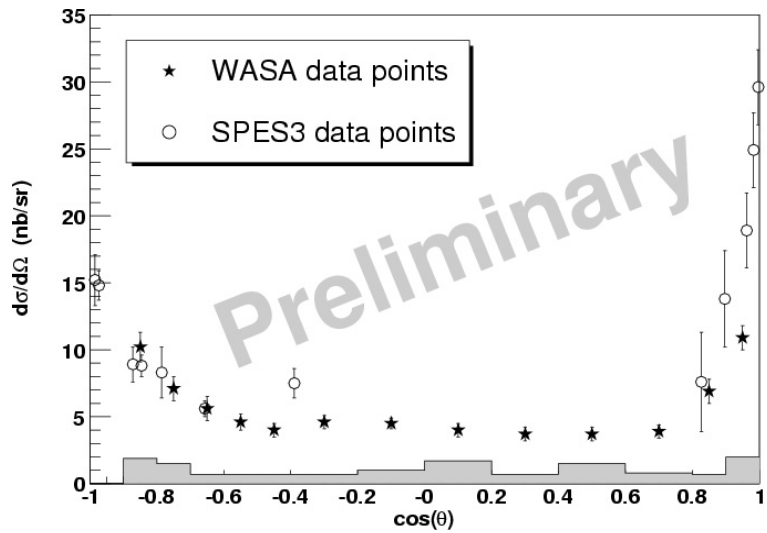

---

[14] The bin size is apparent from the histogram showing the systematic uncertainties in figure 3.

*Figure 3: The new WASA data points (stars) and the SPES3 data (circles) from [5]. The error bars represent statistical uncertainties and the grey area shows the systematic uncertainties for each WASA point.*

Generally, our data agree with SPES3 data within uncertainties, except for very small angles where we did not observe the sharp a rise seen in [5]. For theory comments on the angular distribution, I refer to the contribution by Dr. Khemchandani.

# A study of the $pd \to {}^3He\omega$ reaction

K. P. Khemchandani [15]

*Departamento de Física Teórica and IFIC, Centro Mixto Universidad de Valencia-CSIC, Institutos de Investigación de Paterna, Aptd. 22085, 46071 Valencia, Spain.*

An interest in the $pd \to {}^3He\omega$ reaction has been raised due to the on-going experimental investigation of the same by the CELSIUS/WASA collaboration at two beam energies, viz., 1360 MeV and 1450 MeV [1]. A measurement of full angular distribution at these two energies, which are about 17 MeV and 64 MeV above threshold respectively, has been carried out. Amongst the three earlier measurements [2,3,4] of the same reaction in the threshold region; [2] lead to only preliminary results, [3] remains unpublished and the analysis in [4], where an anomalous suppression of squared amplitude was claimed in the threshold region, has been questioned in [5]. Besides, the authors in [4] assumed isotropic angular distributions, expecting S-wave dominance in the threshold region. Contradictorily, the results from [3] showed a very anisotropic angular distribution at 1450 MeV but it was argued in [4] that these results are above the so called "suppression region". In dearth of availability of good and well analyzed data on the $pd \to {}^3He\omega$ reaction, the results from [1], with one measurement falling in the suppression region and one being above it, can be very helpful in obtaining some new information.

A theoretical study of this reaction has been carried out using a two step model as shown in Fig.1. This mechanism favors sharing of the large momentum transfer which is about ~ 1 GeV at both the energies. The T-matrix for this production mechanism can be written as

$$\langle |T_{pd \to {}^3He\omega}| \rangle = i\frac{3}{2} \frac{dP_1}{(2\pi)^3} \frac{dP_2}{(2\pi)^3} \sum_{int\, m's} \langle pn|d \rangle \langle \pi^+ d|T_{pp \to \pi d}|pp \rangle \frac{1}{K_\pi^2 - m_\pi^2 + i\varepsilon} \langle \omega p|T_{\pi n \to \omega p}|\pi^+ n \rangle \langle {}^3He|pd \rangle.$$

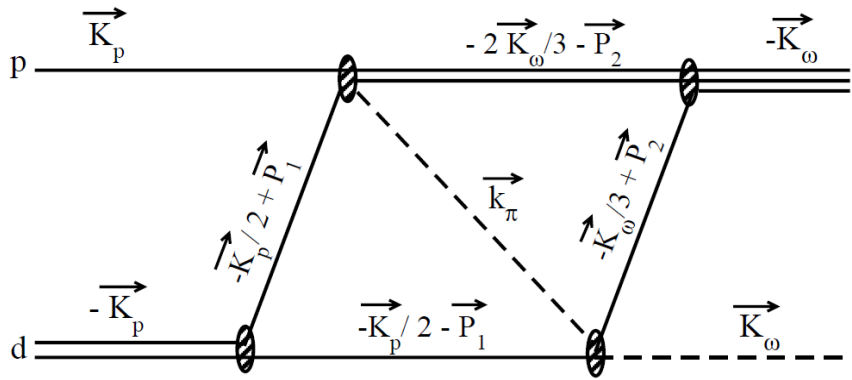

Figure 1. The two step production mechanism for the $pd \to {}^3He\omega$ reaction.

---

[15] kanchan@ific.uv.es

The $pp \to \pi d$ vertex has been written in terms of a parameterized T-matrix [6], the deuteron wave function has been written in terms of the Paris potential [7] and for the $\langle {}^3He|pd \rangle$ overlap function, the parameterization available from [8] has been used. Finally, the $\pi N \to \omega N$ sub-process has been written in terms of the Giessen model [9], which is an effective Lagrangian model solved by taking seven coupled channels in account, viz., $\gamma N$, $\pi N$, $2\pi N$, $\eta N$, $\omega N$, $K\Lambda$ and $K\Sigma$, for simultaneous analysis of all the data up to 2 GeV in terms of 11 isospin 1/2 resonances. It was shown in [9] that the data on $\pi N \to \omega N$ reaction could be explained well by taking up to l=3 partial waves in to account and that S-wave alone did not suffice even in the very near-threshold region, indicating important role of multiple resonances. Hence, above calculations have also been done by taking all partial waves up to l=3 in to account. The result of the calculations performed in the plane wave approach, at the beam energy of 1450 MeV, is shown in Fig.2 with the preliminary data available from the CELSIUS/WASA group [1]. There seems to be a disagreement between the data and these calculations at the extreme angles.

Since this data is close to the threshold, its important to check the effect of the $\omega - {}^3He$ final state interaction and this work is in progress presently. A preliminary calculation, at the beam energy = 1360 MeV, in the plane wave approach shows forward peaked angular distributions. Also, a calculation at various energies from threshold to about 1500 MeV of beam energy has been done which does indeed show a suppression in the threshold region. However, it remains to check if the suppression arises due to the input $\pi N \to \omega N$ T-matrix or from purely kinematical reasons.

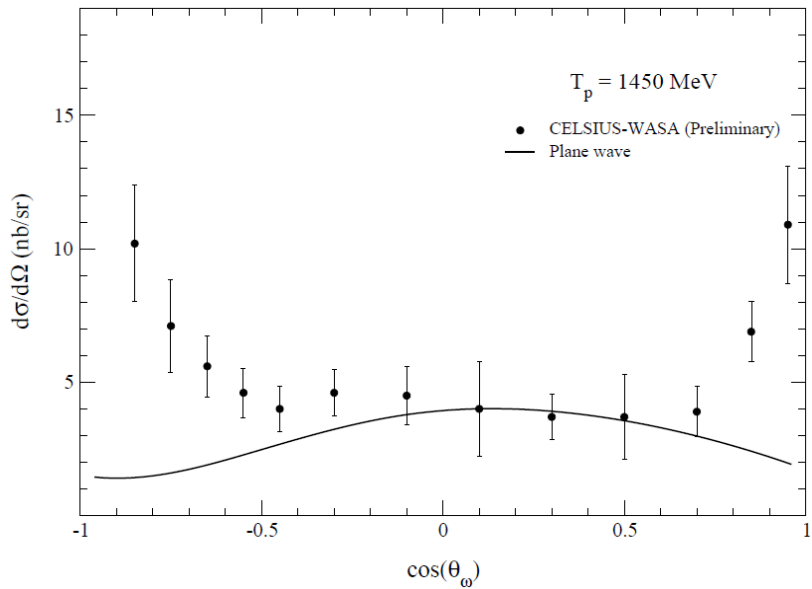

Figure 2. The angular distribution for the $pd \to {}^3He\,\omega$ reaction at a beam energy = 1450 MeV. The error bars on the data [1] include systematic as well as statistical uncertainties.

Finally, I must thank V. Shkylar for providing the t-matrices for the $\pi N \to \omega N$ and $\omega N \to \omega N$ reactions and C. Hanhart for useful discussions.

# σ- Channel Low-Mass Enhancement in Double- Pionic Fusion*

M. Bashkanov[a], H. Clement[a], O. Khakimova[a], F. Kren[a], A. Pricking[a], T. Skorodko[a] and G.J. Wagner[a] for the CELSIUS-WASA Collaboration

(a) Physikalisches Institut der Universität Tübingen, Germany

The double-pionic fusion process has been measured at CELSIUS-WASA for the first time exclusively and kinematically complete in the region of the ΔΔ excitation for the basic reaction $pn \to d\pi^0\pi^0$ at different energies as well as for $pd \to {}^3He\,\pi^0\pi^0$ and $pd \to {}^3He\,\pi^+\pi^-$ at T p = 0.9 GeV.

In all cases we observe a huge low-mass enhancement in the σ-channel of the ππ invariant mass distribution. This enhancement at the ππ threshold is much larger than anticipated from previous inclusive measurements, where it was named ABC effect. The analysis of all differential data shows that a ΔΔ system is generated in the intermediate step with a strong mutual attraction indicating even a quasibound system.

The observed invariant mass and angular distributions reveal the ABC-effect to be a σ channel phenomenon associated with the formation of a ΔΔ system in the intermediate state. The most pronounced feature is the enormous low-mass enhancement in the observed $\pi^0\pi^0$ invariant mass distributions, which in the $\pi^0\pi^0$ channel is larger than in the $\pi^+\pi^-$ channel, since the latter also contains isovector contributions [1]. In contrast to previous inclusive measurements and theoretical predictions we observe no high-mass enhancement.

The differential distributions for the $\pi^0\pi^0$ channels can be well described, if a strong attraction between two Δs in the intermediate state or even a bound ΔΔ system is assumed - as predicted previously [2], [3]. Such an ansatz is capable of describing not only the data on d and ³He, but also the results of previous inclusive measurements on ⁴He as well as the resonance-like energy dependences of the respective total cross sections. The latter are in favor of a substantial binding between the two Δs.

* supported by BMBF, DFG (Europ. Graduate School) and COSY-FFE

# Excitation of the Roper Resonance in Single and Double-Pion Production*

T. Skorodko[(a)], M. Bashkanov [(a)], H. Clement[(a)], O. Khakimova[(a)], F. Kren[(a)], A. Pricking[(a)] and G.J. Wagner[(a)] for the CELSIUS-WASA Collaboration

(a) Physikalisches Institut der Universität Tübingen, Germany

The Roper resonance has been a puzzle ever since its detection in πN phase shifts [1]. In most investigations no apparent resonance signatures could be found in the observables. Not only its nature has been a matter of permanent debate, also its resonance parameters show a big scatter in their values. New evaluations of data show the pole of the Roper resonance to be nearly 100 MeV below its canonical value of 1440 MeV with a width not much different from that of neighboring baryon states. Its decay into Nσ is found to be the most dominant process in contrast to previous findings.

In the $pp \to np\pi^+$ reaction measured at CELSIUS-WASA at incident energies above 1GeV a pronounced resonance structure is observed at $M_{n\pi^+} \approx$ 1355 MeV with Γ ≈ 140MeV. These numbers agree very favourably with recent SAID πN phase shift results [2] for the Roper pole as well as with the very recent BES results [3] from $J/\Psi \to \bar{N}N^*$

With the pole position being roughly 100 MeV below the previously believed value of the N*(1440), also its decay branchings (defined at the pole position) change dramatically. From near-threshold two-pion production, where Roper excitation is the only process of significance, we observe - in contrast to other findings - the decay $N^* \to \Delta\pi$ to be much smaller than the $N^* \to N\sigma$ decay. The latter being the dominant decay mode points to a breathing mode nature of the Roper resonance.

*supported by BMBF (06 TU 261), DFG (Europ. Grad. School) and COSY-FFE

# List of participants:

1. **Fabio Ambrosino**, INFN-Napoli
2. **Mikhail Bashkanov**, Tuebingen University
3. **Reinhard Beck**, Bonn University
4. **Marcin Berlowski**, INS Warsaw
5. **Caterina Bloise**, LN Frascati
6. **Bugra Borasoy**, Bonn University
7. **Fabio Bossi**, LN Frascati
8. **Hans Calen**, Uppsala University
9. **Tiziana Capussela**, INFN-Napoli
10. **Heinz Clement**, Tuebingen University
11. **Eryk Czerwinski**, Jagiellonian University
12. **Rafal Czyzykiewicz**, Jagiellonian University
13. **Michael Doering**, Universidad de Valencia
14. **David Duniec**, Uppsala University
15. **Rafel Escribano**, Autonoma University Barcelona
16. **Kjell Fransson**, Uppsala University
17. **Göran Fäldt**, Uppsala University
18. **Paolo Gauzzi**, Roma "La Sapienza"
19. **Karim Ghorbani**, Lund University
20. **Christoph Hanhart**, FZ-Juelich
21. **Satoru Hirenzaki**, Nara Women's University
22. **Bo Höistad**, Uppsala University
23. **Marek Jacewicz**, Uppsala University
24. **Daisuke Jido**, Yukawa Institute of Kyoto University
25. **Bengt Karlsson**, Uppsala University
26. **Kanchan P. Khemchandani**, Universidad de Valencia
27. **Alfons Khoukaz**, Münster University
28. **Pawel Klaja**, Jagiellonian University
29. **Bernd Krusche**, University of Basel
30. **Andrzej Kupsc**, Uppsala University
31. **Michael Lang**, Bonn University
32. **Mariola Lesiak**, FZ-Juelich
33. **Kalle Lindberg**, Stockholms University
34. **Hartmut Machner**, FZ-Juelich
35. **Ulf Meissner**, FZ-Juelich/Bonn Uni.
36. **Bagio Di Micco**, LN Frascati
37. **Pawel Moskal**, Jagiellonian University
38. **Kanzo Nakayama**, Athens University (USA)
39. **Marianna Nanova**, Giessen University
40. **Mauro Napsuciale**, University of Guanajuato

41. **Alexander Nikolaev** , Mainz University
42. **Jouni Niskanen**, University of Helsinki
43. **Robin Nissler**, Bonn University
44. **Walter Oelert**, FZ-Juelich
45. **Pilar Ortola** , *Secretary* , Universidad de Valencia
46. **Eulogio Oset**, Universidad de Valencia
47. **Michael Ostrick**, Mainz University
48. **Christian Pauly** , FZ-Juelich
49. **Teresa Pena**, IST Lisbon
50. **Henrik Pettersson** , Uppsala University
51. **Joanna Przerwa**, Jagiellonian University
52. **Luis Roca**, Universidad de Murcia
53. **Matthias Rost** ,Mainz University
54. **Susan Schadmand**, FZ-Juelich
55. **Karin Schönning**, Uppsala University
56. **Vitaliy Shklyar** , Giessen University
57. **Alexander Sibirtsev**, FZ-Juelich
58. **Tatiana Skorodko**, Tuebingen University
59. **Jerzy Smyrski**, Jagiellonian University
60. **Aleksander Starostin**, UCLA
61. **Joanna Stepaniak**, INS Warsaw
62. **Alfred Svarc**, Rudjer Boskovic Institute
63. **Per-Erik Tegner**, Stockholms University
64. **Ulla Tengblad**, Uppsala University
65. **Lothar Tiator**, Mainz University
66. **Roberto Versaci**, LN Frascati
67. **Colin Wilkin** ,University College London
68. **Magnus Wolke**, FZ-Juelich
69. **Irina Zartova**, Stockholms University
70. **Jozef Zlomanczuk**, Uppsala University